\documentclass[a4paper,11pt]{article}

\pdfoutput=1

\usepackage{amsmath,amssymb,mathtools}
\usepackage{color}
\usepackage{graphicx}
\usepackage{subfigure}
\usepackage{cite}
\usepackage[colorlinks=true,linkcolor=blue, citecolor=red, urlcolor=blue, bookmarks]{hyperref}
\usepackage{multirow,makecell}
\usepackage{textcomp}
\usepackage{wasysym}
\usepackage{ulem}

\usepackage[utf8]{inputenc}
\usepackage[T1]{fontenc}

\usepackage{simpler-wick} 

\usepackage[text={17.1cm,24.6cm},centering]{geometry} 

\numberwithin{equation}{section}

\def \be {\begin{equation}}
\def \ee {\end{equation}}
\def \ba {\begin{array}}
\def \ea {\end{array}}
\def \bea{\begin{eqnarray}}
\def \eea{\end{eqnarray}}
\def \nn {\nonumber}

\def \a {\alpha}
\def \b {\beta}
\def \g {\gamma}
\def \c {\gamma}
\def \G {\Gamma}
\def \d {\delta}
\def \D {\Delta}
\def \e {\epsilon}
\def \ve {\varepsilon}

\def \n {\nu}

\def \l {\lambda}
\def \lam {\lambda}

\def \s {\sigma}

\def \r {\rho}

\def \O {\Omega}
\def \th {\theta}

\def \vph {\varphi}
\def \vphi {\varphi}

\def \z {\zeta}

\def \cA {\mathcal A}
\def \cB {\mathcal B}
\def \cC {\mathcal C}
\def \cD {\mathcal D}

\def \cF {\mathcal F}

\def \cM {\mathcal M}

\def \p {\partial}

\def \f {\frac}

\def \lt {\left}
\def \rt {\right}

\def \lra {\leftrightarrow}
\def \sr {\sqrt}
\def \td {\tilde}

\def \inf {\infty}

\def \lag {\langle}
\def \rag {\rangle}

\def \ep {\mathrm{e}}
\def \ii {\mathrm{i}}

\def \tr {\textrm{tr}}
\def \Tr {{\textrm{Tr}}}

\def \and {{~\textrm{and}~}}

\def \univ {{\textrm{univ}}}

\def \per {\mathop{\textrm{per}}}

\begin{document}

\title{
\textbf{Corrections to universal R\'enyi entropy in quasiparticle excited states of quantum chains}
}
\author{
Jiaju Zhang$^{1,2}$ 
and
M. A. Rajabpour$^{3}$ 
}
\date{}
\maketitle
\vspace{-10mm}
\begin{center}
{\it
$^{1}$Center for Joint Quantum Studies and Department of Physics, School of Science, Tianjin University,\\135 Yaguan Road, Tianjin 300350, China\\\vspace{1mm}
$^{2}$SISSA and INFN, Via Bonomea 265, 34136 Trieste, Italy\\\vspace{1mm}
$^{3}$Instituto de Fisica, Universidade Federal Fluminense,\\
      Av. Gal. Milton Tavares de Souza s/n, Gragoat\'a, 24210-346, Niter\'oi, RJ, Brazil
}
\vspace{10mm}
\end{center}

\begin{abstract}
  We investigate the energy eigenstate R\'enyi entropy of generic bipartition in the fermionic, bosonic, and spin-1/2 XY chains. When the gap of the theory is large or all the momenta of the excited quasiparticles are large, the R\'enyi entropy takes a universal form, which is independent of the model, the quasiparticle momenta, and the subsystem connectedness. We calculate analytically the R\'enyi entropy in the extremely gapped limit and find different additional contributions to the universal R\'enyi entropy in various models. The corrections to the universal R\'enyi entropy cannot be neglected when the momentum differences of the excited quasiparticles are small. The R\'enyi entropy derived in the extremely gapped limit is still valid in the slightly gapped and even critical chains as long as all the momenta of the excited quasiparticles are large. In the case of double interval in the XY chain we find new universal results and their corrections. We call the result universal even though it is only valid for double interval in the spin-1/2 XY chain. In the case of the bosonic chain in the extremely massive limit we find analytically a novel formula for the R\'enyi entropy written as the permanent of a certain matrix. We support all of our analytical results with numerical calculations.
\end{abstract}

\baselineskip 18pt
\thispagestyle{empty}
\newpage


\tableofcontents

\section{Introduction}

In the last couple of decades different measures of quantum entanglement have been used extensively to study different phases of the matter \cite{Amico:2007ag,Eisert:2008ur,calabrese2009entanglement,Laflorencie:2015eck}.
For a pure state, a good measure of entanglement is the entanglement entropy, which is just the von Neumann entropy of the reduced density matrix (RDM) and can be calculated as the analytical continuation of the R\'enyi entropy.
The entanglement and R\'enyi entropies have been studied for various models in different situations.
In this paper we will focus on one-dimensional quantum chains, whose subsystems can be one single interval, double disjoint intervals, or more disjoint intervals.
Most of the early studies of the entanglement and R\'enyi entropies were focused on the ground state, for the single-interval cases see%
\cite{Bombelli:1986rw,Srednicki:1993im,Callan:1994py,Holzhey:1994we,peschel1999density,Peschel:1999DensityMatrices,%
Chung2000Densitymatrix,chung2001density,Vidal:2002rm,peschel2003calculation,Latorre:2003kg,jin2004quantum,Korepin:2004zz,Plenio:2004he,Calabrese:2004eu,%
Cramer:2005mx,Casini:2005rm,Casini:2005zv,Casini:2009sr,Calabrese:2009qy,peschel2009reduced,peschel2012special}
and for multi-interval cases see \cite{Casini:2004bw,Furukawa:2008uk,Casini:2008wt,Facchi:2008Entanglement,Caraglio:2008pk,Casini:2009vk,Calabrese:2009ez,Alba:2009ek,%
Igloi:2009On,Fagotti:2010yr,Headrick:2010zt,Calabrese:2010he,Alba:2011fu,Rajabpour:2011pt,Coser:2013qda,DeNobili:2015dla,Coser:2015dvp,%
Dupic:2017hpb,Ruggiero:2018hyl,Arias:2018tmw,Brightmore:2019Entanglement}.
One can also find various studies of the excited state entanglement and R\'enyi entropies in%
\cite{Alba:2009th,Alcaraz:2011tn,Berganza:2011mh,Pizorn2012Universality,Essler2013ShellFilling,Berkovits2013Twoparticle,Taddia:2013Entanglement,%
Storms2014Entanglement,Palmai:2014jqa,Calabrese:2014Entanglement,Molter2014Bound,Taddia:2016dbm,Castro-Alvaredo:2018dja,Castro-Alvaredo:2018bij,Murciano:2018cfp,%
Castro-Alvaredo:2019irt,Castro-Alvaredo:2019lmj,Jafarizadeh:2019xxc,Capizzi:2020jed,You:2020osa,Haque:2020Entanglement,Wybo:2020fiz}.
The excited states are natural generalizations of the ground state and are interesting for their own rights.
The study of the entanglement in the individual excited states can be helpful in better understanding the equilibrium and dynamical properties of many-body quantum systems that can be well approximated by collective excitations, known as quasiparticles.
They can be also useful in investigating the thermal states especially in lower temperatures. Moreover, the excite state R\'enyi entropy can be also measured in experiments following the setup in \cite{Jurcevic:2014qfa,Jurcevic:2015inc}.

It is usually more desirable to look into the universal behaviors of the entanglement and R\'enyi entropies.
The ground state entanglement entropy in a gapped model is usually proportional to the area of the subsystem \cite{Eisert:2008ur}, while the ground sate entanglement entropy in a critical quantum chain whose continuum limit gives a two-dimensional conformal field theory takes a universal logarithmic formula  \cite{Holzhey:1994we,Vidal:2002rm,jin2004quantum,Korepin:2004zz,Calabrese:2004eu}.
In a highly excited state the entanglement entropy usually behaves like the thermal entropy and has the form of the volume law \cite{Deutsch:2013Microscopic,Santos:2012Weak,Beugeling:2015Global,Garrison:2015lva}.
The average entanglement entropy in the fermionic chains is also conjectured to take a particular universal volume law \cite{Page:1993df,Vidmar:2017uux,Vidmar:2017pak,Huang:2019dxk,Vidmar:2018rqk,Hackl:2018tyl}.
Recently, a novel form of universal entanglement and R\'enyi entropies in the excited state of quasiparticles was discovered in \cite{Castro-Alvaredo:2018dja,Castro-Alvaredo:2018bij,Castro-Alvaredo:2019irt,Castro-Alvaredo:2019lmj}, see also for earlier partial results \cite{Pizorn2012Universality,Berkovits2013Twoparticle,Molter2014Bound}, which we will revisit in this work.
We have reported some of the results in the letter \cite{Zhang:2020vtc}, and will present the details with additional novel results in this work.

\begin{figure}[tp]
  \centering
  \includegraphics[width=0.6\textwidth]{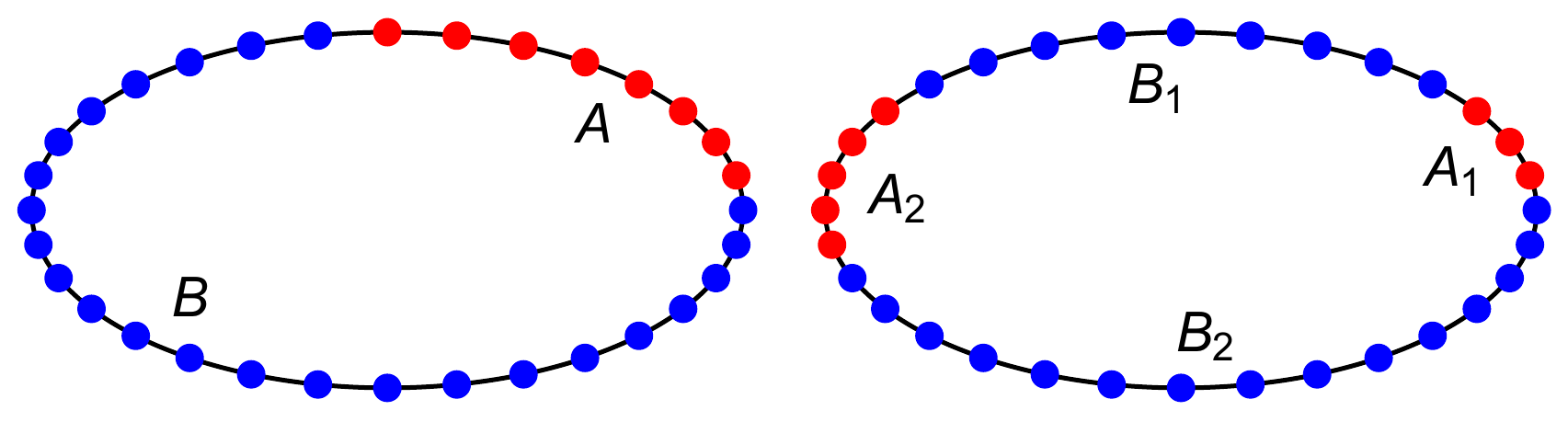}\\
  \caption{A subsystem made of a single interval (left) and a subsystem made of double disjoint intervals (right) on a circular chain of $L$ sites.
  For the single interval $A$ the length is $|A|=\ell$, and it is convenient to define $x=\f{\ell}{L}$.
  For the double interval $A=A_1\cup A_2$ the lengths are $|A_1|=\ell_1$, $|A_2|=\ell_2$, $|B_1|=d_1$, $|B_2|=d_2$, and one can define the parameters $x_1=\f{\ell_1}{L}$, $x_2=\f{\ell_2}{L}$, $y_1=\f{d_1}{L}$, $y_2=\f{d_2}{L}$, $x=x_1+x_2$, $y=y_1+y_2$.}
  \label{subsystems}
\end{figure}

In this paper we consider the R\'enyi entropy of a subsystem $A$ of $\ell$ sites on a circular chain of $L$ sites in the scaling limit $L\to+\inf$ and $\ell\to+\inf$ with finite ratio $x=\f{\ell}{L}$.
The subsystem can be consecutive or be composed of double or more disjoint intervals as shown in figure~\ref{subsystems}.
The whole system has the density matrix $\r_K=|K\rag\lag K|$. Then one can integrate out the degrees of freedom of the complement of $A$, which we denote by $B=\bar A$, and obtain the RDM $\r_{A,K}=\tr_{B}\r_K$.
Finally, the R\'enyi entropy of the state $|K\rag$ is
\be
S_{A,K}^{(n)} = -\f{1}{n-1} \log \tr_A \r_{A,K}^n.
\ee
As in \cite{Alcaraz:2011tn,Berganza:2011mh}, we will use $\cF_{A,K}^{(n)}$ to denote the difference between $S_{A,K}^{(n)}$ and the R\'enyi entropy $S_{A,G}^{(n)}$ for the ground state $|G\rag$ as
\be
S_{A,K}^{(n)} = S_{A,G}^{(n)} - \f{1}{n-1} \log \cF_{A,K}^{(n)}.
\ee
Explicitly, one can write
\be
\cF_{A,K}^{(n)} = \f{\tr_A\r_{A,K}^n}{\tr_A\r_{A,G}^n}.
\ee

In an integrable many-body quantum system, an excited state can be written in a simple form with respect to the excited quasiparticles.
A general quasiparticle excited state takes the form $|K\rag = |k_1^{r_1}k_2^{r_2}\cdots k_s^{r_s}\rag$, where the quasiparticle with momentum $k_i$ is excited $r_i$ times with $i=1,2,\cdots,s$. The total number of exited particles are $R=\sum_{i=1}^s r_i$.
We use $k_i$ which is an integer or a half-integer as the momentum. This is different from the convention in \cite{Castro-Alvaredo:2018dja,Castro-Alvaredo:2018bij,Castro-Alvaredo:2019irt,Castro-Alvaredo:2019lmj}, where $p_i=\f{2\pi k_i}{L}$ is used as the momentum. Actually, $k_i$ is the total number of waves in the circular chain, and $p_i$ is the physical momentum.
The universal R\'enyi entropy of the state $|k_1^{r_1}k_2^{r_2}\cdots k_s^{r_s}\rag$ is \cite{Castro-Alvaredo:2018dja,Castro-Alvaredo:2018bij,Castro-Alvaredo:2019irt,Castro-Alvaredo:2019lmj}
\be \label{cFAk1r1k2r2cdotsksrsnsup}
\cF_{A,p_1^{r_1}p_2^{r_2}\cdots p_s^{r_s}}^{(n),\univ} = \prod_{i=1}^s \Big\{ \sum_{p=0}^{r_i} [ C_{r_i}^p x^p (1-x)^{{r_i}-p} ]^n \Big\},
\ee
where $C_{r_i}^p$ is the binomial coefficient.
The above equation is derived in the limit that $L\to+\inf$, $\ell\to+\inf$, $k_i\to+\inf$ with finite fixed $x=\f{\ell}{L}$, $p_i=\f{2\pi k_i}{L}$.%
\footnote{We thank Olalla Castro-Alvaredo, Cecilia De Fazio, Benjamin Doyon and Istv\'an Sz\'ecs\'enyi for  explaining to us the precise limit they have used in \cite{Castro-Alvaredo:2018dja,Castro-Alvaredo:2018bij,Castro-Alvaredo:2019irt}.}
Note that we use $\cF_{A,p_1^{r_1}p_2^{r_2}\cdots p_s^{r_s}}^{(n),\univ}$ instead of $\cF_{A,k_1^{r_1}k_2^{r_2}\cdots k_s^{r_s}}^{(n),\univ}$ to denote the universal R\'enyi entropy to emphasize the different limit that was used in \cite{Castro-Alvaredo:2018dja,Castro-Alvaredo:2018bij,Castro-Alvaredo:2019irt}.
The universal R\'enyi entropy is valid when either the correlation length $\f{1}{\D}$ of the model or the maximal de Broglie wavelength of the quasiparticles is much smaller than the sizes of the subsystems but one does not need to require both, i.e.\ that it requires \cite{Castro-Alvaredo:2018dja,Castro-Alvaredo:2018bij,Castro-Alvaredo:2019irt}
\be \label{conditionCDDS}
\min\Big[ \f1\D, \max_i\Big(\f{L}{|k_i|}\Big) \Big] \ll \min( \ell, L-\ell ).
\ee
In other words, one needs either the gap $\D$ of the model is large or all the momenta of the excited quasiparticles are large.
The universal R\'enyi entropy is independent of the model, the momenta of the quasiparticles, and the subsystem connectedness.

In this paper we will calculate the R\'enyi entropy of single and double intervals in the quasiparticle excited state $|K\rag = |k_1^{r_1}k_2^{r_2}\cdots k_s^{r_s}\rag$ with general momenta $k_i$ in the fermionic, bosonic and XY chains.
We calculate analytically the R\'enyi entropy in the single-particle, double-particle and triple-particle states in the extremely gapped limit by writing the excited states in terms of subsystem excitations, which we call subsystem mode method, and find different additional contributions to the universal R\'enyi entropy (\ref{cFAk1r1k2r2cdotsksrsnsup}) in different models, and the new results with correction terms match perfectly with the numerical results calculated from other methods.
The corrections to the universal R\'enyi entropy cannot be neglected when the momentum differences of the excited quasiparticles are small and are negligible when all the momentum differences are large
\be \label{conditionZR}
\min_{i_1 \neq i_2} | k_{i_1} - k_{i_2} | \gg 1.
\ee
We also compare the  R\'enyi entropy with additional correction terms derived in the extremely gapped limit with the numerical results in the corresponding slightly gapped and critical models and find that the new R\'enyi entropy is still valid as long as all the momenta of the excited quasiparticles are large.
The validity of the universal R\'enyi entropy (\ref{cFAk1r1k2r2cdotsksrsnsup}) found in \cite{Castro-Alvaredo:2018dja,Castro-Alvaredo:2018bij,Castro-Alvaredo:2019irt,Castro-Alvaredo:2019lmj} requires both (\ref{conditionCDDS}) and (\ref{conditionZR}), while the validity of the R\'enyi entropy with corrections we report in this paper only requires the condition (\ref{conditionCDDS}).
In a slightly gapped or critical model, the validity of the universal R\'enyi entropy requires that both all the quasiparticle momenta and all the momentum differences are large, while the validity of the R\'enyi entropy with additional corrections only requires all the momenta are large.

The remaining part of the paper is arranged as follows:
In section~\ref{secSum} we summarize the main results of this paper.
In section~\ref{secF} we investigate the single-interval and double-interval R\'enyi entropies in the single-particle, double-particle, and triple-particle states in the fermionic chain.
In section~\ref{secB} we consider the R\'enyi entropies in the bosonic chain.
In section~\ref{secXY} we consider the R\'enyi entropies in the XY chain.
We conclude with discussions in section~\ref{secDis}.
In appendix~\ref{appMWF}, we review the wave function method to calculate the R\'enyi entropy in the bosonic chain and we also further adapt the method to the extremely gapped limit.

\section{Summary of results} \label{secSum}

In this section, we summarize the main results of this paper which will be presented in full detail in sections~\ref{secF}, \ref{secB} and \ref{secXY}.

In section~\ref{secF}, we investigate the R\'enyi entropy in the excited states of quasiparticles in the fermionic chain.
We focus on the cases of the single interval and double interval.
The generalization to multiple intervals is easy and we will not show details in this paper.
We consider the single-particle state $|k\rag$, double-particle state $|k_1k_2\rag$ and triple-particle state $|k_1k_2k_3\rag$.
Analytically we calculate the R\'enyi entropy $\cF_{A,K}^{(n)}$, and we find the additional corrections $\d\cF_{A,K}^{(n)}$ to the universal R\'enyi entropy $\cF_{A,P}^{(n),\univ}$ (\ref{cFAk1r1k2r2cdotsksrsnsup})
\be
\cF_{A,K}^{(n)} = \cF_{A,P}^{(n),\univ} + \d \cF_{A,K}^{(n)}.
\ee
Explicitly, we get $\cF_{A,k}^{(n)}=\cF_{A,p}^{(n),\univ}$ with general $n$, for which there is no additional contribution.
We also get the R\'enyi entropy $\cF_{A,k_1k_2}^{(n)}$ with $n=2,3,\cdots,7$ and $\cF_{A,k_1k_2k_3}^{(n)}$ with $n=2,3,4,5$.
We compare the universal R\'enyi entropy and the corrected R\'enyi entropy with the numerical results calculated using the correlation functions%
\cite{chung2001density,Vidal:2002rm,peschel2003calculation,Latorre:2003kg,Alba:2009th,Alcaraz:2011tn,Berganza:2011mh}.
We find perfect matches between the corrected R\'enyi entropy with the numerical results.
In the limit of small momentum differences the additional terms are significant and cannot be neglected, while in the limit of large momentum differences the additional terms are negligible.
We check that R\'enyi entropy with corrections is still valid for slightly gapped and even critical fermionic chains as long as all the quasiparticle momenta are large.

In section~\ref{secB}, we investigate the single-interval and double-interval R\'enyi entropies in the multi-particle state with the same momenta $|k^r\rag$, double-particle state $|k_1k_2\rag$, triple-particle state $|k_1^2k_2\rag$ and triple-particle state $|k_1k_2k_3\rag$ in the bosonic chain.
In the extremely gapped limit, we calculate the analytical R\'enyi entropy using two different methods: one is by writing the excited states in terms of subsystem excitations, i.e.\ the subsystem mode method, similar to that in the fermionic chain, and the other is the wave function method.
Explicitly, we get $\cF_{A,k^r}^{(n)}=\cF_{A,p^r}^{(n),\univ}$ with general $n$, for which there is no additional contribution.
We also get the analytical R\'enyi entropy $\cF_{A,k_1k_2}^{(n)}$ with $n=2,3,\cdots,7$, $\cF_{A,k_1^2k_2}^{(n)}$ with $n=2,3,\cdots,7$ and $\cF_{A,k_1k_2k_3}^{(n)}$ with $n=2,3,4,5$, which are still valid in the slightly gapped bosonic chains as long as all the quasiparticle momenta are large.

In section~\ref{secXY}, we investigate the single-interval and double-interval R\'enyi entropies in the single-particle state $|k\rag$ and double-particle state $|k_1k_2\rag$ in the XY chain.
The XY chain can be mapped to the fermionic chain that we investigate in section~\ref{secF}, but they have different local degrees of freedoms.
The single-interval R\'enyi entropy in the XY chain is the same as that in the fermionic chain, while the double-interval R\'enyi entropy is very different.
There is no additional contribution to the universal double-interval R\'enyi entropy in the single-particle state $|k\rag$.
It is remarkable that we find a new universal double-interval R\'enyi entropy in the double-particle state $|k_1k_2\rag$, for which there are also additional corrections when the momentum difference $|k_1-k_2|$ is small.
We find that the analytical R\'enyi entropy, which is the new universal R\'enyi entropy plus the additional corrections terms, shows perfect match with the numerical results in the extremely gapped limit.
When the momentum difference $|k_1-k_2|$ is large, the R\'enyi entropy approaches the new universal R\'enyi entropy instead of the old one.
The double-interval R\'enyi entropy is still valid in the slightly gapped XY chains as long as all the quasiparticle momenta are large.
But in the critical XY chains we find that all the known analytical results do not match the numerical ones, for which we do not have a good explanation.

The derived corrected R\'enyi entropy depends on the model, the quasiparticle momenta, and the subsystem connectedness.
The corrected R\'enyi entropy depends on the momenta only through the momentum differences.
It is exact in the extremely gapped limit of the fermionic, bosonic, and XY chains for arbitrary sizes of the whole system and the subsystem.
In the fermionic and XY chains, the universal R\'enyi entropy derived in the extremely gapped limit is still exactly valid in the critical but non-relativistic model with $\g=0,\l=1$.

One highlight of the paper is that we establish the determinant formula (\ref{TheFormulaF}) in the extremely gapped fermionic chain and the permanent formula (\ref{TheFormulaB}) in the extremely gapped bosonic chain, which are very efficient for both analytical and numerical calculations of the R\'enyi entropy in the quasiparticle excited state.

\section{Fermionic chain}\label{secF}

We consider the Hamiltonian of fermionic chain of $L$ sites
\be \label{fermionicchain}
H = \sum_{j=1}^L \Big[ \lam \Big( a_j^\dag a_j - \f12 \Big) - \f12 ( a_j^\dag a_{j+1} + a_{j+1}^\dag a_j ) - \f{\g}{2} ( a_j^\dag a_{j+1}^\dag + a_{j+1} a_j ) \Big],
\ee
with periodic or antiperiodic boundary conditions for the spinless fermions $a_j$, $a_j^\dag$.
It can be diagonalized as
\be
H = \sum_k \ve_k \Big( c_k^\dag c_k - \f12 \Big), ~~
\ve_k = \sr{ \Big(\l - \cos\f{2\pi k}{L}\Big)^2 + \g^2 \sin^2\f{2\pi k}{L} },
\ee
by the successive Fourier transformation and Bogoliubov transformation \cite{Lieb:1961fr,katsura1962statistical,pfeuty1970one}
\be
b_k = \f{1}{\sr{L}}\sum_{j=1}^L\ep^{\ii j \vph_k}a_j, ~~
b_k^\dag = \f{1}{\sr{L}}\sum_{j=1}^L\ep^{-\ii j \vph_k}a_j^\dag,
\ee
\be
c_k = b_k \cos\f{\th_k}{2} + \ii b_{-k}^\dag \sin\f{\th_k}{2}, ~~
c_k^\dag = b_k^\dag \cos\f{\th_k}{2} - \ii b_{-k} \sin\f{\th_k}{2},
\ee
with the definitions
\be \label{vphikepiithk}
\vph_k = \f{2\pi k}{L}, ~~
\ep^{\ii\th_k}=\f{\l-\cos\vph_k+\ii\g\sin\vph_k}{\ve_k}.
\ee
In this paper we only consider the cases that $L$'s are even integers, and we also only consider the states in the  Neveu-Schwarz (NS) sector, i.e.\ the antiperiodic boundary conditions for the fermionic modes $a_j$, $a_j^\dag$.
In fact, all the results we obtain in this paper are still valid for the states in the Ramond sector too.
We have the half-integer momenta
\be
k =\f{1-L}{2}, \cdots, -\f12, \f12, \cdots, \f{L-1}{2}.
\ee
The ground state $|G\rag$ is annihilated by all the lowering operators
\be
c_k|G\rag=0,
\ee
and the excited states are generated by applying the raising operators with different momenta on the ground state
\be \label{XYk1k2cdotsksrag}
|k_1k_2\cdots k_s\rag = c_{k_1}^\dag c_{k_2}^\dag \cdots c_{k_s}^\dag |G\rag.
\ee

We consider the extremely gapped limit $\l\to+\inf$ of the fermionic chain. The Hamiltonian becomes
\be
H = \l \sum_{j=1}^L \Big( a_j^\dag a_j - \f12 \Big),
\ee
and the fermions $a_j,a_j^\dag$ at different positions decouple from each other.
The Bogoliubov angle is vanishing $\th_k=0$, and we have
\be
c_k = \f{1}{\sr{L}}\sum_{j=1}^L\ep^{\ii j \vph_k}a_j, ~~
c_k^\dag = \f{1}{\sr{L}}\sum_{j=1}^L\ep^{-\ii j \vph_k}a_j^\dag.
\ee
The ground state is also annihilated by all the local lowering operators $a_j$
\be
a_j|G\rag=0, ~ j=1,2,\cdots,L.
\ee

\subsection{Single interval}

\begin{figure}[tp]
  \centering
  \includegraphics[height=0.3\textwidth]{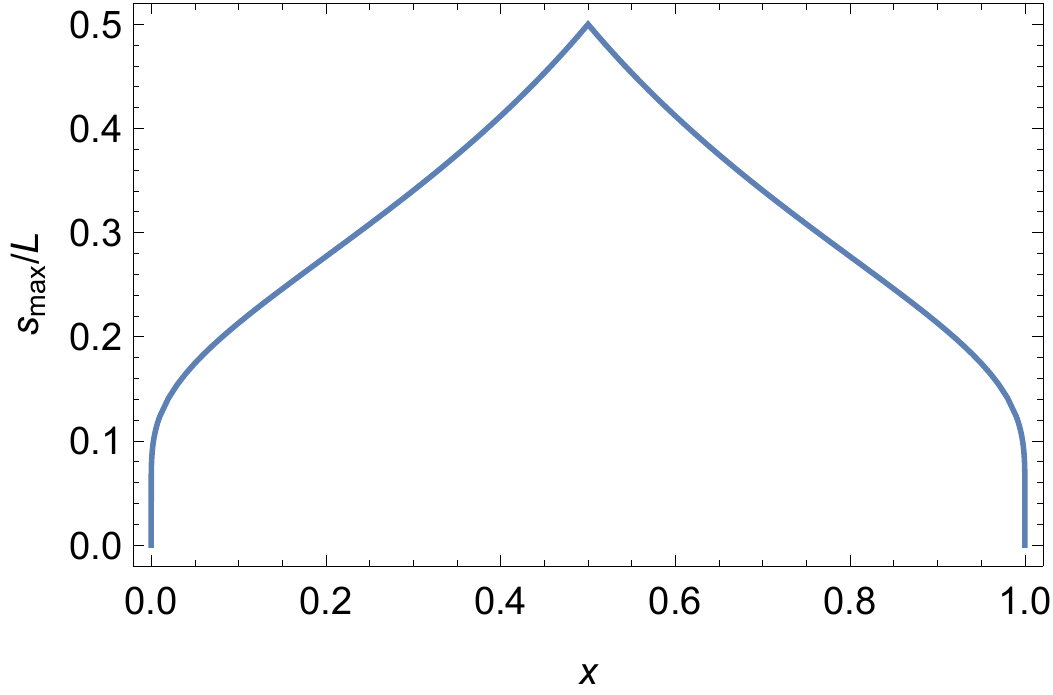}\\
  \caption{The necessary upper bound (\ref{upperbound}) of the number of the excited quasiparticles for the universal entanglement entropy (\ref{SAuniv}) to be valid.}
  \label{upperboundplot}
\end{figure}

We consider an interval with $\ell$ consecutive sites $A=[1,\ell]$ on the periodic fermionic chain with $L$ sites.
In the extremely gapped limit $\l\to+\inf$, the ground state is just a direct product state
\be
|G\rag = |G_A\rag |G_B\rag,
\ee
and the ground state R\'enyi entropy is vanishing
\be
S_{A,G}^{(n)} = 0.
\ee
In the limit $L\to+\inf$, $\ell\to+\inf$, $k_i\to+\inf$ with fixed $x=\f\ell{L}$, $p_i=\f{2\pi k_i}{L}$, the R\'enyi entropy for the excited state $|k_1k_2\cdots k_s\rag$ takes the universal form \cite{Castro-Alvaredo:2018dja,Castro-Alvaredo:2018bij,Castro-Alvaredo:2019irt,Castro-Alvaredo:2019lmj}
\be \label{FAuniv}
\cF_{A,p_1p_2\cdots p_s}^{(n),\univ} = [ x^n + (1-x)^n ]^s.
\ee
The universal R\'enyi entropy leads to the universal entanglement entropy
\be \label{SAuniv}
S_{A,p_1p_2\cdots p_s}^\univ = s [ - x \log x - (1-x)\log(1-x) ].
\ee
In general this does not apply to an excited state with arbitrary $s$ as there is an upper bound for the entanglement entropy
\be
S_{A,p_1p_2\cdots p_s}^\univ \leq \min(\ell,L-\ell) \log 2.
\ee
It gives an upper bound of the number of the excited quasiparticles
\be \label{upperbound}
\f{s}{L} \leq \f{\min(x,1-x)\log2}{- x \log x - (1-x)\log(1-x)},
\ee
which we show in figure~\ref{upperboundplot}.
This is not surprising, since as stated in \cite{Castro-Alvaredo:2018dja} there should be finite number of excited quasiparticles for the universal entanglement and R\'enyi entropies to be valid.
In the following, we will relax the constraint for the momenta $k_i$ and find the nontrivial additional contributions $\d \cF_{A,k_1k_2\cdots k_s}^{(n)}$ to the universal R\'enyi entropy
\be
\cF_{A,k_1k_2\cdots k_s}^{(n)} = \cF_{A,p_1p_2\cdots p_s}^{(n),\univ}  + \d \cF_{A,k_1k_2\cdots k_s}^{(n)}.
\ee

The entanglement and R\'enyi entropies in the fermionic chain can be calculated numerically using the two-point correlation functions \cite{chung2001density,Vidal:2002rm,peschel2003calculation,Latorre:2003kg,Alba:2009th,Alcaraz:2011tn,Berganza:2011mh}.
We denote the general excited state $|k_1k_2\cdots k_s\rag$ by the set of excited momenta $K=\{k_1,k_2,\cdots,k_s\}$.
One can define the Majorana modes as
\be
d_{2j-1} = a_j + a_j^\dag, ~~
d_{2j} = \ii ( a_j - a_j^\dag ).
\ee
For the interval $A=[1,\ell]$ on the fermionic chain in excited state $|K\rag$, one defines the $2\ell \times 2\ell$ correlation matrix
\be
\lag d_{m_1} d_{m_2} \rag_K = \d_{m_1m_2} + \G^K_{m_1m_2}, ~
m_1,m_2 = 1,2,\cdots,2\ell.
\ee
The $\G^K$ matrix which is antisymmetric and purely imaginary has entries
\be
\G^K_{2j_1-1,2j_2-1}=\G^K_{2j_1,2j_2}=f_{j_2-j_1}^K, ~~
\G^K_{2j_1-1,2j_2}=-\G^K_{2j_2,2j_1-1}=g_{j_2-j_1}^K,
\ee
where we defined the functions
\be
f^K_j = -\f{2\ii}{L} \sum_{k \in K}\sin(j \vph_k), ~~
g^K_j = -\f{\ii}{L} \sum_{k \notin K} \cos( j\vph_k-\th_k ) + \f{\ii}{L} \sum_{k \in K} \cos( j\vph_k-\th_k ).
\ee
In terms of the $2\ell$ eigenvalues $\pm\g^K_j$, $j=1,2,\cdots,\ell$, of the correlation matrix $\G^K$, the entanglement and R\'enyi entropies of the length $\ell$ interval in state $K$ is
\bea
&& S_{A,K} = \sum_{j=1}^{\ell} \Big( - \f{1+\g_j^K}{2}\log\f{1+\g_j^K}{2} - \f{1-\g_j^K}{2}\log\f{1-\g_j^K}{2}\Big), \nn\\
&& S_{A,K}^{(n)} = -\f{1}{n-1}\sum_{j=1}^{\ell} \log \Big[ \Big(\f{1+\g_j^K}{2}\Big)^n + \Big(\f{1-\g_j^K}{2}\Big)^n \Big].
\eea
In the extremely gapped limit $\l\to+\inf$, there is vanishing Bogoliubov angle $\th_k=0$. One can also define the $\ell\times\ell$ correlation matrix $C_A^K$ with the entries
\be
[C_A^K]_{j_1j_2}=\lag c^\dag_{j_1} c_{j_2} \rag_K = h_{j_2-j_1}^K, ~
j_1,j_2 = 1,2,\cdots,\ell,
\ee
and the function
\be
h_j^K = \f1L \sum_{k\in K} \ep^{\ii j\vphi_k},
\ee
and calculate the entanglement and R\'enyi entropies from the eigenvalues $\n^K_j$, $j=1,2,\cdots,\ell$, of the correlation matrix $C_A^K$ as
\bea \label{SAKSAKn}
&& S_{A,K} = \sum_{j=1}^{\ell} \big[ - \n^K_j \log \n^K_j - (1-\n^K_j) \log (1-\n^K_j) \big], \nn\\
&& S_{A,K}^{(n)} = -\f{1}{n-1}\sum_{j=1}^{\ell} \log \big[ (\n^K_j)^n + (1-\n^K_j)^n \big].
\eea

For the following analytical calculations of the R\'enyi entropy it is convenient to define the subsystem modes in the extremely gapped limit
\bea
&& c_{A,k} = \f{1}{\sr{L}}\sum_{j\in A} \ep^{\ii j \vph_k}a_j, ~~
   c_{A,k}^\dag = \f{1}{\sr{L}}\sum_{j\in A} \ep^{-\ii j \vph_k}a_j^\dag, \nn\\
&& c_{B,k} = \f{1}{\sr{L}}\sum_{j\in B} \ep^{\ii j \vph_k}a_j, ~~
   c_{B,k}^\dag = \f{1}{\sr{L}}\sum_{j\in B} \ep^{-\ii j \vph_k}a_j^\dag.
\eea
There are anti-commutation relations
\be
\{ c_{A,k}, c_{A,k}^\dag \} = x, ~~
\{ c_{B,k}, c_{B,k}^\dag \} = 1-x,
\ee
and for $k_1\neq k_2$ we have
\be
\{ c_{A,k_1}, c_{A,k_2}^\dag \} = - \{ c_{B,k_1}, c_{B,k_2}^\dag \} = \a_{k_1-k_2},
\ee
with
\be \label{alphakdefinition}
\a_k = \f{1}{L} \sum_{j=1}^\ell \ep^{\f{2\pi\ii j k}{L}} = \ep^{\f{\pi\ii(\ell+1)k}{L}} \f{\sin\f{\pi\ell k}{L}}{L\sin\f{\pi k}{L}}.
\ee

\subsubsection{Single-particle state $|k\rag$} \label{subsubsecSPS}

In the single-particle state $|k\rag$ of the extremely gapped fermionic chain, we write the density matrix of the whole system in terms of the subsystem modes
\be
\r_k = ( c_{A,k}^\dag + c_{B,k}^\dag ) |G\rag \lag G| ( c_{A,k} + c_{B,k} ).
\ee
Tracing out the degrees of freedom of $B$, we get the RDM
\be
\r_{A,k} = c_{A,k}^\dag |G_A\rag\lag G_A| c_{A,k} + \lag c_{B,k} c_{B,k}^\dag \rag_{G}   |G_A\rag\lag G_A|.
\ee
We have used $\tr_B(c_{A,k}^\dag|G\rag \lag G|c_{B,k})=\tr_B(c_{B,k}^\dag|G\rag \lag G|c_{A,k})=0$.
Then we get
\be
\tr_A\r_{A,k}^n = \lag c_{A,k} c_{A,k}^\dag \rag_{G}^n + \lag c_{B,k} c_{B,k}^\dag \rag_{G}^n = x^n+(1-x)^n.
\ee
There is no additional contribution to the universal R\'enyi entropy in the single-particle state $|k\rag$.
The analytical results match the numerical ones, which has been checked in \cite{Castro-Alvaredo:2018dja,Castro-Alvaredo:2018bij}, and we will not show the details here.

Since the subsystem modes play a crucial role in the analytical calculations we call the procedure the subsystem mode method.

\subsubsection{Double-particle state $|k_1k_2\rag$} \label{singleintervalk1k2}

In the double-particle state $|k_1k_2\rag$ with general momenta $k_1,k_2$, we write the density matrix of the whole system as
\be
\r_{k_1k_2} = ( c_{A,k_1}^\dag + c_{B,k_1}^\dag ) ( c_{A,k_2}^\dag + c_{B,k_2}^\dag ) |G\rag \lag G| ( c_{A,k_2} + c_{B,k_2} ) ( c_{A,k_1} + c_{B,k_1} ),
\ee
then we get the RDM
\bea \label{rAk1k2Fermion}
&& \r_{A,k_1k_2} = c_{A,k_1}^\dag c_{A,k_2}^\dag |G_A\rag\lag G_A| c_{A,k_2} c_{A,k_1}
              + \lag c_{B,k_1} c_{B,k_1}^\dag \rag_{G} c_{A,k_2}^\dag |G_A\rag\lag G_A| c_{A,k_2} \nn\\
&& \phantom{\r_{A,k_1k_2} =}
              + \lag c_{B,k_2} c_{B,k_2}^\dag \rag_{G} c_{A,k_1}^\dag |G_A\rag\lag G_A| c_{A,k_1}
              - \lag c_{B,k_1} c_{B,k_2}^\dag \rag_{G} c_{A,k_1}^\dag |G_A\rag\lag G_A| c_{A,k_2} \nn\\
&& \phantom{\r_{A,k_1k_2} =}
              - \lag c_{B,k_2} c_{B,k_1}^\dag \rag_{G} c_{A,k_2}^\dag |G_A\rag\lag G_A| c_{A,k_1}
              + \lag c_{B,k_2} c_{B,k_1} c_{B,k_1}^\dag c_{B,k_2}^\dag \rag_{G} |G_A\rag\lag G_A|.
\eea
The sign of each term can be determined by $\tr \r_{k_1k_2} = \tr_A \r_{A,k_1k_2}$.
Note the minus signs due to the anti-commutation relations of the modes.
Then one can calculate $\tr_A\r_{A,k_1k_2}^n$ with $n=2,3,4,5,6,7$.
Explicitly, we get the nontrivial additional contributions to the universal R\'enyi entropy (\ref{FAuniv})
\be \label{ours}
\d\cF_{A,k_1k_2}^{(2)} = 8 x ( 1 - x) \a_{12}^2 + 4 \a_{12}^4,
\ee
\be
\d\cF_{A,k_1k_2}^{(3)} = -3 (1 - 6 x + 6 x^2) \a_{12}^2 + 9 \a_{12}^4,
\ee
\bea
&& \d\cF_{A,k_1k_2}^{(4)} = - 4 (1 - 6 x + 12 x^2 - 16 x^3 + 18 x^4 - 12 x^5 + 4 x^6) \a_{12}^2 \nn\\
&& \phantom{\d\cF_{A,k_1k_2}^{(4)} =}
                         + 8 (1 + 3 x^2 - 6 x^3 + 3 x^4) \a_{12}^4
                         + 8 ( 1 + 2 x - 2 x^2) \a_{12}^6
                         + 4 \a_{12}^8,
\eea
\bea
&& \d\cF_{A,k_1k_2}^{(5)} = - 5 (1 - 8 x + 28 x^2 - 60 x^3 + 80 x^4 - 60 x^5 + 20 x^6) \a_{12}^2 \nn\\
&& \phantom{\d\cF_{A,k_1k_2}^{(5)} =}
                         + 10 (1 - 5 x + 20 x^2 - 30 x^3 + 15 x^4) \a_{12}^4
                         + 100 x (1 - x) \a_{12}^6
                         + 25 \a_{12}^8,
\eea
as well as $\d\cF_{A,k_1k_2}^{(n)}$ with $n=6,7$ which we will not show in this paper.
In the above expressions we have defined the shorthand $\a_{12}\equiv|\a_{k_1-k_2}|$ with the function $\a_k$ defined in (\ref{alphakdefinition}).
We see that the results become increasingly complex for higher R\'enyi indices $n$.

\begin{figure}[t]
  \centering
  \includegraphics[height=0.3\textwidth]{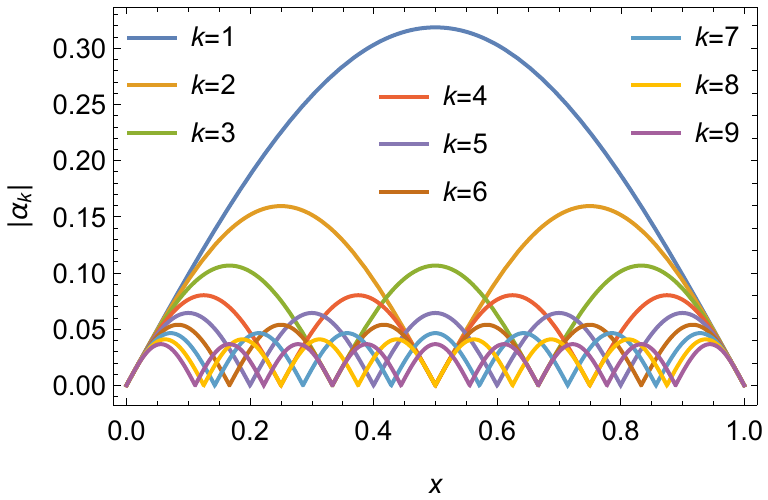}\\
  \caption{The absolute value of the function $\a_k$ (\ref{alphakdefinition}) decreases with the increase of $k$. We have set $L=64$.}
  \label{alphakplot}
\end{figure}

\begin{figure}[t]
  \centering
  \includegraphics[height=0.6\textwidth]{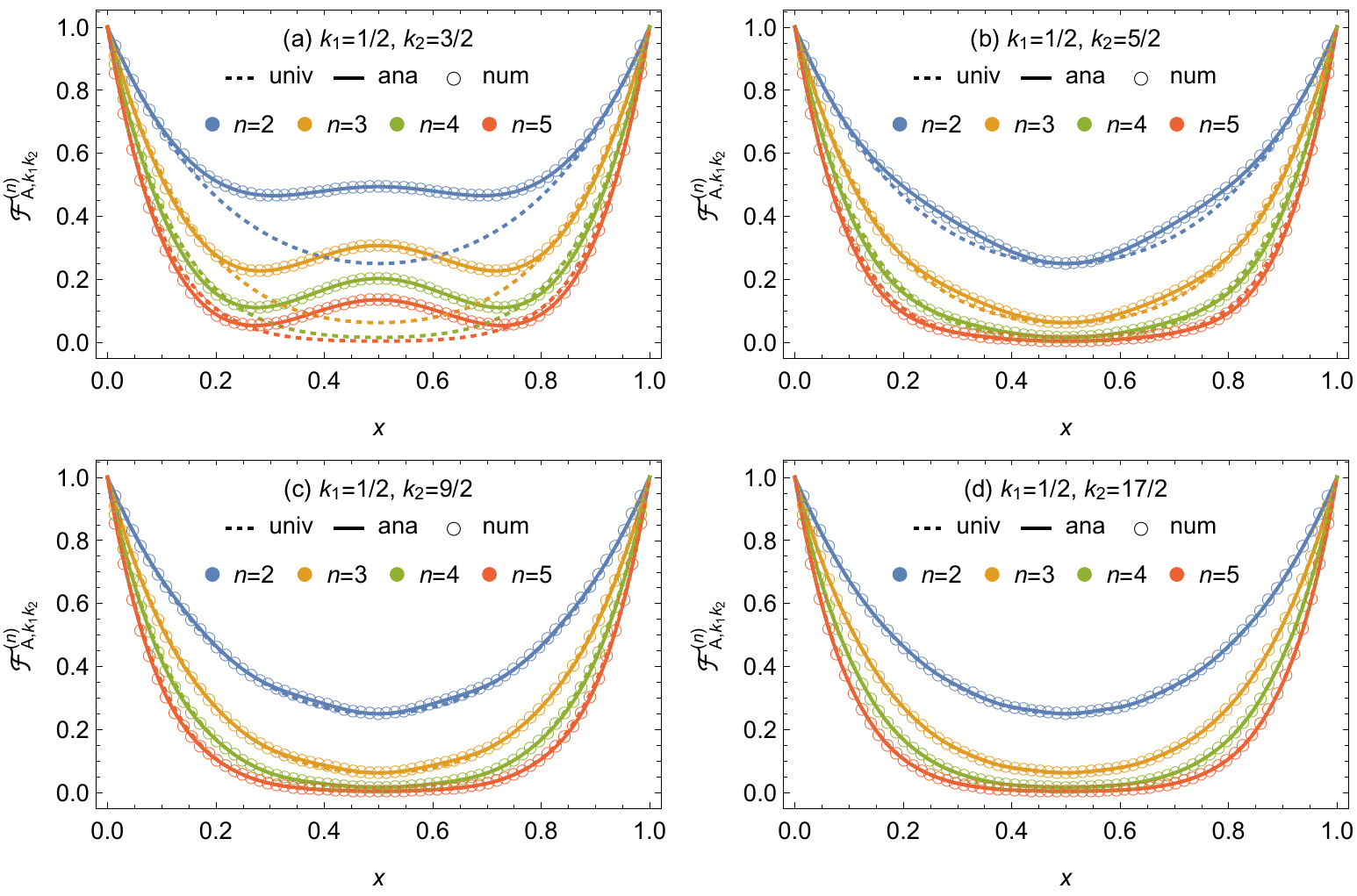}\\
  \caption{The universal R\'enyi entropy (``univ'', dotted lines), and the analytical (``ana'', solid lines) and numerical (``num'', empty circles) results of the single-interval R\'enyi entropy in the double-particle state $|k_1k_2\rag$ in the extremely gapped fermionic chain. We use different colors to represent different R\'enyi indices $n$. We have set $\l=+\inf$, $L=64$.}\label{FermionAk1k2}
\end{figure}

We plot the absolute value of the function $\a_k$ (\ref{alphakdefinition}) in figure~\ref{alphakplot}, and we see that it is nonnegligible for a small $k$ but it goes to zero rapidly by increasing the $k$. Note that we have already taken the limit $L\to+\inf$, $\ell\to+\inf$ with fixed $x=\f{\ell}{L}$.
We compare the results of the universal R\'enyi entropy, the analytical results with additional correction terms and the numerical results for the state $|k_1k_2\rag$ in figure~\ref{FermionAk1k2}.
We find perfect matches of the new R\'enyi entropy with additional corrections with the numerical results.
For a small momentum difference $|k_1-k_2|$ the additional terms are significant.
For a large momentum difference $|k_1-k_2|$ the additional terms can be neglected.

In fact, the result $\cF_{A,k_1k_2}^{(2)}=\cF_{A,p_1p_2}^{(2),\univ}+\d \cF_{A,k_1k_2}^{(2)}$ with the universal R\'enyi entropy (\ref{FAuniv}) and the additional corrections (\ref{ours}) could be retrieved in the supplementary material of \cite{Castro-Alvaredo:2018dja} by relaxing the constraint for the quasiparticle momenta.
We identify the parameters $r,\vphi,p_1,p_2$ in the supplementary material of \cite{Castro-Alvaredo:2018dja} as $r=x$, $\vphi=\pi$, $p_1=\f{2\pi k_1}{L}$, $p_2=\f{2\pi k_2}{L}$. Note that in our convention $k_1-k_2$ is a general integer that can be either small or large.
We get from Eq.\ (33) therein
\be
N=L^2,
\ee
from Eq.\ (47) therein
\be \label{theirs1}
\f{1}{N^2}\Tr_A( (\r_A^{(1)})^2 ) = ( x^2 - \a_{12}^2  )^2,
\ee
from Eq.\ (49) therein
\be \label{theirs2}
\f{1}{N^2}\Tr_A( (\r_A^{(2)})^2 ) = [ (1-x)^2 - \a_{12}^2  ]^2,
\ee
and from Eq.\ (54) therein
\be \label{theirs3}
\f{1}{N^2}\Tr_A( (\r_A^{(3)})^2 ) = 2x^2(1-x)^2 +2 ( 1 + 2 x - 2 x^2) \a_{12}^2 + 2 \a_{12}^4.
\ee
Summing (\ref{theirs1}), (\ref{theirs2}), and (\ref{theirs3}) we get the R\'enyi entropy consistent with (\ref{ours})
\be
\cF_{A,k_1k_2}^{(2)} = [ x^2 + (1-x)^2 ]^2 + 8 x ( 1 - x) \a_{12}^2 + 4 \a_{12}^4.
\ee
For $k_1-k_2$ being a fixed finite integer, $\a_{12}$ is not vanishing in the limit $L\to+\inf$, $\ell\to+\inf$ with fixed $x=\f{\ell}{L}$
\be
\a_{12} \to \Big| \f{\sin[\pi (k_1-k_2) x]}{\pi (k_1-k_2)} \Big|.
\ee
So generally the additional terms cannot be neglected.
On the other hand, for a large difference of the momenta $|k_1-k_2| \to +\inf$ in the limit $L\to+\inf$, $\ell\to+\inf$ with fixed $x=\f{\ell}{L}$,
\be
\a_{12} \to 0,
\ee
and the universal entanglement and R\'enyi entropies in \cite{Castro-Alvaredo:2018dja,Castro-Alvaredo:2018bij,Castro-Alvaredo:2019irt,Castro-Alvaredo:2019lmj} are valid.

\subsubsection{Triple-particle state $|k_1k_2k_3\rag$} \label{singleintervalk1k2k3}

\begin{figure}[t]
  \centering
  \includegraphics[height=0.6\textwidth]{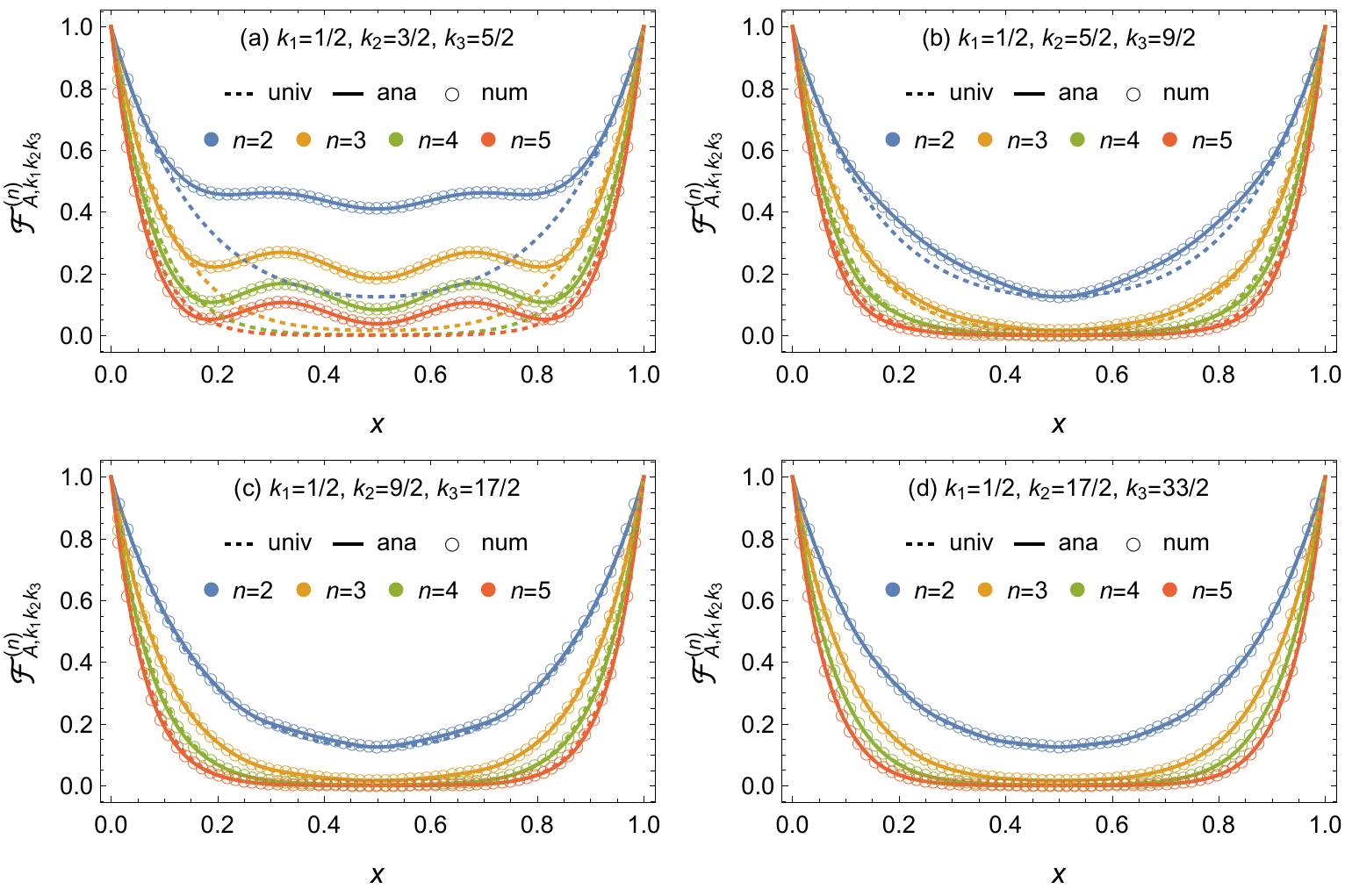}\\
  \caption{The universal R\'enyi entropy (dotted lines), and the analytical (solid lines) and numerical (empty circles) results of the single-interval R\'enyi entropy in the triple-particle state $|k_1k_2k_3\rag$ in the extremely gapped fermionic chain.
  We use different colors for different R\'enyi indices $n$'s.
  We have set $\l=+\inf$, $L=64$.}\label{FermionAk1k2k3}
\end{figure}

The calculations of the R\'enyi entropy in the triple-particle state $|k_1k_2k_3\rag$ with general momenta $k_1,k_2,k_3$ are similar to the above. We will not show the details here.
We get the additional contributions to the universal R\'enyi entropy
\bea
&& \d\cF_{A,k_1k_2k_3}^{(2)} = 8 x ( 1 - x) (1 - 2 x + 2 x^2) (\a_{12}^2 + \a_{13}^2 + \a_{23}^2)
      + 8 (1 - 2 x) (1 + 2 x - 2 x^2) \a_{12} \a_{13} \a_{23} \nn\\
&& \phantom{\d\cF_{A,k_1k_2k_3}^{(2)} =}
      + 4 (1 - 2 x + 2 x^2) (\a_{12}^2 + \a_{13}^2 + \a_{23}^2)^2
      + 16 (1 - 2 x) \a_{12} \a_{13} \a_{23} (\a_{12}^2 + \a_{13}^2 + \a_{23}^2) \nn\\
&& \phantom{\d\cF_{A,k_1k_2k_3}^{(2)} =}
      + 32 \a_{12}^2 \a_{13}^2 \a_{23}^2,
\eea
as well as $\d\cF_{A,k_1k_2k_3}^{(n)}$ with $n=3,4,5$ which we will not show in this paper.
Note that we have defined the shorthand $\a_{i_1i_2}\equiv|\a_{k_{i_1}-k_{i_2}}|$ with the function (\ref{alphakdefinition}).
We compare the results of the universal R\'enyi entropy, the analytical results with additional corrections and the numerical results for the state $|k_1k_2k_3\rag$ in the figure~\ref{FermionAk1k2k3}.
We find perfect matches between the new analytical results and the numerical ones.

\subsubsection{Slightly gapped and critical fermionic chains}

\begin{figure}[p]
  \centering
  \includegraphics[width=0.99\textwidth]{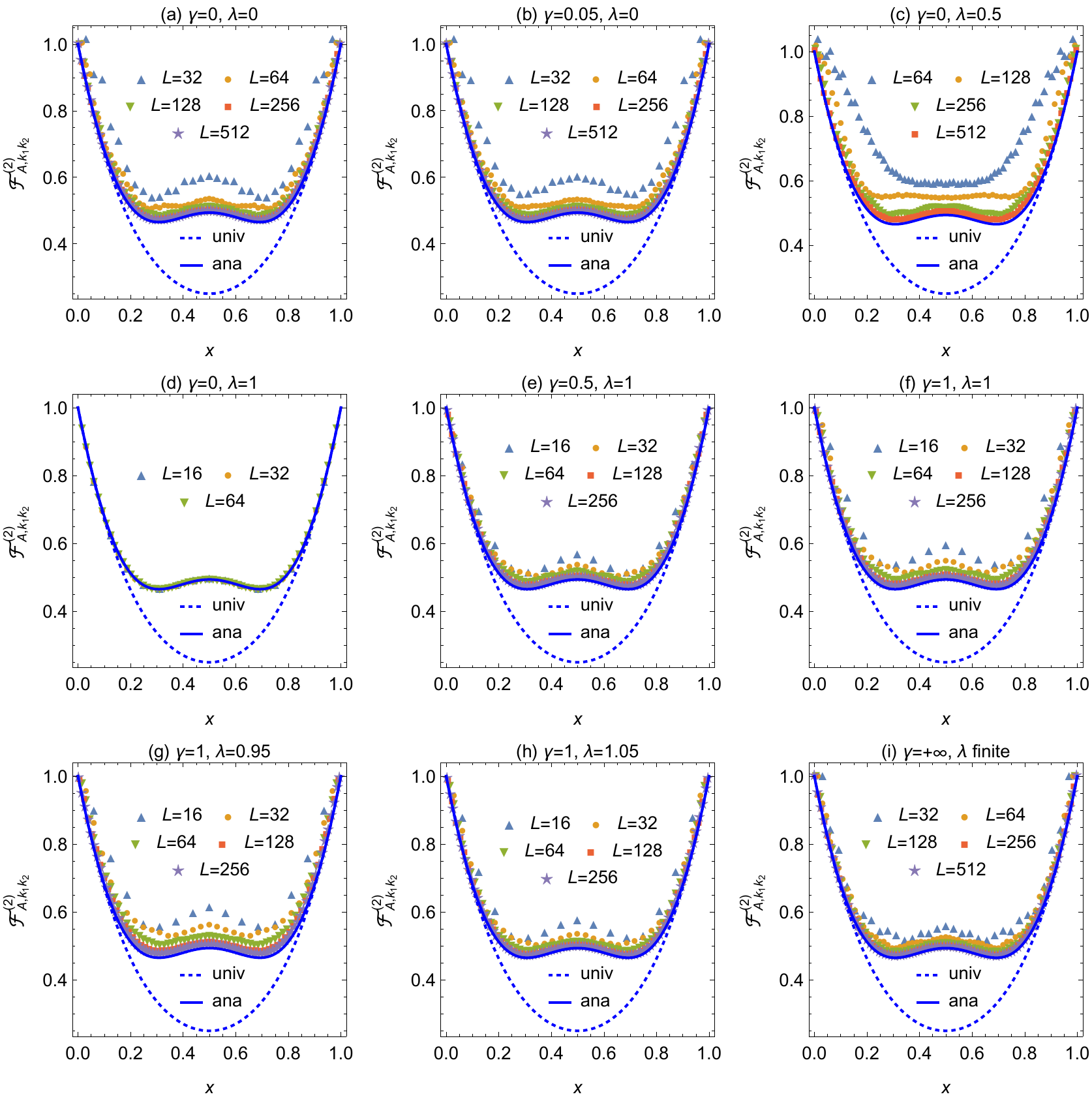}\\
  \caption{The universal R\'enyi entropy (dotted lines), the analytical single-interval R\'enyi entropy in the extremely gapped fermionic chain (solid lines), and the numerical single-interval R\'enyi entropy in the slightly gapped and critical fermionic chains (symbols) in the double-particle state $|k_1k_2\rag$. We have set the momenta $(k_1,k_2)=(\f12,\f32)+\f{L}{8}$, so that $k_i\to+\inf$ in the limit $L\to+\inf$, which is essential for the critical chains but is not important for the slightly gapped chains. For the analytical R\'enyi entropy we have set $L=+\inf$.}\label{FermionAk1k2slightlygapped}
\end{figure}

\begin{figure}[p]
  \centering
  \includegraphics[width=0.99\textwidth]{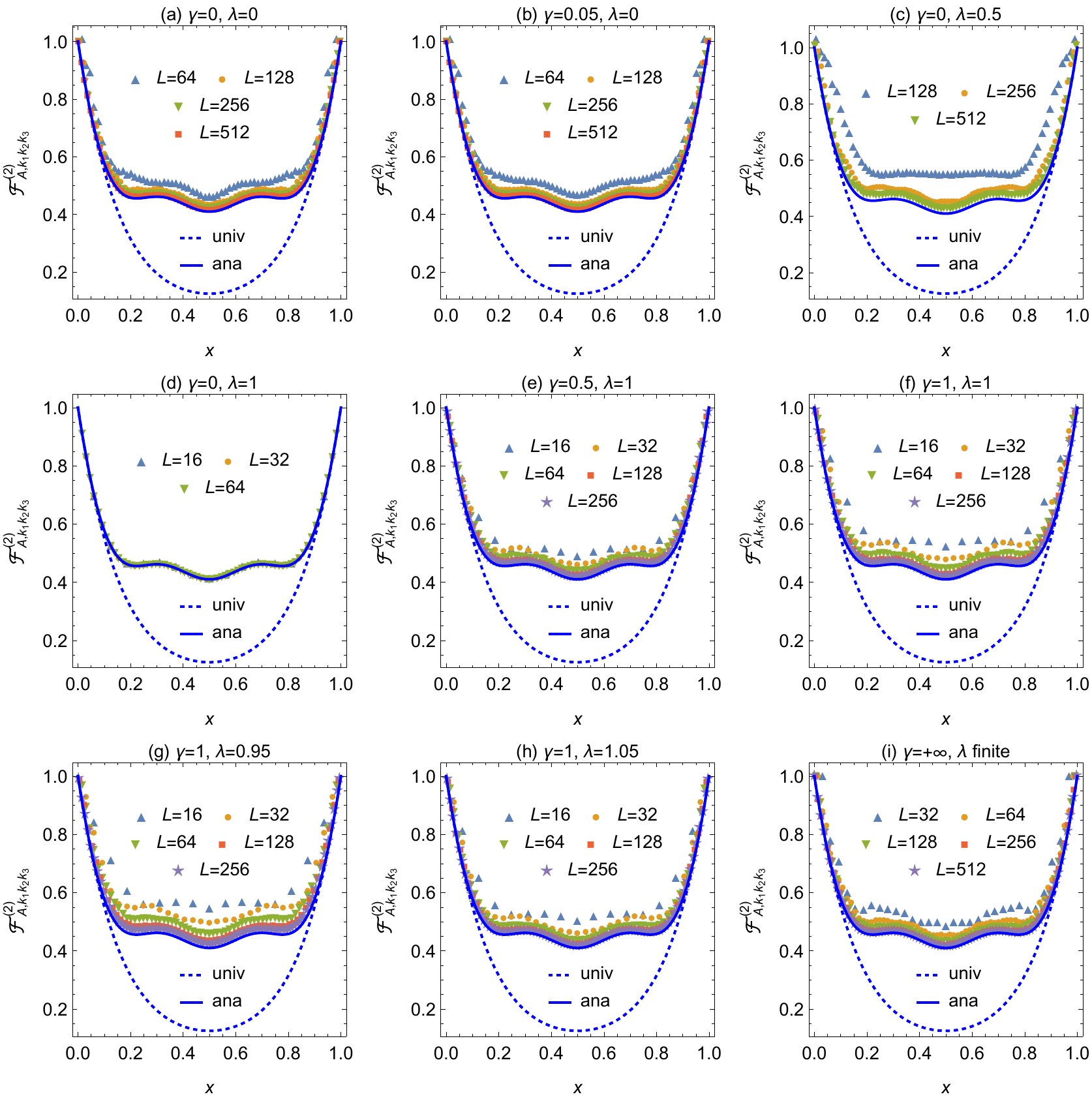}\\
  \caption{The universal R\'enyi entropy (dotted lines), the analytical single-interval R\'enyi entropy in the extremely gapped fermionic chain (solid lines), and the numerical single-interval R\'enyi entropy (symbols) in the slightly gapped and critical fermionic chains in the triple-particle state $|k_1k_2k_3\rag$. We have set the momenta $(k_1,k_2,k_3)=(\f12,\f32,\f52)+\f{L}{8}$. For the analytical R\'enyi entropy we have set $L=+\inf$.}\label{FermionAk1k2k3slightlygapped}
\end{figure}

We have calculated the analytical expressions of the R\'enyi entropy in the extremely gapped fermionic chain, and found a good matching with the numerical results.
We wonder if the results apply to excited states of quasiparticles in the slightly gapped and even critical fermionic chains.
We compare the results of the universal R\'enyi entropy, the results with additional corrections, and the numerical results in the double-particle state $|k_1k_2\rag$ and the triple-particle state $|k_1k_2k_3\rag$ in the slightly gapped and critical fermionic chains in the figures~\ref{FermionAk1k2slightlygapped} and \ref{FermionAk1k2k3slightlygapped}.
We see that the results of the R\'enyi entropy with the additional corrections are universal in the sense that they are still valid as long as all the momenta of the excited quasiparticles and the size of the system $L$ are large.

Especially, for the special fermionic chain with $\g=0,\l=1$, which is critical but non-relativistic, the new R\'enyi entropy with additional terms is exact even for  small $L$, $\ell$.
This is because for $\g=0,\l=1$ the Bogoliubov angle defined in (\ref{vphikepiithk}) is also vanishing $\th_k=0$, the same as that in the extremely gapped fermionic chain.

In section \ref{secDis} we will quantify the convergence of the results of slightly gapped cases to the extremely gapped cases in more detail.

\subsection{Double interval}

We consider the double interval on a circular fermionic chain as shown in figure~\ref{subsystems}. The universal R\'enyi entropy does not depend on the connectedness of the subsystem, and so the double-interval R\'enyi entropy in the quasiparticle excited states only depends on $x=x_1+x_2$, not on $x_1-x_2$ or $y_1$.
We will show both analytically and numerically that generally the R\'enyi entropy would depend on all the independent parameters $x_1,x_2,y_1$ if one relaxes the constraints on the momenta $k_i$.

For the double-interval with $A=A_1\cup A_2$ and $B=B_1\cup B_2$, in the extremely gapped limit we can still define the subsystem modes
\bea
&& c_{A,k} = \f{1}{\sr{L}}\sum_{j\in A} \ep^{\ii j \vph_k}a_j, ~~
   c_{A,k}^\dag = \f{1}{\sr{L}}\sum_{j\in A} \ep^{-\ii j \vph_k}a_j^\dag, \nn\\
&& c_{B,k} = \f{1}{\sr{L}}\sum_{j\in B} \ep^{\ii j \vph_k}a_j, ~~
   c_{B,k}^\dag = \f{1}{\sr{L}}\sum_{j\in B} \ep^{-\ii j \vph_k}a_j^\dag.
\eea
There are anti-commutation relations
\be
\{ c_{A,k}, c_{A,k}^\dag \} = x, ~~
\{ c_{B,k}, c_{B,k}^\dag \} = y,
\ee
and for $k_1\neq k_2$ we have
\be
\{ c_{A,k_1}, c_{A,k_2}^\dag \} = - \{ c_{B,k_1}, c_{B,k_2}^\dag \} = \b_{k_1-k_2},
\ee
where we have defined
\be \label{betakdefinition}
\b_k = \f{1}{L} \sum_{j\in A_1\cup A_2} \ep^{\f{2\pi\ii j k}{L}}
     = \f{\ep^{\f{\pi \ii k (\ell_1+1)}{L}}}{L\sin\f{\pi k}{L}}
       \Big( \sin\f{\pi k \ell_1}{L}
           + \ep^{\f{\pi\ii k(\ell_1+2d_1+\ell_2)}{L}} \sin\f{\pi k \ell_2}{L} \Big).
\ee

\subsubsection{Single-particle state $|k\rag$}

The calculations and the results are the same as those for the single-interval R\'enyi entropy in subsection (\ref{subsubsecSPS}). There is no additional contribution to the universal double-interval R\'enyi entropy in the single particle state $|k\rag$.

\subsubsection{Double-particle state $|k_1k_2\rag$} \label{fermionDIDPS}

\begin{figure}[p]
  \centering
  \includegraphics[height=0.975\textwidth]{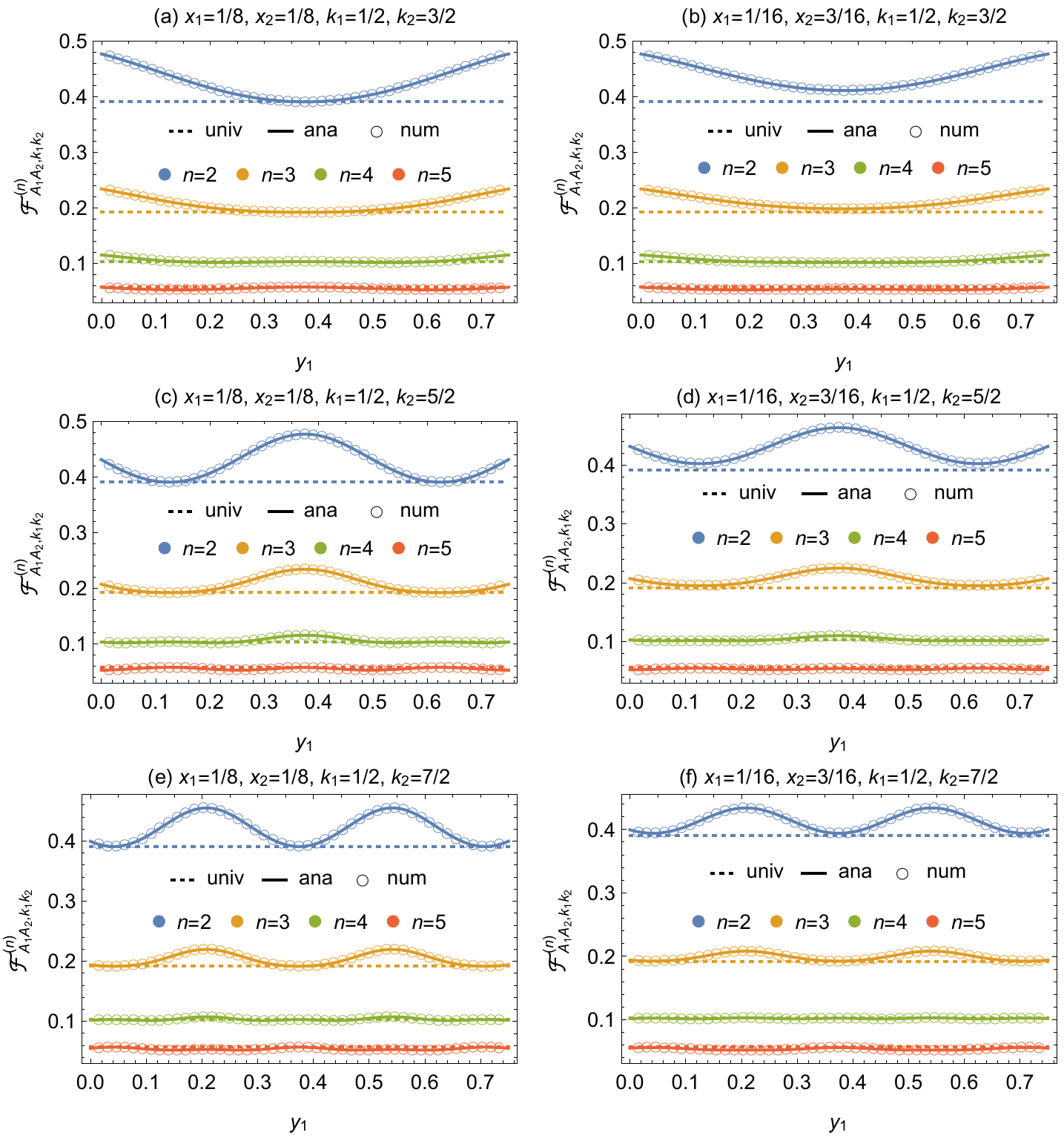}\\
  \caption{The universal R\'enyi entropy (dotted lines), and the analytical (solid lines) and numerical (empty circles) results of the double-interval R\'enyi entropy in the double-particle state $|k_1k_2\rag$ in the extremely gapped fermionic chain. We have set $\l=+\inf$, $L=64$.}\label{FermionA1A2k1k2}
\end{figure}

The calculations of the double-interval R\'enyi entropy in the double-particle state $|k_1k_2\rag$ are similar to those for the single-interval R\'enyi entropy in subsection \ref{singleintervalk1k2}.
The results are just the ones in subsection~\ref{singleintervalk1k2} after the following substitution
\be
| \a_{k_1-k_2} | \to | \b_{k_1-k_2} |,
\ee
where $\b_k$ is defined in (\ref{betakdefinition}).
For example, we have
\be
\d\cF_{A_1A_2,k_1k_2}^{(2)} = 8 x ( 1 - x) | \b_{k_1-k_2} |^2 + 4 | \b_{k_1-k_2} |^4,
\ee
\be
\d\cF_{A_1A_2,k_1k_2}^{(3)} = -3 (1 - 6 x + 6 x^2) | \b_{k_1-k_2} |^2 + 9 | \b_{k_1-k_2} |^4.
\ee
We compare the analytical and numerical results in figure~\ref{FermionA1A2k1k2}.
We see the necessity of the additional correction terms when the momentum difference $|k_1-k_2|$ is small.

\subsubsection{Triple-particle state $|k_1k_2k_3\rag$} \label{fermionDITPS}

\begin{figure}[p]
  \centering
  \includegraphics[height=0.975\textwidth]{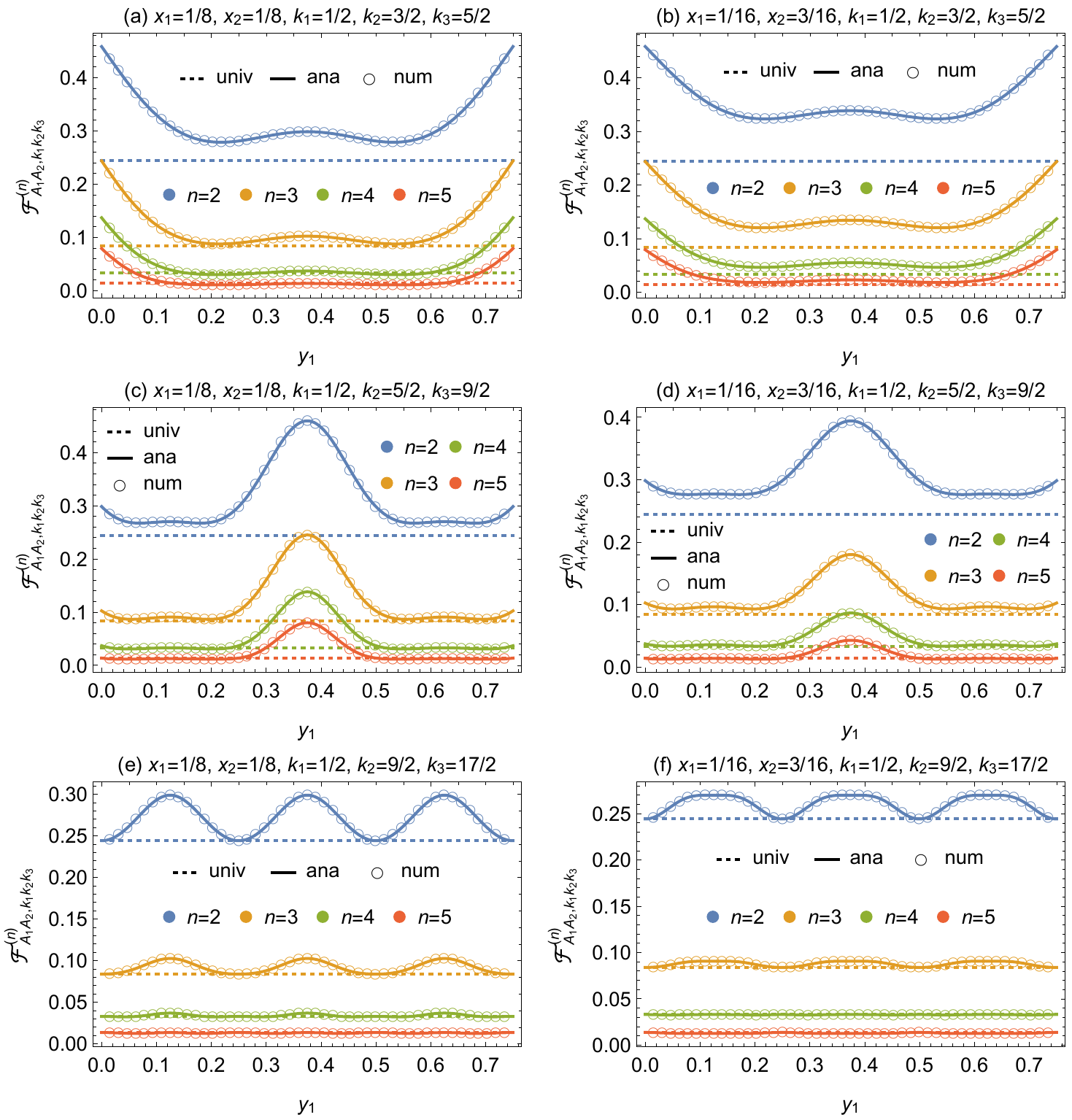}\\
  \caption{The universal R\'enyi entropy (dotted lines), and the analytical (solid lines) and numerical (empty circles) results of the double-interval R\'enyi entropy in the triple-particle state $|k_1k_2k_3\rag$ in the extremely gapped fermionic chain. We have set $\l=+\inf$, $L=64$.}\label{FermionA1A2k1k2k3}
\end{figure}

The calculations of the double-interval R\'enyi entropy in the triple-particle state $|k_1k_2k_3\rag$ are similar to those presented before.
We get the additional contributions to the universal R\'enyi entropy
\bea
&& \d\cF_{A_1A_2,k_1k_2k_3}^{(2)} =
     8 x ( 1 - 3 x + 4 x^2 - 2 x^3) \g_{k_1k_2k_3}
   + 4 (1 - 6 x^2 + 4 x^3) \d_{k_1k_2k_3}\nn\\
&& \phantom{\d\cF_{A_1A_2,k_1k_2k_3}^{(2)} =}
   + 4 (1 - 2 x + 2 x^2) \g_{k_1k_2k_3}^2
   + 8 (1 - 2 x) \g_{k_1k_2k_3} \d_{k_1k_2k_3} + 8 \d_{k_1k_2k_3}^2,
\eea
\bea
&& \d\cF_{A_1A_2,k_1k_2k_3}^{(3)} =
    - 3 (1 - 9 x + 27 x^2 - 36 x^3 + 18 x^4) \g_{k_1k_2k_3}
    + 27 x (1 - 3 x + 2 x^2) \d_{k_1k_2k_3} \nn\\
&& \phantom{\d\cF_{A_1A_2,k_1k_2k_3}^{(3)} =}
    + 9 (1 - 3 x + 3 x^2) \g_{k_1k_2k_3}^2
    + 27 (1 - 2 x) \g_{k_1k_2k_3} \d_{k_1k_2k_3}
    + 27 \d_{k_1k_2k_3}^2,
\eea
with the definitions
\bea \label{gk1k2k3dk1k2k3}
&& \g_{k_1k_2k_3} = |\b_{k_1-k_2}|^2 + |\b_{k_1-k_3}|^2 + |\b_{k_2-k_3}|^2, \nn\\
&& \d_{k_1k_2k_3} = \b_{k_1-k_2}\b_{k_2-k_3}\b_{k_3-k_1} + \b_{k_1-k_3}\b_{k_3-k_2}\b_{k_2-k_1}.
\eea
We also get $\d\cF_{A_1A_2,k_1k_2k_3}^{(n)}$ with $n=4,5$, which we will not show in this paper.
We compare the analytical and numerical results in the figure~\ref{FermionA1A2k1k2k3}.
There are perfect matches between the results of the R\'enyi entropy with the numerical ones.

\subsubsection{Slightly gapped and critical fermionic chains}

\begin{figure}[p]
  \centering
  \includegraphics[width=0.99\textwidth]{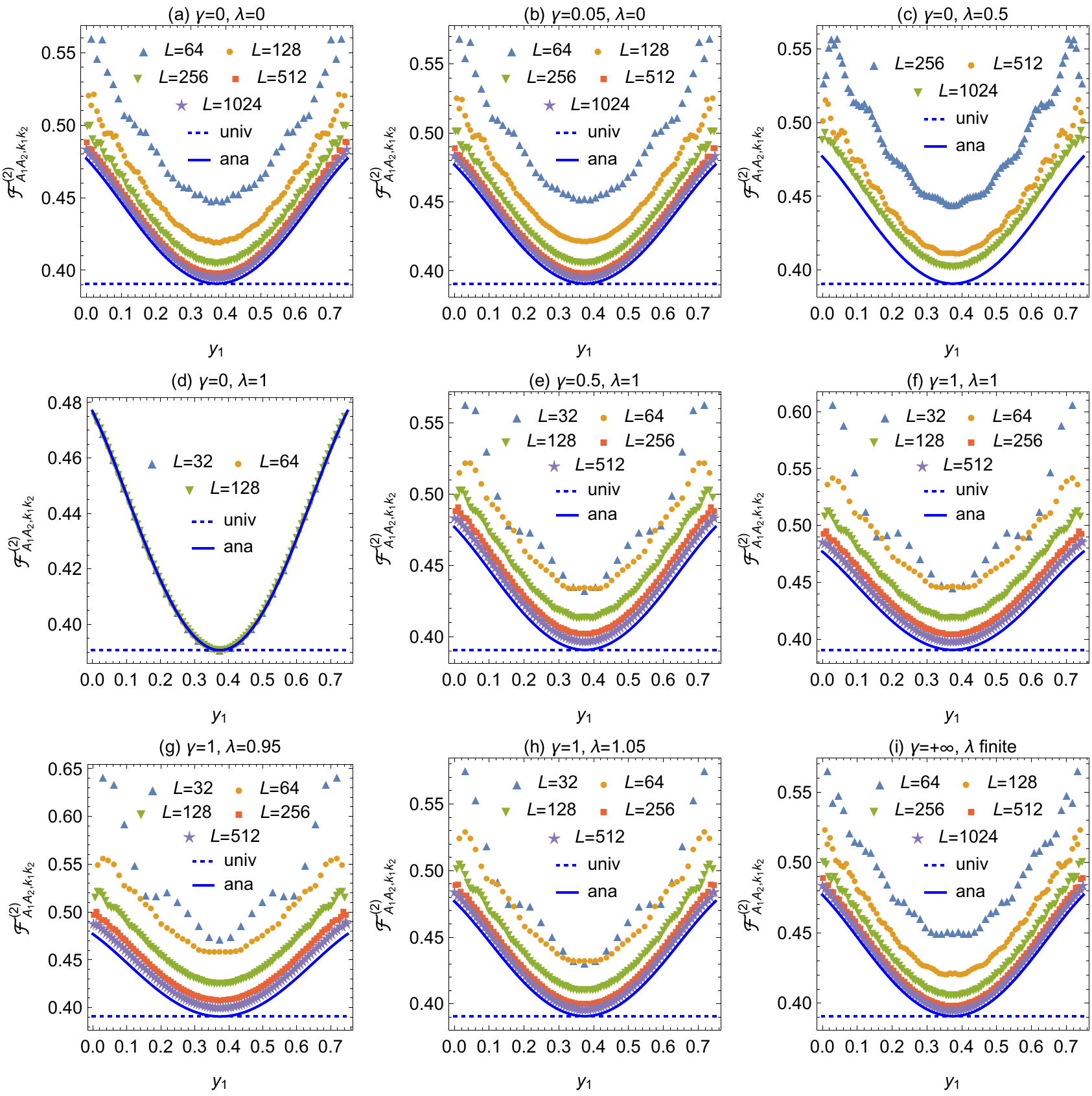}\\
  \caption{The universal R\'enyi entropy (dotted lines), the analytical double-interval R\'enyi entropy in the extremely gapped fermionic chain (solid lines), and the numerical double-interval R\'enyi entropy in the slightly gapped and critical fermionic chains (symbols) in the double-particle state $|k_1k_2\rag$. We have set the momenta $(k_1,k_2)=(\f12,\f32)+\f{L}{8}$, $x_1=x_2=\f18$. For the analytical R\'enyi entropy we have set $L=+\inf$.}\label{FermionA1A2k1k2slightlygapped}
\end{figure}

\begin{figure}[p]
  \centering
  \includegraphics[width=0.99\textwidth]{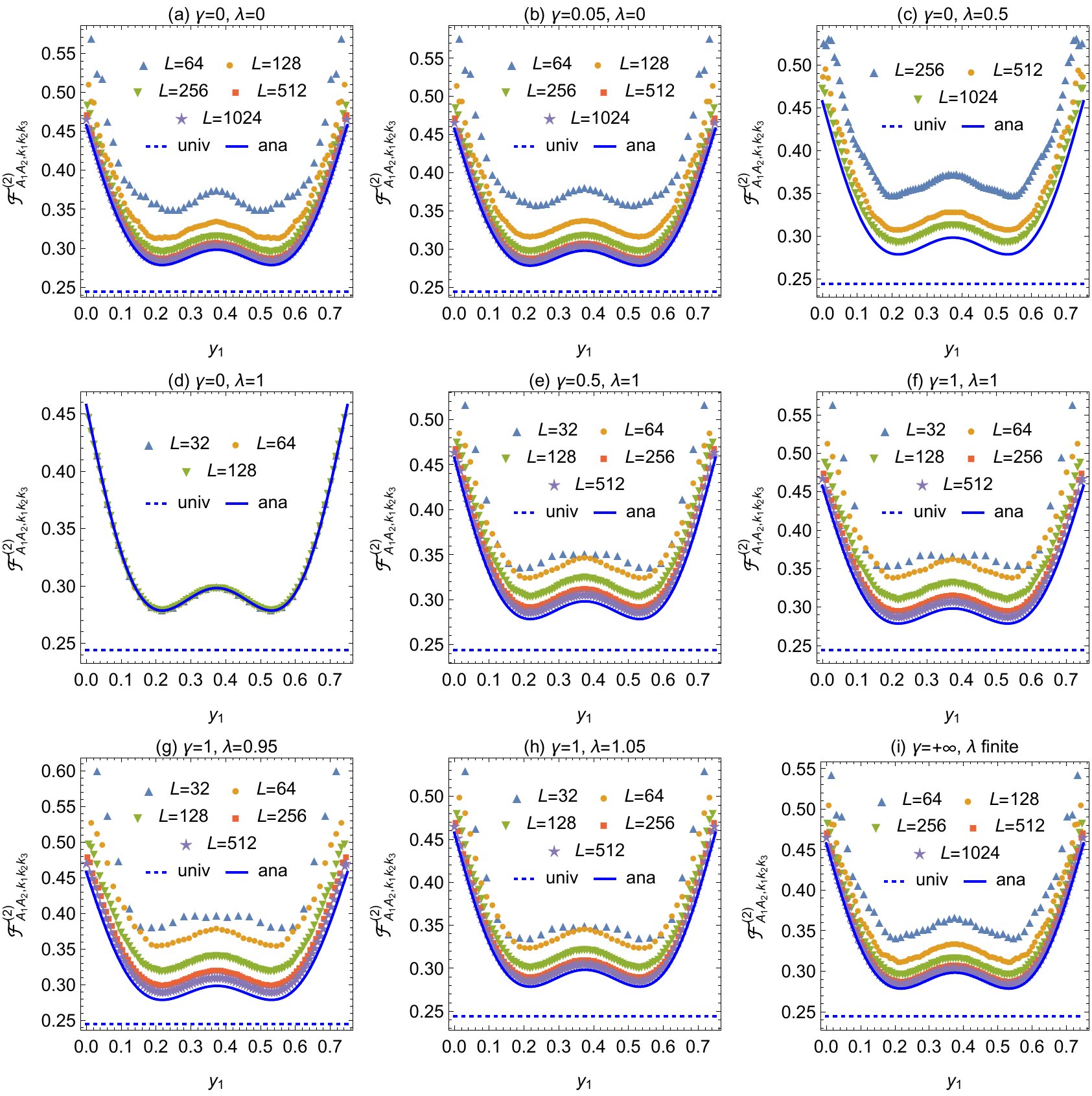}\\
  \caption{The universal R\'enyi entropy (dotted lines), the analytical double-interval R\'enyi entropy in the extremely gapped fermionic chain (solid lines), and the numerical double-interval R\'enyi entropy in the slightly gapped and critical fermionic chains (symbols) in the triple-particle state $|k_1k_2k_3\rag$. We have set the momenta $(k_1,k_2,k_3)=(\f12,\f32,\f52)+\f{L}{8}$, $x_1=x_2=\f18$. For the analytical R\'enyi entropy we have set $L=+\inf$.}\label{FermionA1A2k1k2k3slightlygapped}
\end{figure}

We calculate the double-interval R\'enyi entropy in the slightly gapped and critical fermionic chains, and find that the results with additional terms are still valid in the limit that all the quasiparticle momenta are large.
The results are shown in the figures~\ref{FermionA1A2k1k2slightlygapped} and \ref{FermionA1A2k1k2k3slightlygapped}.
As before, the universal R\'enyi entropy is exact for the fermionic chain with $\g=0,\l=1$.

\subsection{Multiple intervals}

The generalization from double interval to multiple intervals is easy.
We just need to change $\b_k$ defined in (\ref{betakdefinition}) to
\be \label{betakdefinitionX}
\b_k = \f{1}{L} \sum_{j\in A} \ep^{\f{2\pi\ii j k}{L}},
\ee
where $A$ can be made of arbitrary number of disjoint intervals.

\subsubsection{R\'enyi entropy}

In the single-particle state $|k\rag$, there is no additional contribution to the universal R\'enyi entropy.
In the double-particle state $|k_1k_2\rag$ and triple-particle state $|k_1k_2k_3\rag$, the expressions of the R\'enyi entropy are the same as those in the subsection~\ref{fermionDIDPS} and subsection~\ref{fermionDITPS}, respectively, and we just need to use the definition of $\b_k$ presented in (\ref{betakdefinitionX}).
It is easy to compare the analytical and numerical results of the multi-interval R\'enyi entropy, but we will not show the details here.

\subsubsection{Average R\'enyi entropy}

In the extremely gapped limit,the interactions between the fermions at the neighboring sites could be omitted.
We check that our results of the single-interval and double-interval R\'enyi entropies satisfy the position-momentum duality \cite{Lee:2014nra,Carrasco:2017eul}
\be \label{duality}
S_{A,K}^{(n)} = S_{K,A}^{(n)}.
\ee
We have used integers to label the sites, while the momenta in the NS sector are half-integers.
We can still change $A\lra K$, because $S_{A,K}^{(n)}$ only depend on the momentum differences rather than the momenta themselves.
In other words, $S_{A,K}^{(n)}=S_{A,K+k_0}^{(n)}$ for an arbitrary constant $k_0$.
When the momenta in the set $K$ are half-integers, the duality (\ref{duality}) can be understood as
\be
S_{A,K}^{(n)} = S_{K+\f12,A}^{(n)}.
\ee
Explicitly, we are able to check that
\bea
&& \cF_{j,k_1k_2}^{(n)} = \cF_{k_1k_2,j}^{(n)}, ~~ n=2,3,\cdots,7, \nn\\
&& \cF_{j_1j_2,k_1k_2}^{(n)} = \cF_{k_1k_2,j_1j_2}^{(n)}, ~~ n=2,3,\cdots,7, \nn\\
&& \cF_{j,k_1k_2k_3}^{(n)} = \cF_{k_1k_2k_3,j}^{(n)}, ~~ n=2,3,4,5, \nn\\
&& \cF_{j_1j_2,k_1k_2k_3}^{(n)} = \cF_{k_1k_2k_3,j_1j_2}^{(n)}, ~~ n=2,3,4,5.
\eea
For example, we have
\bea
&& \cF_{j,k_1k_2}^{(2)} = \cF_{k_1k_2,j}^{(2)} = \frac{L^2-4 L+8}{L^2}, \nn\\
&&\cF_{j,k_1k_2k_3}^{(4)} = \cF_{k_1k_2k_3,j}^{(4)} = \frac{L^4-12 L^3+54 L^2-108 L+162}{L^4}.
\eea

Using the duality (\ref{duality}), it is also convenient to calculate the average R\'enyi entropy \cite{Page:1993df,Vidmar:2017uux,Vidmar:2017pak,Vidmar:2018rqk,Hackl:2018tyl,Jafarizadeh:2019xxc}
\be
\lag S_{A}^{(n)} \rag = \f{1}{2^L} \sum_K S_{A,K}^{(n)} = \f{1}{2^L} \sum_K S_{K,A}^{(n)}.
\ee
For $A$ consisting of one (general $n$), two ($n=2,3,\cdots,7$), and three ($n=2,3,4,5$) successive sites, i.e.\ single interval $A$ with $|A|=\ell=1,2,3$, we can check numerically the average R\'enyi entropy
\be
\lim_{L \to +\inf} \lag S_{A}^{(n)} \rag  = |A| \log 2.
\ee
This is just the finding in \cite{Vidmar:2017uux}.

It is easy to check that the expression (\ref{cFAk1r1k2r2cdotsksrsnsup}) does not satisfy the position-momentum duality (\ref{duality}).
For example, there are
\be \label{dualityjp1p2k1k22pijL}
\cF_{j,p_1p_2}^{(n),\univ} = \Big[ \f{1}{L^n} + \Big(1-\f{1}{L}\Big)^n \Big]^2
\neq
\cF_{k_1k_2,\f{2\pi j}{L}}^{(n),\univ} = \Big(\f{2}{L}\Big)^n + \Big(1-\f{2}{L}\Big)^n.
\ee
This is not surprising, as the universal R\'enyi entropy is expected to be valid only in the limit $L\to+\inf$. On the other hand, in the limit $L\to+\inf$, the position-momentum duality is satisfied trivially for $\cF_{j,p_1p_2}^{(n),\univ}$ and $\cF_{k_1k_2,\f{2\pi j}{L}}^{(n),\univ}$ in (\ref{dualityjp1p2k1k22pijL}).
As can be seen in figure~\ref{upperboundplot}, the universal R\'enyi entropy is not valid also when too many quasiparticles are excited.
Moreover, the universal R\'enyi entropy is not valid when the momentum differences are small.
So one cannot calculate the average R\'enyi entropy from the universal R\'enyi entropy.

\subsection{R\'enyi entropy with general index $n$ and entanglement entropy} \label{FermionEE}

One can use the result (\ref{SAKSAKn}) from the correlation matrix method and the position-momentum duality (\ref{duality}) and calculate the excited state R\'enyi entropy analytically.
For a subsystem $A$ in the state $|K\rag=|k_1k_2\cdots k_s\rag$ with the excitations of $|K|=s$ different quasiparticles, we define the $|K|\times|K|$ matrix $C_K^A$ with entries
\be
[C_K^A]_{i_1i_2} = h_{k_{i_2}-k_{i_1}}^A, ~ i_1,i_2=1,2,\cdots,s,
\ee
with the function
\be
h_{k}^A = \f{1}{L} \sum_{j \in A} \ep^{\f{2\pi\ii j k}{L}}.
\ee
Note that $h_{0}^A=x$ and for $k\neq 0$ we have $h_{k}^A=\b_k$ that is defined in (\ref{betakdefinitionX}).
The excited state R\'enyi entropy is just
\be \label{TheFormulaF}
\cF_{A,K}^{(n)} = \det [ (C_K^A)^n + (1-C_K^A)^n ].
\ee
The formula is very efficient for both analytical and numerical calculations.
We confirmed that it always leads to the same analytical results as we obtained by writing the excited states in terms of subsystem excitations.

In the single-particle state $|k\rag$ there is $C_k^A=x$, and we reproduce the universal R\'enyi entropy with no additional contribution $\cF_{A,k}^{(n)} = \cF_{A,p}^{(n),\univ} = x^n+(1-x)^n$.
In a general multi-particle state $|k_1k_2\cdots k_s\rag$ that all the momentum differences are large, $C_{k_1k_2\cdots k_s}^A=xI_s$ with $I_s$ being an $s\times s$ identity matrix, and we then get easily the most general universal R\'enyi entropy in the fermionic chain $\cF_{A,p_1p_2\cdots p_s}^{(n),\univ} = [x^n+(1-x)^n]^s$.

In the double-particle state $|k_1k_2\rag$ with general momenta $k_1, k_2$ there is
\be
C_{k_1k_2}^A = \lt(
\ba{cc}
x & \b_{k_2-k_1} \\ \b_{k_1-k_2} & x
\ea
\rt),
\ee
whose eigenvalues are
\be
\n_1 = x + |\b_{k_1-k_2}|, ~~
\n_2 = x - |\b_{k_1-k_2}|.
\ee
We get the double-particle state R\'enyi entropy with general index $n$
\be
\cF_{A,k_1k_2}^{(n)} = [ (x + |\b_{k_1-k_2}|)^n + (1 - x - |\b_{k_1-k_2}|)^n ]
                       [ (x - |\b_{k_1-k_2}|)^n + (1 - x + |\b_{k_1-k_2}|)^n ].
\ee
We take the $n\to1$ analytical continuation and get the entanglement entropy
\bea
&& S_{A,k_1k_2} = - (x + |\b_{k_1-k_2}|) \log (x + |\b_{k_1-k_2}|)
               - (1 - x - |\b_{k_1-k_2}|) \log (1 - x - |\b_{k_1-k_2}|) \nn\\
&& \phantom{S_{A,k_1k_2} =}
               - (x - |\b_{k_1-k_2}|) \log (x - |\b_{k_1-k_2}|)
               - (1 - x + |\b_{k_1-k_2}|) \log (1 - x + |\b_{k_1-k_2}|).
\eea
We see nontrivial additional contributions to the universal double-particle state R\'enyi entropy (\ref{SAuniv}).

In the triple-particle state $|k_1k_2k_3\rag$ with general momenta $k_1,k_2,k_3$ there is
\be
C_{k_1k_2k_3}^A = \lt(
\ba{ccc}
x            & \b_{k_2-k_1} & \b_{k_3-k_1} \\
\b_{k_1-k_2} & x            & \b_{k_3-k_2} \\
\b_{k_1-k_3} & \b_{k_2-k_3} & x
\ea
\rt),
\ee
whose eigenvalues $\n_i$ with $i=1,2,3$ are solutions to the equation
\be
\n^3 - 3 x \n^2 + ( 3x^2 - \g_{k_1k_2k_3} ) \n - ( x^3 - \g_{k_1k_2k_3} x + \d_{k_1k_2k_3} ) = 0,
\ee
with the definitions $\g_{k_1k_2k_3}$ and $\d_{k_1k_2k_3}$ in (\ref{gk1k2k3dk1k2k3}).
The R\'enyi entropy with general index $n$ and the entanglement entropy are just
\bea
&& \cF_{A,k_1k_2k_3}^{(n)} = \prod_{i=1}^3 [ \n_i^n + (1-\n_i)^n ], \nn\\
&& S_{A,k_1k_2k_3} = \sum_{i=1}^3 [ - \n_i \log \n_i - (1-\n_i) \log (1-\n_i) ].
\eea

\begin{figure}[tp]
  \centering
  \includegraphics[height=0.3\textwidth]{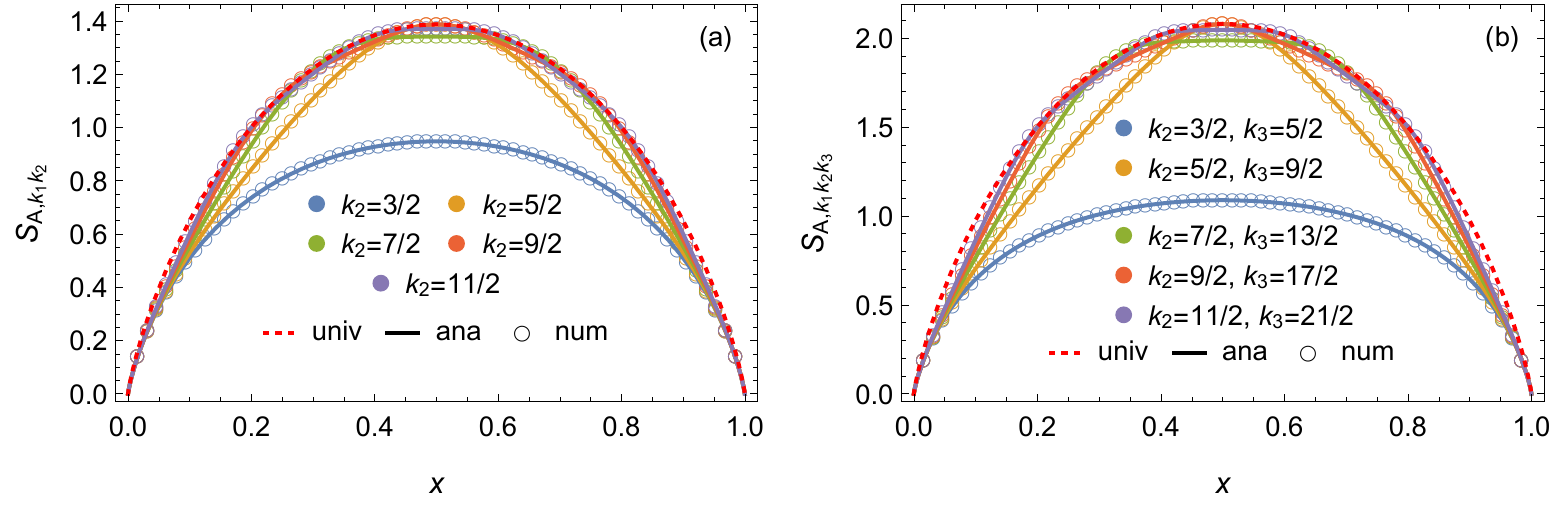}\\
  \caption{The universal entanglement entropy (dotted lines), and the analytical (solid lines) and numerical (circles) results of the entanglement entropy in the double-particle state $|k_1k_2\rag$ (left) and the triple-particle state $|k_1k_2k_3\rag$ (right) in the extremely gapped fermionic chain.
  We use different colors for different momenta.
  We have set $\l=+\inf$, $L=64$, $k_1=\f12$.}\label{FermionEEAk1k2k1k2k3}
\end{figure}

We compare the universal entanglement entropy, the analytical and numerical entanglement entropy in the double-particle state $|k_1k_2\rag$ and the triple-particle state $|k_1k_2k_3\rag$ in the extremely gapped fermionic chain in figure~\ref{FermionEEAk1k2k1k2k3}.

The analytical entanglement and R\'enyi entropies in an excited state with a larger number of quasiparticles can be calculated similarly, but we will not report the results in this paper.

\section{Bosonic chain}\label{secB}

We consider the circular bosonic chain
\be \label{bosonicchain}
H = \f{1}{2} \sum_{j=1}^L \big[ p_j^2 + m^2 q_j^2 + (q_j-q_{j+1})^2 \big],
\ee
which is just the locally coupled harmonic chain and is also the discretization of the two-dimensional massive scalar field theory.
We have the periodic boundary condition $q_{L+1}=q_{1}$.
The mass $m$ is just the gap of the model.
We take the number of sites $L$ as an even integer.
It can be diagonalized by the Fourier transformation
\be
q_j = \f{1}{\sr{L}} \sum_k \ep^{-\f{2\pi\ii j k}{L}}\vph_k, ~~
p_j = \f{1}{\sr{L}} \sum_k \ep^{-\f{2\pi\ii j k}{L}}\pi_k,
\ee
with the momentum
\be
k=1-\f{L}{2},\cdots,-1,0,1,\cdots,\f{L}{2}-1,\f{L}{2}.
\ee
The Hamiltonian becomes
\be
H = \f12 \sum_k ( \pi_k^\dag \pi_k + \ve_k^2\vph_k^\dag\vph_k ),
\ee
with the frequency
\be \label{vek}
\ve_k = \sr{m^2+4\sin^2\f{\pi k}{L}}.
\ee
One can define the ladder operators
\be \label{bKbkdag}
b_k=\sr{\f{\ve_k}{2}}\Big( \vph_k + \f{\ii}{\ve_k} \pi_k \Big), ~~
b_k^\dag=\sr{\f{\ve_k}{2}}\Big( \vph_k^\dag - \f{\ii}{\ve_k} \pi_k^\dag \Big).
\ee
The Hamiltonian becomes
\be
H = \sum_k \ve_k \Big( b_k^\dag b_k +\f12 \Big).
\ee
The ground state $|G\rag$ is annihilated by all the lowering operators
\be
b_k | G \rag = 0.
\ee
The excited states are constructed by applying various raising operators on the ground state
\be \label{k1r1k2r2cdotsksrsrag}
|k_1^{r_1}k_2^{r_2}\cdots k_s^{r_s}\rag = \f{( b^\dag_{k_1} )^{r_1} ( b^\dag_{k_2} )^{r_2} \cdots ( b^\dag_{k_s} )^{r_s}}{\sr{r_1!r_2!\cdots r_s!}} | G \rag.
\ee
In the extremely massive limit $m\to+\inf$, the bosonic chain approaches $L$ decoupled oscillators
\be
H = \f12 \sum_{j=1}^L ( p_j^2 + m^2 x_j^2 ),
\ee
and for each oscillator one can define the local ladder operators
\be \label{ajajdag}
a_j =\sr{\f{m}{2}}\Big( q_j + \f{\ii}{m} p_j \Big), ~~
a_j^\dag =\sr{\f{m}{2}}\Big( q_j - \f{\ii}{m} p_j \Big).
\ee
In the limit $m\to+\inf$, the lowering and raising operators (\ref{bKbkdag}) become
\be
b_k= \f{1}{\sr{L}} \sum_{j=1}^L \ep^{\f{2\pi\ii j k}{L}} a_j, ~~
b_k^\dag= \f{1}{\sr{L}} \sum_{j=1}^L \ep^{-\f{2\pi\ii j k}{L}} a_j^\dag.
\ee
The ground state is also annihilated by the lowering operators at each site
\be
a_j |G\rag = 0, ~ j=1,2,\cdots,L.
\ee

\subsection{Single interval}

We consider an interval with $\ell$ consecutive sites $A=[1,\ell]$ on the periodic bosonic chain with $L$ sites.
In the extremely gapped limit $m\to+\inf$, the ground state is just a direct product state
\be
|G\rag = |G_A\rag |G_B\rag,
\ee
and the R\'enyi entropy is vanishing
\be
S_{A,G}^{(n)} = 0.
\ee
In the excited state $|k_1^{r_1}k_2^{r_2}\cdots k_s^{r_s}\rag$ with large $k_i$, the universal R\'enyi entropy discussed in \cite{Castro-Alvaredo:2018dja,Castro-Alvaredo:2018bij,Castro-Alvaredo:2019irt,Castro-Alvaredo:2019lmj} is just (\ref{cFAk1r1k2r2cdotsksrsnsup}).
We will relax the constraints for the momenta $k_i$ and show nontrivial additional contributions $\d \cF_{A,k_1^{r_1}k_2^{r_2}\cdots k_s^{r_s}}^{(n)}$ to the universal R\'enyi entropy
\be
\cF_{A,k_1^{r_1}k_2^{r_2}\cdots k_s^{r_s}}^{(n)} =  \cF_{A,p_1^{r_1}p_2^{r_2}\cdots p_s^{r_s}}^{(n),\univ} + \d \cF_{A,k_1^{r_1}k_2^{r_2}\cdots k_s^{r_s}}^{(n)}.
\ee
For later analytical calculations, it is convenient to define in the extremely gapped limit
\bea
&& b_{A,k} = \f{1}{\sr{L}}\sum_{j\in A} \ep^{\ii j \vph_k}a_j, ~~
   b_{A,k}^\dag = \f{1}{\sr{L}}\sum_{j\in A} \ep^{-\ii j \vph_k}a_j^\dag, \nn\\
&& b_{B,k} = \f{1}{\sr{L}}\sum_{j\in B} \ep^{\ii j \vph_k}a_j, ~~
   b_{B,k}^\dag = \f{1}{\sr{L}}\sum_{j\in B} \ep^{-\ii j \vph_k}a_j^\dag.
\eea
There are commutation relations
\be
[ b_{A,k}, b_{A,k}^\dag ] = x, ~~
[ b_{B,k}, b_{B,k}^\dag ] = 1-x,
\ee
and for $k_1\neq k_2$ there are
\be
[ b_{A,k_1}, b_{A,k_2}^\dag ] = - [ b_{B,k_1}, b_{B,k_2}^\dag ] = \a_{k_1-k_2}.
\ee
with the definition of $\a_k$ presented in (\ref{alphakdefinition}).
The following analytical calculations in the harmonic chain are similar to those in the fermionic chain, with the difference of changing the anti-commutation relations to the commutation ones.

For a general gap $m$, the excited state R\'enyi entropy can be calculated numerically from the method of the wave function \cite{Castro-Alvaredo:2018dja,Castro-Alvaredo:2018bij}.
In the extremely gapped limit $m\to+\inf$ the wave function method can be also used to calculate the analytical R\'enyi entropy. We will come back to this method at the end of this section. In the following in analogy with the fermionic calculations we calculate the
R\'enyi entropies using the subsystem mode method.

\subsubsection{Multi-particle state with equal momenta $|k^r\rag$}

For the state $|k^r\rag$ with $r$ quasiparticles of equal momenta $k$, we write the density matrix of the whole system as
\be
\r_{k^r} = \f{1}{r!} ( b_{A,k}^\dag + b_{B,k}^\dag )^r |G\rag \lag G| ( b_{A,k} + b_{B,k} )^r,
\ee
then we get the RDM
\be
\r_{A,k^r} = \f{1}{r!} \sum_{p=0}^r ( C_r^p )^2 \lag ( b_{B,k} )^{r-p} ( b_{B,k}^\dag )^{r-p} \rag_{G} (b_{A,k}^\dag)^p |G_A\rag\lag G_A| (b_{A,k})^p.
\ee
where $C_r^p$ is the binomial coefficient. Then we obtain
\be
\tr_A\r_{A,k^r}^n = \sum_{p=0}^r \f{( C_r^p )^{2n}}{(r!)^n}
                    \big[ \lag ( b_{A,k} )^p ( b_{A,k}^\dag )^p \rag_{G} \big]^n
                    \big[ \lag ( b_{B,k} )^{r-p} ( b_{B,k}^\dag )^{r-p} \rag_{G} \big]^n
                  = \sum_{p=0}^r [ C_r^p x^p (1-x)^{r-p} ]^n.
\ee
There is no additional contribution to the universal R\'enyi entropy in the state $|k^r\rag$.

\subsubsection{Double-particle state $|k_1k_2\rag$}\label{bosonsingleintervalk1k2}

\begin{figure}[tp]
  \centering
  \includegraphics[height=0.6\textwidth]{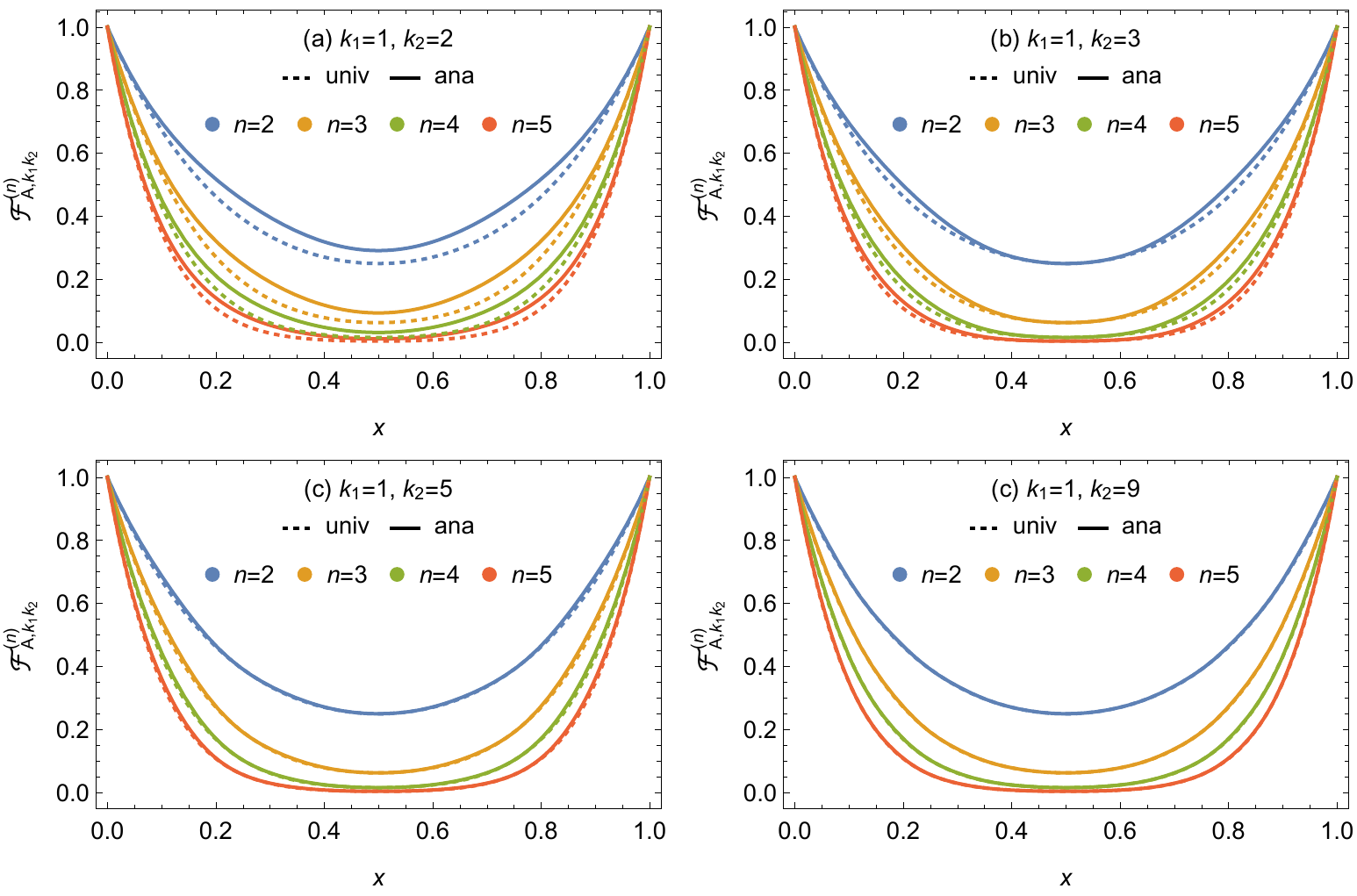}\\
  \caption{The universal R\'enyi entropy (dotted lines) and the single-interval R\'enyi entropy with corrections (solid lines) in the double-particle state $|k_1k_2\rag$ in the extremely gapped bosonic chain. We have set $m=+\inf$, $L=64$.}\label{BosonAk1k2}
\end{figure}

For the excited state with two different quasiparticles $|k_1k_2\rag$ with general nonequal $k_1,k_2$, we write the density matrix of the whole system as
\be
\r_{k_1k_2} = ( b_{A,k_1}^\dag + b_{B,k_1}^\dag ) ( b_{A,k_2}^\dag + b_{B,k_2}^\dag ) |G\rag \lag G| ( b_{A,k_2} + b_{B,k_2} ) ( b_{A,k_1} + b_{B,k_1} ),
\ee
and then we get the RDM%
\footnote{It is interesting to compare the double-particle state RDMs in the fermionic chain (\ref{rAk1k2Fermion}) and the bosonic chain (\ref{rAk1k2Boson}). Note the sign differences due to the difference of the anti-commutation and commutation relations.}
\bea \label{rAk1k2Boson}
&& \r_{A,k_1k_2} = b_{A,k_1}^\dag b_{A,k_2}^\dag |G_A\rag\lag G_A| b_{A,k_2} b_{A,k_1}
              + \lag b_{B,k_1} b_{B,k_1}^\dag \rag_{G} b_{A,k_2}^\dag |G_A\rag\lag G_A| b_{A,k_2} \nn\\
&& \phantom{\r_{A,k_1k_2} =}
              + \lag b_{B,k_2} b_{B,k_2}^\dag \rag_{G} b_{A,k_1}^\dag |G_A\rag\lag G_A| b_{A,k_1}
              + \lag b_{B,k_1} b_{B,k_2}^\dag \rag_{G} b_{A,k_1}^\dag |G_A\rag\lag G_A| b_{A,k_2} \nn\\
&& \phantom{\r_{A,k_1k_2} =}
              + \lag b_{B,k_2} b_{B,k_1}^\dag \rag_{G} b_{A,k_2}^\dag |G_A\rag\lag G_A| b_{A,k_1}
              + \lag b_{B,k_2} b_{B,k_1} b_{B,k_1}^\dag b_{B,k_2}^\dag \rag_G |G_A\rag\lag G_A|.
\eea
Finally we get the nontrivial additional contributions to the universal R\'enyi entropy
\be
\d\cF_{A,k_1k_2}^{(2)} = 
        4 (1 - 2 x)^2 \a_{12}^2
      + 4 \a_{12}^4 ,
\ee
\be \label{dcFAk1k23B}
\d\cF_{A,k_1k_2}^{(3)} = 
        3 (1 - 2 x)^2 (1 + 2 x - 2 x^2) \a_{12}^2
      - 3 (1 - 8 x + 8 x^2) \a_{12}^4 ,
\ee
\bea
&& \d\cF_{A,k_1k_2}^{(4)} = 
        4 (1 - 2 x)^2 (1 - 2 x + 6 x^2 - 8 x^3 + 4 x^4) \a_{12}^2 \nn\\
&& \phantom{\d\cF_{A,k_1k_2}^{(5)} =}
      + 8 (1 - 8 x + 27 x^2 - 38 x^3 + 19 x^4) \a_{12}^4
      + 16 (1 - 2 x)^2 \a_{12}^6 + 4 \a_{12}^8,
\eea
as well as $\d\cF_{A,k_1k_2}^{(n)}$ with $n=5,6,7$ which we will not show in this paper.
Note that $\a_{12}=|\a_{k_1-k_2}|$ is defined in (\ref{alphakdefinition}).
We get the same analytical results using the wave function method in the extremely gapped limit.
We compare the results of the universal R\'enyi entropy and the analytical results with additional terms for the state $|k_1k_2\rag$ in the figure~\ref{BosonAk1k2}.
For a small momentum difference $|k_1-k_2|$, the additional terms cannot be neglected, while for a large momentum difference, the additional terms are negligible.

\subsubsection{Triple-particle state $|k_1^2k_2\rag$}\label{bosonsingleintervalk12k2}

\begin{figure}[tp]
  \centering
  \includegraphics[height=0.6\textwidth]{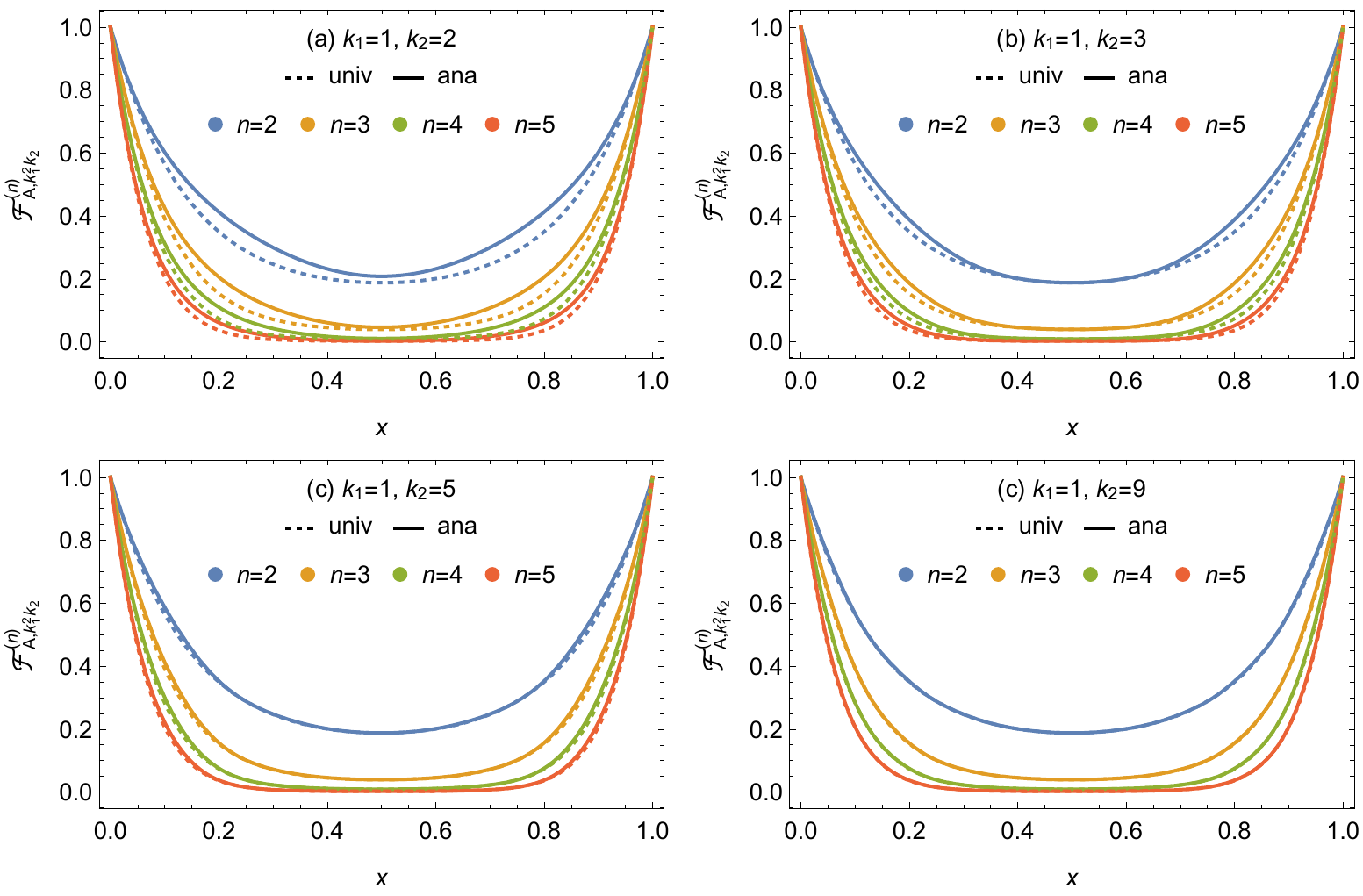}\\
  \caption{The universal R\'enyi entropy (dotted lines) and the single-interval R\'enyi entropy with corrections (solid lines) in the triple-particle state $|k_1^2k_2\rag$ in the extremely gapped bosonic chain. We have set $m=+\inf$, $L=64$.}\label{BosonAk12k2}
\end{figure}

In the triple-particle state $|k_1^2k_2\rag$ with general nonequal momenta $k_1,k_2$, the calculations are similar and we will not show the details here.
We get the additional contributions to the universal R\'enyi entropy
\be
\d\cF_{A,k_1^2k_2}^{(2)} = 
        8 (1 -2 x)^2 (1 - 3 x + 3 x^2) \a_{12}^2
      + 4 (5 - 18 x + 18 x^2) \a_{12}^4,
\ee
\bea
&& \d\cF_{A,k_1^2k_2}^{(3)} = 
        6 (1 -2 x)^2 ( 1 - 15 x^2 + 30 x^3 - 15 x^4) \a_{12}^2
      \nn\\
&& \phantom{\d\cF_{A,k_1^2k_2}^{(3)}=} - 6 (2 - 33 x + 132 x^2 - 198 x^3 + 99 x^4) \a_{12}^4
      - 12 ( 2 - 3 x) ( 1 - 3 x) \a_{12}^6,
\eea
as well as $\d\cF_{A,k_1^2k_2}^{(n)}$ with $n=4,5,6,7$
which we will not show in this paper.
We get the same analytical results using the wave function method in the extremely gapped limit.
We compared the results of the universal R\'enyi entropy and the results with additional terms for the state $|k_1^2k_2\rag$ in the figure~\ref{BosonAk12k2}.

\subsubsection{Triple-particle state $|k_1k_2k_3\rag$}

\begin{figure}[tp]
  \centering
  \includegraphics[height=0.6\textwidth]{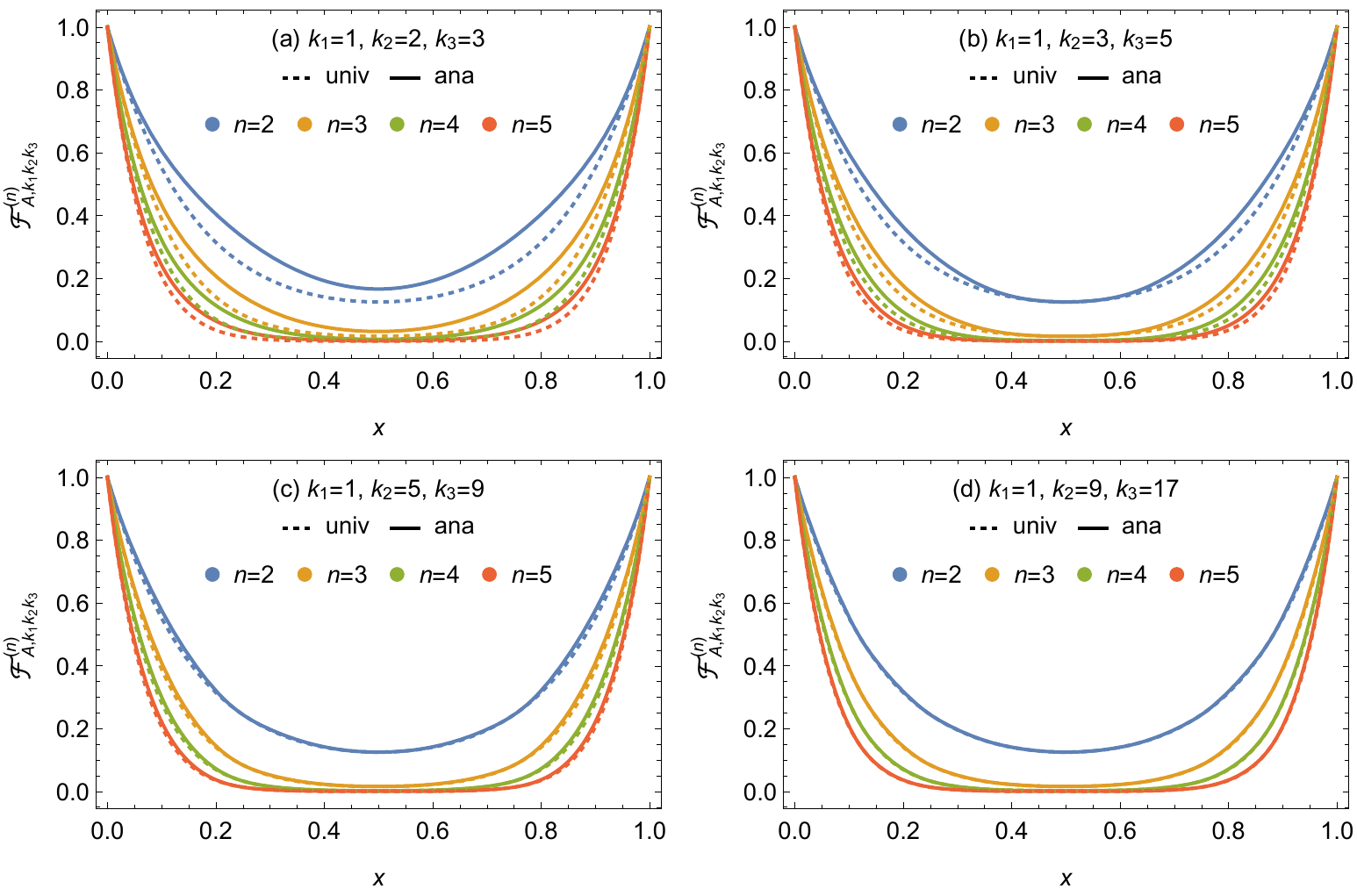}\\
  \caption{The universal R\'enyi entropy (dotted lines) and the single-interval R\'enyi entropy with corrections (solid lines) in the triple-particle state $|k_1k_2k_3\rag$ in the extremely gapped bosonic chain. We have set $m=+\inf$, $L=64$.}\label{BosonAk1k2k3}
\end{figure}

In the triple-particle state $|k_1k_2k_3\rag$ with general nonequal $k_1,k_2,k_3$, we get the additional contributions to the universal R\'enyi entropy
\bea
&& \d\cF_{A,k_1k_2k_3}^{(2)} =
        4 (1 - 2 x)^2 (1 - 2 x + 2 x^2) (\a_{12}^2 + \a_{13}^2 + \a_{23}^2)
      - 16 ( 1 - 2 x)^3 \a_{12} \a_{13} \a_{23}  \nn\\
&& \phantom{\d\cF_{A,k_1k_2k_3}^{(2)} =}
      + 16 (1 - 2 x)^2 (\a_{12}^2 \a_{13}^2 + \a_{12}^2 \a_{23}^2 + \a_{13}^2 \a_{23}^2)
      + 4 (1 - 2 x + 2 x^2) (\a_{12}^4 + \a_{13}^4 + \a_{23}^4) \nn\\
&& \phantom{\d\cF_{A,k_1k_2k_3}^{(2)} =}
      - 32 ( 1 - 2 x) \a_{12} \a_{13} \a_{23} (\a_{12}^2 + \a_{13}^2 + \a_{23}^2)
      + 80 \a_{12}^2 \a_{13}^2 \a_{23}^2,
\eea
\bea
&& \d\cF_{A,k_1k_2k_3}^{(3)} =
       3 (1 - 2 x)^2 ( 1 + 2 x - 2 x^2) (1 - 3 x + 3 x^2) (\a_{12}^2 + \a_{13}^2 + \a_{23}^2) \nn\\
&& \phantom{\d\cF_{A,k_1k_2k_3}^{(3)} =}
       - 54 x ( 1 - x)( 1 - 2 x)^3 \a_{12} \a_{13} \a_{23}
       - 6 (1 - 2 x)^2 (1 - 13 x + 13 x^2) (\a_{12}^2 \a_{13}^2  \nn\\
&& \phantom{\d\cF_{A,k_1k_2k_3}^{(3)} =} + \a_{12}^2 \a_{23}^2 + \a_{13}^2 \a_{23}^2)
       - 3 (1 - 3 x + 3 x^2) (1 - 8 x + 8 x^2) (\a_{12}^4 + \a_{13}^4 + \a_{23}^4) \nn\\
&& \phantom{\d\cF_{A,k_1k_2k_3}^{(3)} =}
       + 6 ( 1 - 2 x) (7 - 40 x + 40 x^2) \a_{12} \a_{13} \a_{23} (\a_{12}^2 + \a_{13}^2 + \a_{23}^2)
       - 144 (1 - 5 x  \nn\\
&& \phantom{\d\cF_{A,k_1k_2k_3}^{(3)} =} + 5 x^2) \a_{12}^2 \a_{13}^2 \a_{23}^2
       - 24 (1 - 2 x)^2 (\a_{12}^4 \a_{13}^2 + \a_{12}^2 \a_{13}^4 + \a_{12}^4 \a_{23}^2 + \a_{13}^4 \a_{23}^2 \nn\\
&& \phantom{\d\cF_{A,k_1k_2k_3}^{(3)} =} + \a_{12}^2 \a_{23}^4 + \a_{13}^2 \a_{23}^4)
       + 96 ( 1 - 2 x) \a_{12} \a_{13} \a_{23} (\a_{12}^2 \a_{13}^2 + \a_{12}^2 \a_{23}^2 + \a_{13}^2 \a_{23}^2) \nn\\
&& \phantom{\d\cF_{A,k_1k_2k_3}^{(3)} =}
       - 24 \a_{12}^2 \a_{13}^2 \a_{23}^2 (\a_{12}^2 + \a_{13}^2 + \a_{23}^2),
\eea
as well as $\d\cF_{A,k_1k_2k_3}^{(n)}$ with $n=4,5$
which we will not show the details here.
We get the same analytical results using the wave function method in the extremely gapped limit.
We compare the results of the universal R\'enyi entropy and the results with additional contributions for the state $|k_1k_2k_3\rag$ in the figure~\ref{BosonAk1k2k3}.

\subsubsection{Slightly gapped bosonic chain}

\begin{figure}[tp]
  \centering
  \includegraphics[width=0.99\textwidth]{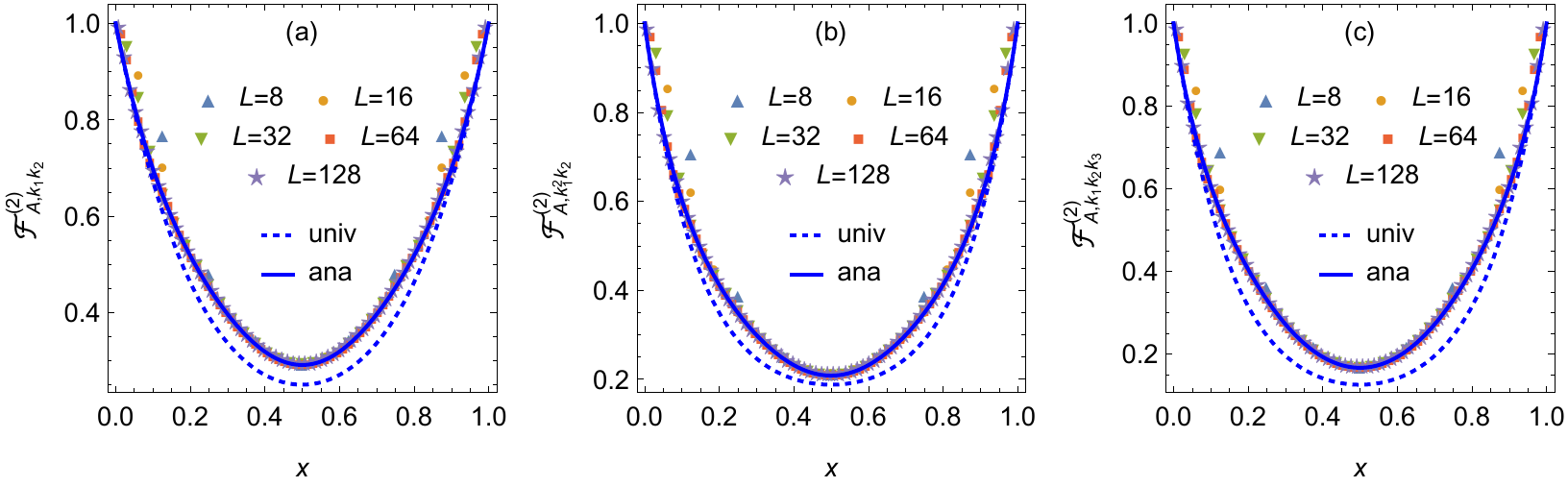}\\
  \caption{The universal R\'enyi entropy (dotted lines), the analytical single-interval R\'enyi entropy in the extremely gapped bosonic chain (solid lines), and the numerical single-interval R\'enyi entropy in the slightly gapped bosonic chain (symbols) in the double-particle state $|k_1k_2\rag$ (left), the triple-particle state $|k_1^2k_2\rag$ (middle) and the triple-particle state $|k_1k_2k_3\rag$ (right).
  We have set $m=10^{-4}$, $(k_1,k_2,k_3)=(1,2,3)+\f{L}{8}$. For the analytical results we have set $L=+\inf$.}\label{BosonAslightlygapped}
\end{figure}

We have calculated the analytical expressions of the R\'enyi entropy in the extremely gapped harmonic chain.
We compare the results of the universal R\'enyi entropy, the results with additional corrections in the extremely gapped bosonic chain, and the numerical results in the double-particle state $|k_1k_2\rag$, the triple-particle state $|k_1^2k_2\rag$ and the triple-particle state $|k_1k_2k_3\rag$ in the slightly gapped harmonic chain in the figure~\ref{BosonAslightlygapped}.
We see that the results of the R\'enyi entropy with the additional correction terms in the extremely gapped bosonic chain are still valid as long as all the momenta of the excited quasiparticles are large.

\subsection{Double interval}

We consider the double interval on a circular chain as shown in figure~\ref{subsystems}.
The calculations are the same as those of the single-interval case.
We can derive analytically the results using either the subsystem mode method or the wave function method.
We will not show the details here.

\subsubsection{Multi-particle state with equal momenta $|k^r\rag$}

There is no additional contribution to the universal double-interval R\'enyi entropy in the multi-particle state with equal momenta $|k^r\rag$.

\subsubsection{Double-particle state $|k_1k_2\rag$}

\begin{figure}[p]
  \centering
  \includegraphics[height=0.975\textwidth]{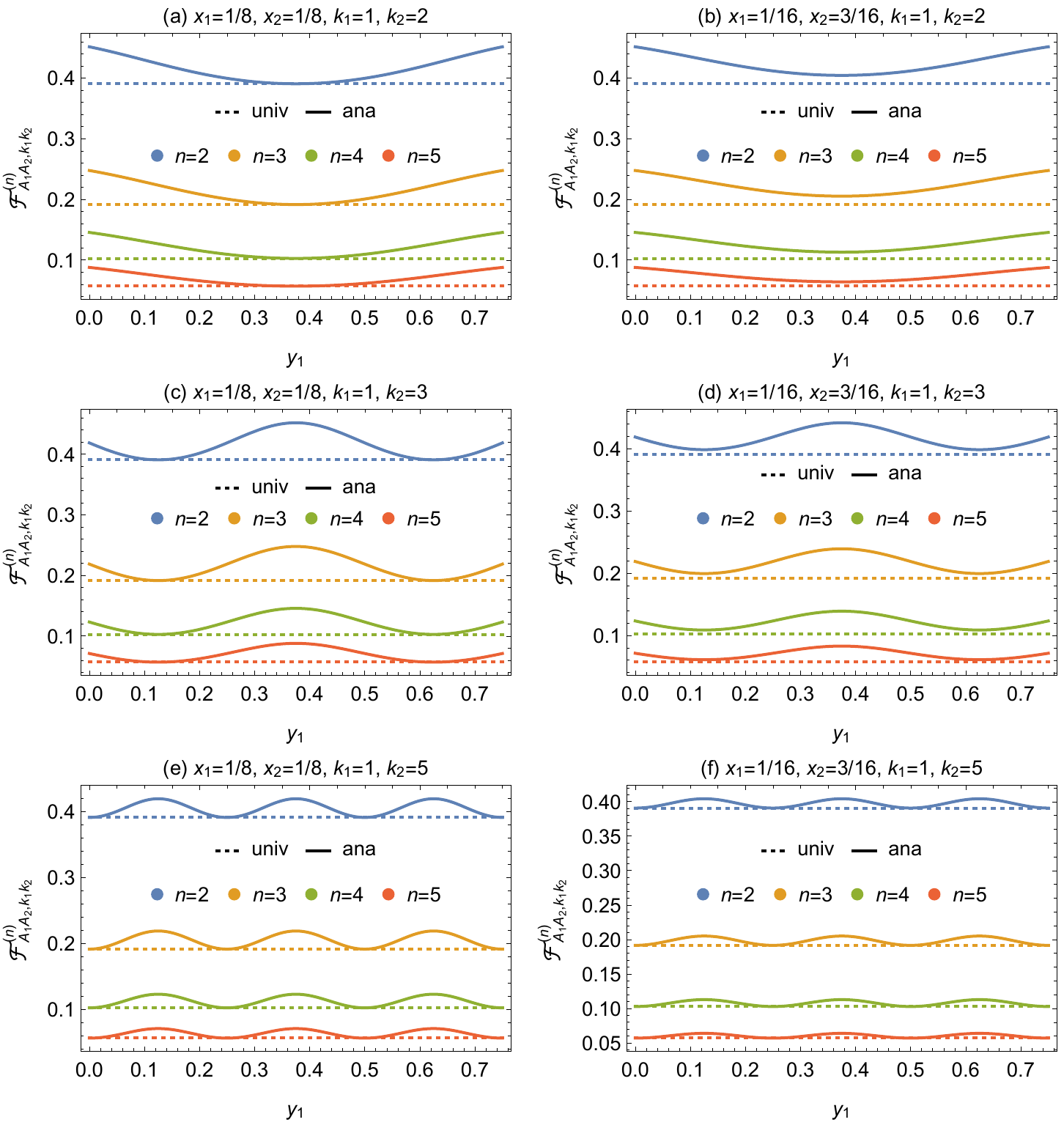}\\
  \caption{The universal R\'enyi entropy (dotted lines) and the double-interval R\'enyi entropy with corrections (solid lines) in the double-particle state $|k_1k_2\rag$ in the extremely gapped bosonic chain. We have set $m=+\inf$, $L=64$.}\label{BosonA1A2k1k2}
\end{figure}

We get analytically the additional contributions $\d\cF_{A_1A_2,k_1k_2}^{(n)}$ with $n=2,3,\cdots,7$ to the universal double-interval R\'enyi entropy $\cF_{A_1A_2,p_1p_2}^{(n),\univ}$, in the double-particle state $|k_1k_2\rag$, in the extremely gapped bosonic chain.
The results are just the expressions $\d\cF_{A,k_1k_2}^{(n)}$ presented in the subsection~\ref{bosonsingleintervalk1k2} after the following substitution
\be
| \a_{k_1-k_2} | \to | \b_{k_1-k_2} |.
\ee
We compare the universal double-interval R\'enyi entropy and the R\'enyi entropy with corrections in the figure~\ref{BosonA1A2k1k2}.

\subsubsection{Triple-particle state $|k_1^2k_2\rag$}

\begin{figure}[p]
  \centering
  \includegraphics[height=0.975\textwidth]{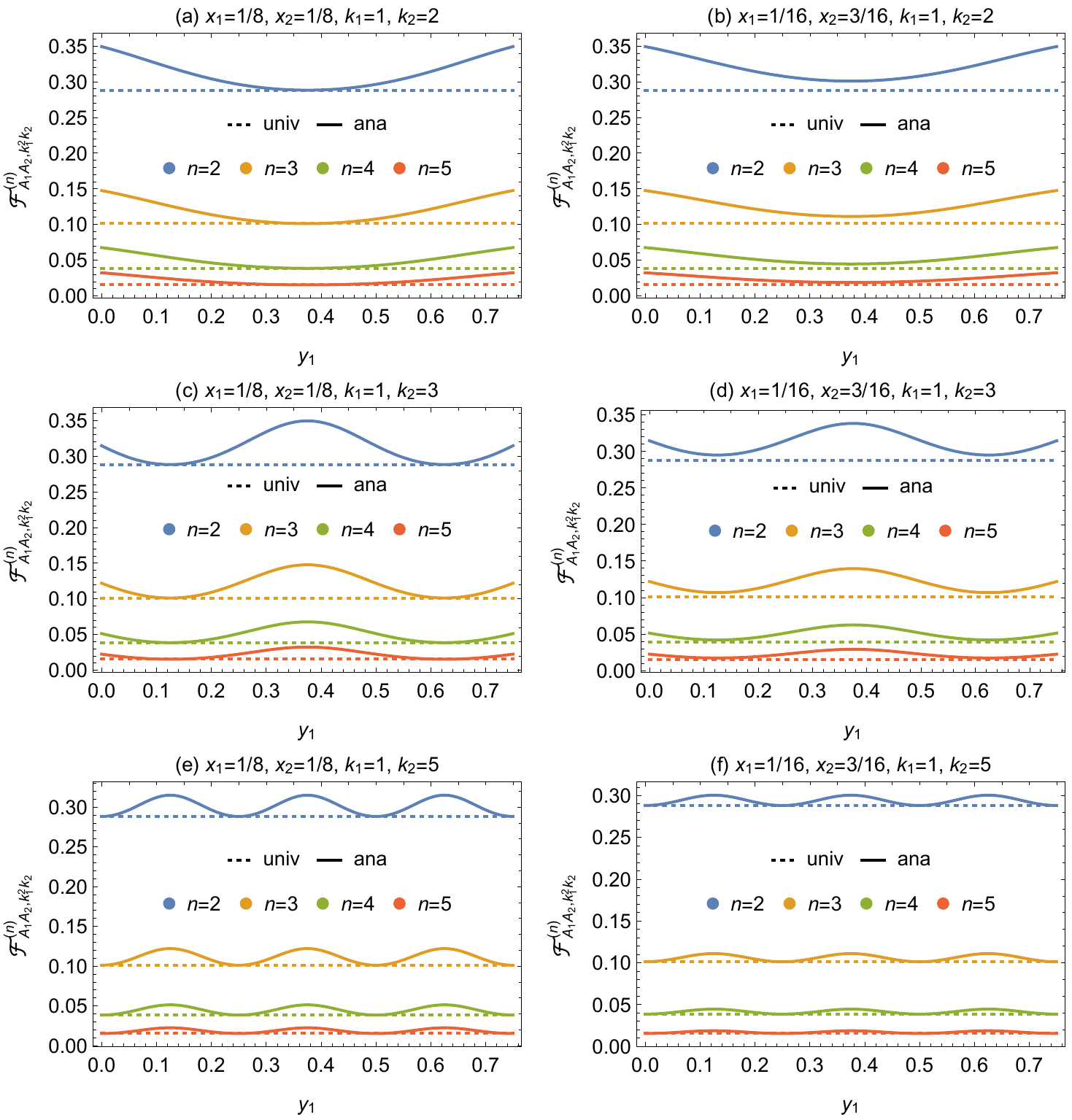}\\
  \caption{The universal R\'enyi entropy (dotted lines) and the double-interval R\'enyi entropy with corrections (solid lines) in the triple-particle state $|k_1^2k_2\rag$ in the extremely gapped bosonic chain. We have set $m=+\inf$, $L=64$.}\label{BosonA1A2k12k2}
\end{figure}

We calculate analytically the additional terms $\d\cF_{A_1A_2,k_1^2k_2}^{(n)}$ with $n=2,3,\cdots,7$ in the triple-particle state $|k_1^2k_2\rag$.
The expressions are just the results $\d\cF_{A,k_1^2k_2}^{(n)}$ in subsection~\ref{bosonsingleintervalk12k2} by sending
\be
| \a_{k_1-k_2} | \to | \b_{k_1-k_2} |.
\ee
We compare the universal double-interval R\'enyi entropy and the universal R\'enyi entropy in figure~\ref{BosonA1A2k12k2}.

\subsubsection{Triple-particle state $|k_1k_2k_3\rag$}

\begin{figure}[p]
  \centering
  \includegraphics[height=0.975\textwidth]{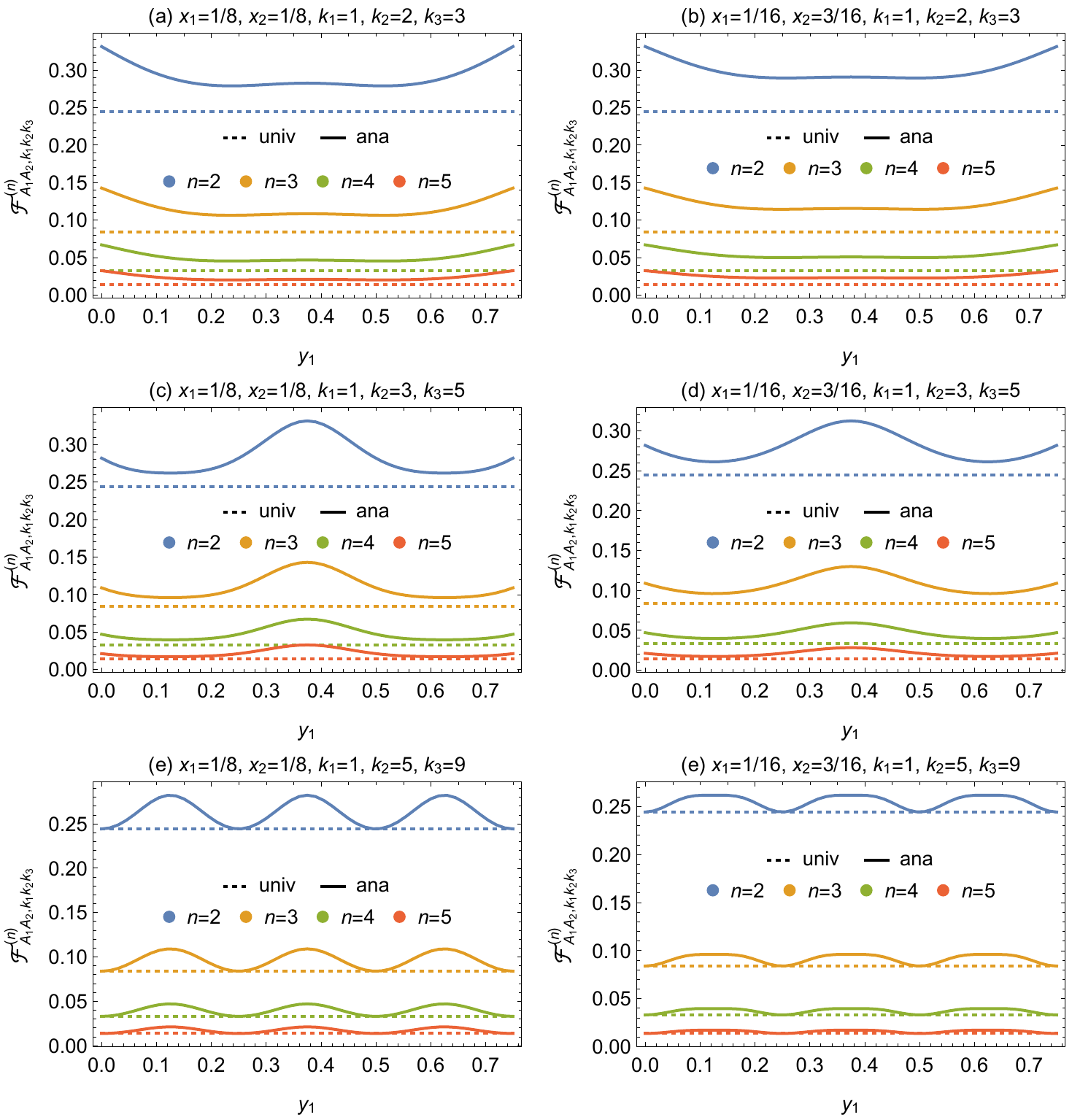}\\
  \caption{The universal R\'enyi entropy (dotted lines) and the double-interval R\'enyi entropy with corrections (solid lines) in the triple-particle state $|k_1k_2k_3\rag$ in the extremely gapped bosonic chain. We have set $m=+\inf$, $L=64$.}\label{BosonA1A2k1k2k3}
\end{figure}

We calculate the additional contributions to the universal double-interval R\'enyi entropy in the triple-particle state $|k_1k_2k_3\rag$ as
\bea
&& \d\cF_{A_1A_2,k_1k_2k_3}^{(2)} =
   4 (1 - 2 x)^2 (1 - 2 x + 2 x^2) \g_{k_1k_2k_3}
 - 8 (1 - 2 x)^3 \d_{k_1k_2k_3}  \nn\\
&& \phantom{\d\cF_{A_1A_2,k_1k_2k_3}^{(2)} =}
 + 4 (1 - 2 x + 2 x^2) \g_{k_1k_2k_3}^2
 + 8 (1 - 6 x + 6 x^2) \e_{k_1k_2k_3}  \nn\\
&& \phantom{\d\cF_{A_1A_2,k_1k_2k_3}^{(2)} =}
 - 16 (1 - 2 x) \g_{k_1k_2k_3} \d_{k_1k_2k_3}
 + 8 \d_{k_1k_2k_3}^2
 + 48 \z_{k_1k_2k_3},
\eea
\bea \label{dcFA1A2k1k2k33B}
&& \d\cF_{A_1A_2,k_1k_2k_3}^{(3)} =
  3 (1 - 2 x)^2 (1 + 2 x - 2 x^2) (1 - 3 x + 3 x^2) \g_{k_1k_2k_3}
  - 27 x (1 - x) (1 - 2 x)^3 \d_{k_1k_2k_3} \nn\\
&& \phantom{\d\cF_{A_1A_2,k_1k_2k_3}^{(3)} =}
  - 3 (1 - 11 x + 35 x^2 - 48 x^3 + 24 x^4) \g_{k_1k_2k_3}^2
  + 12 x (1 - x) (3 - 14 x + 14 x^2) \e_{k_1k_2k_3} \nn\\
&& \phantom{\d\cF_{A_1A_2,k_1k_2k_3}^{(3)} =}
  + 3 (1 - 2 x) (7 - 40 x + 40 x^2) \g_{k_1k_2k_3} \d_{k_1k_2k_3}
  - 24 (1 - 2 x)^2 \g_{k_1k_2k_3} \e_{k_1k_2k_3} \nn\\
&& \phantom{\d\cF_{A_1A_2,k_1k_2k_3}^{(3)} =}
  - 3 (5 - 24 x + 24 x^2) \d_{k_1k_2k_3}^2
  - 12 (1 - 12 x + 12 x^2) \z_{k_1k_2k_3} \nn\\
&& \phantom{\d\cF_{A_1A_2,k_1k_2k_3}^{(3)} =}
  + 48 (1 - 2 x) \d_{k_1k_2k_3} \e_{k_1k_2k_3}
  - 24 \g_{k_1k_2k_3} \z_{k_1k_2k_3},
\eea
with  $\g_{k_1k_2k_3}$ and $\d_{k_1k_2k_3}$ defined in (\ref{gk1k2k3dk1k2k3}) and
\bea \label{ek1k2k3zk1k2k3}
&& \e_{k_1k_2k_3} = |\b_{k_1-k_2}|^2|\b_{k_1-k_3}|^2 + |\b_{k_1-k_2}|^2|\b_{k_2-k_3}|^2 + |\b_{k_1-k_3}|^2|\b_{k_2-k_3}|^2, \nn\\
&& \z_{k_1k_2k_3} = |\b_{k_1-k_2}|^2|\b_{k_1-k_3}|^2|\b_{k_2-k_3}|^2.
\eea
We also calculate analytically $\d\cF_{A_1A_2,k_1k_2k_3}^{(n)}$ with $n=4,5$ which we will not show in this paper.
We compare the universal double-interval R\'enyi entropy and the R\'enyi entropy with corrections in figure~\ref{BosonA1A2k1k2k3}.

\subsubsection{Slightly gapped bosonic chain}

\begin{figure}[tp]
  \centering
  \includegraphics[width=0.99\textwidth]{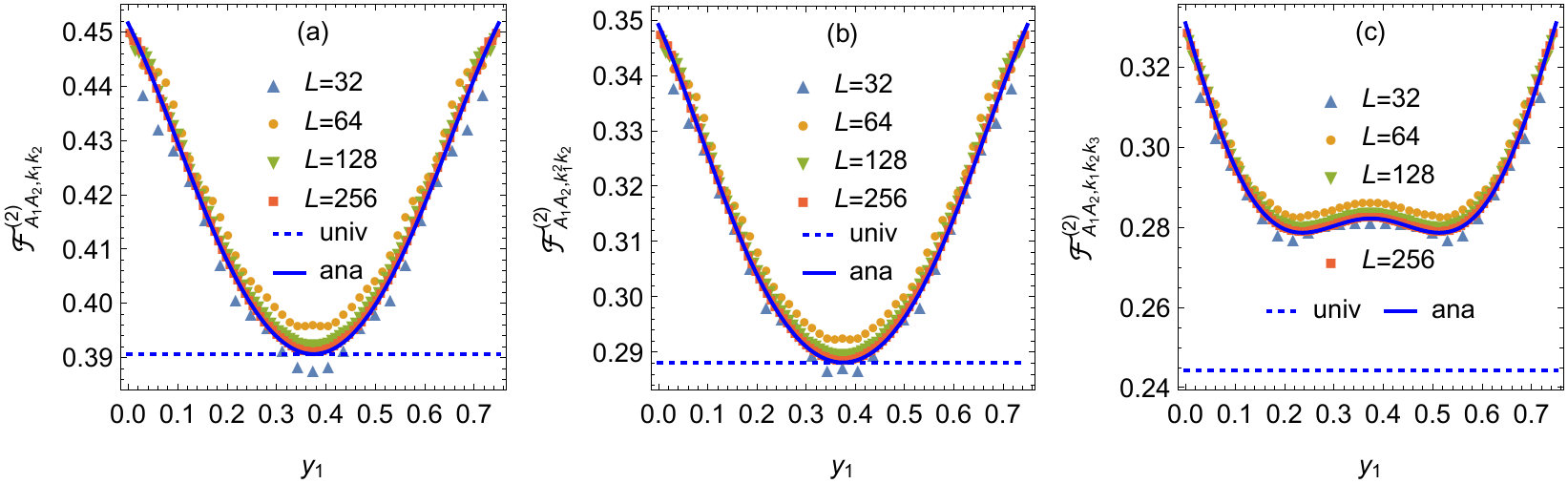}\\
  \caption{The universal R\'enyi entropy (dotted lines), the analytical double-interval  R\'enyi entropy in the extremely gapped bosonic chain (solid lines), and the numerical R\'enyi entropy in the slightly gapped bosonic chain (symbols) in the double-particle state $|k_1k_2\rag$ (left), the triple-particle state $|k_1^2k_2\rag$ (middle) and the triple-particle state $|k_1k_2k_3\rag$ (right). We have set $m=10^{-4}$, $(k_1,k_2,k_3)=(1,2,3)+\f{L}{8}$, $x_1=x_2=\f18$. For the analytical results we have set $L=+\inf$.}\label{BosonA1A2slightlygapped}
\end{figure}

In the slightly gapped bosonic chain, we can calculate the double-interval R\'enyi entropy numerically in the double-particle state $|k_1k_2\rag$, triple-particle state $|k_1^2k_2\rag$ and triple-particle state $|k_1k_2k_3\rag$.
We compare the numerical results with the universal double-interval R\'enyi entropy and the R\'enyi entropy with corrections in the extremely gapped bosonic chain in figure~\ref{BosonA1A2slightlygapped}.
We see that the new R\'enyi entropy with additional corrections in the extremely gapped bosonic chain is still valid in the limit of large momenta.

\subsection{Multiple intervals}

Similar to the extremely gapped fermionic chain, the generalization from double interval to multiple intervals is easy.
We will not show the details here.

\subsection{A permanent formula in the extremely gapped limit}

For the general excited state $|K\rag=|k_1^{r_1}k_2^{r_2}\cdots k_s^{r_s}\rag$, there are totally $|K|=R=\sum_{i=1}^s r_i$ number of excited quasiparticles.
In the extremely gapped limit of the bosonic chain, using the wave function method, in the appendix~\ref{appMWF}, we show that the excited state R\'enyi entropy of a subsystem $A$, which could be either a single interval or general multiple intervals, is just the permanent of a certain matrix as follows:
\be \label{TheFormulaB}
\cF_{A,k_1^{r_1}k_2^{r_2}\cdots k_s^{r_s}}^{(n)} = \f{\per\O_{A,k_1^{r_1}k_2^{r_2}\cdots k_s^{r_s}}^{(n)}}{\prod_{i=1}^s ( r_i! )^n},
\ee
where the definition of the $n|K|\times n|K|$ matrix $\O_{A,k_1^{r_1}k_2^{r_2}\cdots k_s^{r_s}}^{(n)}$ can be found in (\ref{OAKnIJdefinition}) and (\ref{tdVa1k1TtdUa2k2}).
We show the details of the wave function method and the proof of the formula (\ref{TheFormulaB}), i.e.\ (\ref{TheFormulaBapp}), in appendix~\ref{appMWF}.

For example, to calculate $\cF_{A,k}^{(n)}=\cF_{A,p}^{(n),\univ}=x^n+(1-x)^n$ with a general integer $n\geq2$ we need the $n\times n$ matrix
\be \label{OAkn}
\O_{A,k}^{(n)} =
\left(
\begin{array}{ccccc}
 1-x & x   &        &        &  \\
     & 1-x & x      &        &  \\
     &     & \ddots & \ddots &  \\
     &     &        & 1-x    & x \\
 x   &     &        &        & 1-x \\
\end{array}
\right),
\ee
to calculate $\cF_{A,k^2}^{(3)}=\cF_{A,p^2}^{(3),\univ}=x^6+[2x(1-x)]^3+(1-x)^6$ we need the $6\times6$ matrix
\be
\O_{A,k^2}^{(3)} =
\left(
\begin{array}{cccccc}
 1-x & x & 0 & 1-x & x & 0 \\
 0 & 1-x & x & 0 & 1-x & x \\
 x & 0 & 1-x & x & 0 & 1-x \\
 1-x & x & 0 & 1-x & x & 0 \\
 0 & 1-x & x & 0 & 1-x & x \\
 x & 0 & 1-x & x & 0 & 1-x \\
\end{array}
\right),
\ee
and to calculate $\cF_{A,k_1k_2}^{(3)}$ (\ref{dcFAk1k23B}) with general momenta $k_1\neq k_2$ we need the $6\times6$ matrix
\be
\O_{A,k_1k_2}^{(3)} =
\left(
\begin{array}{cccccc}
 1-x           & x             & 0             & -\b_{k_1-k_2} & \b_{k_1-k_2}  & 0 \\
 0             & 1-x           & x             & 0             & -\b_{k_1-k_2} & \b_{k_1-k_2} \\
 x             & 0             & 1-x           &  \b_{k_1-k_2} & 0             & -\b_{k_1-k_2} \\
 -\b_{k_2-k_1} & \b_{k_2-k_1} & 0              & 1-x           & x             & 0 \\
 0             & -\b_{k_2-k_1} & \b_{k_2-k_1}  & 0             & 1-x           & x   \\
 \b_{k_2-k_1}  & 0             & -\b_{k_2-k_1} & x             & 0             & 1-x \\
\end{array}
\right),
\ee
with the definition of $\b_k$ (\ref{betakdefinitionX}).
Note that these examples of $\O_{A,K}^{(n)}$ apply to not only single-interval but also multi-interval cases.

In the limit that all the momentum differences are large, i.e.\ $| k_{i_1} - k_{i_2} | \to +\inf$ for all $i_1\neq i_2$, we have $\b_{k_{i_1} - k_{i_2}} \to 0$, and the excited state R\'enyi entropy should approach the universal R\'enyi entropy (\ref{cFAk1r1k2r2cdotsksrsnsup}).
One can use this idea to prove the existence of the most general universal term as follows:
We first note that for the state $|k^r\rag$ there is the $nr\times nr$ matrix $\O_{A,k^{r}}^{(n)}$ written in the form of  $r\times r$ blocks
\be \label{OAkrn}
\O_{A,k^{r}}^{(n)} =
\left(
\begin{array}{cccc}
 \O_{A,k}^{(n)} & \O_{A,k}^{(n)} & \cdots & \O_{A,k}^{(n)}  \\
 \O_{A,k}^{(n)} & \O_{A,k}^{(n)} & \cdots & \O_{A,k}^{(n)} \\
 \vdots         & \vdots         & \ddots & \vdots \\
 \O_{A,k}^{(n)} & \O_{A,k}^{(n)} & \cdots & \O_{A,k}^{(n)} \\
\end{array}
\right) = J_r \otimes \O_{A,k}^{(n)},
\ee
with the $n\times n$ matrix $\O_{A,k}^{(n)}$ (\ref{OAkn}) and the $r\times r$ matrix $J_r$ all of whose entries are 1.
There is the permanent
\be \label{PermGenralSuper}
\per \O_{A,k^{r}}^{(n)}= (r!)^n \sum_{p=0}^{r} [ C_{r}^p x^p (1-x)^{{r}-p} ]^n.
\ee
Since in the large momentum difference limit we already have $\b_{k_{i_1} - k_{i_2}} \to 0$ for all $i_1\neq i_2$, there is the $n|K|\times n|K|$ matrix for the general state in the form of a direct sum
\be
\O_{A,k_1^{r_1}k_2^{r_2}\cdots k_s^{r_s}}^{(n)} = \bigoplus_{i=1}^s \O_{A,k_i^{r_i}}^{(n)},
\ee
with the $n r_i\times n r_i$ matrix $\O_{A,k_i^{r_i}}^{(n)}$ for each $i$ defined as (\ref{OAkrn}).
It is easy to get the permanent
\be
\per \O_{A,k_1^{r_1}k_2^{r_2}\cdots k_s^{r_s}}^{(n)} = \prod_{i=1}^s \per \O_{A,k_i^{r_i}}^{(n)}.
\ee
This just leads to the universal R\'enyi entropy (\ref{cFAk1r1k2r2cdotsksrsnsup}).

It would be interesting to find the permanent of the matrix $\O_{A,K}^{(n)}$ without the large momentum difference limit and calculate the R\'enyi entropy for general $n$ and then take the $n\to1$ limit to calculate the von Neumann entropy.
In this respect the block circulant structure of the matrix $\O_{A,K}^{(n)}$ might be useful.

\section{XY chain}\label{secXY}

We consider the transverse field XY chain of $L$ sites
\be
H = - \sum_{j=1}^L \Big( \f{1+\g}{4} \s_j^x\s_{j+1}^x + \f{1-\g}{4} \s_j^y\s_{j+1}^y + \f\l2 \s_j^z \Big),
\ee
with periodic or antiperiodic boundary conditions for the Pauli matrices $\s_j^{x,y,z}$.
The XY chain can be mapped to the fermionic chain by some nonlocal transformations, but they have different local degrees of freedom.
It can be diagonalized as
\be
H = \sum_k \ve_k \Big( c_k^\dag c_k - \f12 \Big), ~~
\ve_k = \sr{ \Big(\l - \cos\f{2\pi k}{L}\Big)^2 + \g^2 \sin^2\f{2\pi k}{L} },
\ee
by successive Jordan-Wigner transformation, Fourier transformation and Bogoliubov transformation \cite{Lieb:1961fr,katsura1962statistical,pfeuty1970one}
\be
a_j = \Big(\prod_{i=1}^{j-1}\s_i^z\Big) \s_j^+, ~~
a_j^\dag = \Big(\prod_{i=1}^{j-1}\s_i^z\Big) \s_j^-,
\ee
\be
b_k = \f{1}{\sr{L}}\sum_{j=1}^L\ep^{\ii j \vph_k}a_j, ~~
b_k^\dag = \f{1}{\sr{L}}\sum_{j=1}^L\ep^{-\ii j \vph_k}a_j^\dag,
\ee
\be
c_k = b_k \cos\f{\th_k}{2} + \ii b_{-k}^\dag \sin\f{\th_k}{2}, ~~
c_k^\dag = b_k^\dag \cos\f{\th_k}{2} - \ii b_{-k} \sin\f{\th_k}{2},
\ee
with the definitions
\be \label{sjpmvphkthk}
\s_j^\pm = \f12 ( \s_j^x \pm \ii \s_j^y ), ~~
\vph_k = \f{2\pi k}{L}, ~~
\ep^{\ii\th_k}=\f{\l-\cos\vph_k+\ii\g\sin\vph_k}{\ve_k}.
\ee
In this paper we only consider the cases that $L$ is an even integer, and we only consider the states in the NS sector.
We have the momenta
\be
k =\f{1-L}{2}, \cdots, -\f12, \f12, \cdots, \f{L-1}{2}.
\ee
The ground state $|G\rag$ is annihilated by all the lowering operators
\be
c_k|G\rag=0,
\ee
and the excited states are generated by applying the raising operators with different momenta on the ground state
\be \label{XYk1k2cdotsksrag}
|k_1k_2\cdots k_s\rag = c_{k_1}^\dag c_{k_2}^\dag \cdots c_{k_s}^\dag |G\rag.
\ee
When $s$ is an even integer it is an excited state in the periodic XY chain in terms of the Pauli matrices, and when $s$ is an odd integer it is an excited state in the antiperiodic chain.

We consider the extremely gapped limit $\l\to+\inf$ of the XY chain. The Hamiltonian is
\be
H = - \f\l2 \sum_{j=1}^L \s_j^z,
\ee
in which the spins decouple from each other. The ground state is just the state with spin up at each site in the $\s_j^z$ basis
\be
|G\rag=|\!\!\uparrow\uparrow\cdots\uparrow\rag
\ee
The Bogoliubov angle is vanishing $\th_k=0$, and we have
\be
c_k = \f{1}{\sr{L}}\sum_{j=1}^L\ep^{\ii j \vph_k}a_j, ~~
c_k^\dag = \f{1}{\sr{L}}\sum_{j=1}^L\ep^{-\ii j \vph_k}a_j^\dag.
\ee
Note that the ladder operators $a_j$, $a_j^\dag$ are not actually local modes, although $\s_j^+$, $\s_j^-$ are. The ground state is also annihilated by all the lowering operators $a_j$, $\s_j^+$
\be
a_j|G\rag=\s_j^+|G\rag=0, ~ j=1,2,\cdots,L.
\ee

\subsection{Single interval}

We consider an interval with $\ell$ consecutive sites $A=[1,\ell]$ and its complement $B=[\ell+1,L]$ on the circular XY chain of $L$ sites.
For the analytical calculations of the R\'enyi entropy, it is convenient to define the subsystem modes in the extremely gapped limit
\bea
&& c_{A,k} = \f{1}{\sr{L}}\sum_{j\in A} \ep^{\ii j \vph_k}a_j, ~~
   c_{A,k}^\dag = \f{1}{\sr{L}}\sum_{j\in A} \ep^{-\ii j \vph_k}a_j^\dag, \nn\\
&& c_{B,k} = \f{1}{\sr{L}}\sum_{j\in B} \ep^{\ii j \vph_k}a_j, ~~
   c_{B,k}^\dag = \f{1}{\sr{L}}\sum_{j\in B} \ep^{-\ii j \vph_k}a_j^\dag.
\eea
There are anti-commutation relations
\be
\{ c_{A,k}, c_{A,k}^\dag \} = x, ~~
\{ c_{B,k}, c_{B,k}^\dag \} = 1-x,
\ee
and for $k_1\neq k_2$ we have
\be
\{ c_{A,k_1}, c_{A,k_2}^\dag \} = - \{ c_{B,k_1}, c_{B,k_2}^\dag \} = \a_k,
\ee
with the definition of $\a_k$ presented in (\ref{alphakdefinition}).
We define the string
\be
S_A = \prod_{j\in A}\s_j^z,
\ee
and the operators
\be
\td c_{B,k} = S_A c_{B,k}, ~~ \td c_{B,k}^\dag = S_A c_{B,k}^\dag.
\ee
Note that $\td c_{B,k}$, $\td c_{B,k}^\dag$ are defined locally in $B$, while $c_{B,k}$, $c_{B,k}^\dag$ are not.
Of course, $c_{A,k}$, $c_{A,k}^\dag$ are defined locally in $A$.

\subsubsection{Single-particle state $|k\rag$}

In the single-particle state $|k\rag$, we write the density matrix of the whole system as
\be
\r_k = ( c_{A,k}^\dag + c_{B,k}^\dag ) |G\rag \lag G| ( c_{A,k} + c_{B,k} ).
\ee
Using
\be
\tr_B ( c_{A,k}^\dag |G\rag \lag G| c_{A,k}  ) = c_{A,k}^\dag |G_A\rag\lag G_A| c_{A,k},
\ee
and
\bea
&& \tr_B ( c_{B,k}^\dag |G\rag \lag G| c_{B,k}  ) = \tr_B ( S_A \td c_{B,k}^\dag |G\rag \lag G| S_A \td c_{B,k}  )
                                               = \tr_B ( \td c_{B,k}^\dag |G\rag \lag G| \td c_{B,k}  ) \nn\\
&& \phantom{\tr_B ( c_{B,k}^\dag |G\rag \lag G| c_{B,k}  )}
                                               = \lag \td c_{B,k} \td c_{B,k}^\dag \rag_{G_B} |G_A\rag\lag G_A|
                                               = \lag c_{B,k} c_{B,k}^\dag \rag_{G} |G_A\rag\lag G_A|,
\eea
we get the RDM
\be
\r_{A,k} = c_{A,k}^\dag |G_A\rag\lag G_A| c_{A,k} + \lag c_{B,k} c_{B,k}^\dag \rag_{G}   |G_A\rag\lag G_A|.
\ee
Then we get
\be
\tr_A\r_{A,k}^n = \lag c_{A,k} c_{A,k}^\dag \rag_{G}^n + \lag c_{B,k} c_{B,k}^\dag \rag_{G}^n = x^n+(1-x)^n.
\ee
There is no additional contribution to the R\'enyi entropy in the single-particle state $|k\rag$.

\subsubsection{Double-particle state $|k_1k_2\rag$}

In the double-particle state $|k_1k_2\rag$ with general $k_1,k_2$, we write the density matrix of the whole system as
\be
\r_{k_1k_2} = ( c_{A,k_1}^\dag + c_{B,k_1}^\dag ) ( c_{A,k_2}^\dag + c_{B,k_2}^\dag ) |G\rag \lag G| ( c_{A,k_2} + c_{B,k_2} ) ( c_{A,k_1} + c_{B,k_1} ),
\ee
and we get the RDM
\bea
&& \r_{A,k_1k_2} = c_{A,k_1}^\dag c_{A,k_2}^\dag |G_A\rag\lag G_A| c_{A,k_2} c_{A,k_1}
              + \lag c_{B,k_1} c_{B,k_1}^\dag \rag_{G} c_{A,k_2}^\dag |G_A\rag\lag G_A| c_{A,k_2} \nn\\
&& \phantom{\r_{A,k_1k_2} =}
              + \lag c_{B,k_2} c_{B,k_2}^\dag \rag_{G} c_{A,k_1}^\dag |G_A\rag\lag G_A| c_{A,k_1}
              - \lag c_{B,k_1} c_{B,k_2}^\dag \rag_{G} c_{A,k_1}^\dag |G_A\rag\lag G_A| c_{A,k_2} \nn\\
&& \phantom{\r_{A,k_1k_2} =}
              - \lag c_{B,k_2} c_{B,k_1}^\dag \rag_{G} c_{A,k_2}^\dag |G_A\rag\lag G_A| c_{A,k_1}
              + \lag c_{B,k_2} c_{B,k_1} c_{B,k_1}^\dag c_{B,k_2}^\dag \rag_{G} |G_A\rag\lag G_A|.
\eea
We have used for example:
\bea
&&  \tr_B ( c_{B,k_1}^\dag c_{B,k_2}^\dag |G\rag \lag G| c_{B,k_2} c_{B,k_1} )
= \tr_B ( S_A \td c_{B,k_1}^\dag  S_A \td c_{B,k_2}^\dag |G\rag \lag G|  S_A \td c_{B,k_2}  S_A \td c_{B,k_1} ) \nn\\
&& \phantom{\tr_B ( c_{B,k_1}^\dag c_{B,k_2}^\dag |G\rag \lag G| c_{B,k_2} c_{B,k_1} )}
= \lag \td c_{B,k_2} \td c_{B,k_1} \td c_{B,k_1}^\dag \td c_{B,k_2}^\dag \rag_{G_B} |G_A\rag\lag G_A| \nn\\
&& \phantom{\tr_B ( c_{B,k_1}^\dag c_{B,k_2}^\dag |G\rag \lag G| c_{B,k_2} c_{B,k_1} )}
= \lag c_{B,k_2} c_{B,k_1} c_{B,k_1}^\dag c_{B,k_2}^\dag \rag_{G} |G_A\rag\lag G_A|,
\eea
\bea
&&  \tr_B ( c_{A,k_1}^\dag c_{B,k_2}^\dag |G\rag \lag G| c_{A,k_2} c_{B,k_1})
= \tr_B ( c_{A,k_1}^\dag S_A \td c_{B,k_2}^\dag |G\rag \lag G| c_{A,k_2} S_A \td c_{B,k_1}) \nn\\
&& \phantom{\tr_B ( c_{A,k_1}^\dag c_{B,k_2}^\dag |G\rag \lag G| c_{A,k_2} c_{B,k_1})}
= - \lag \td c_{B,k_1} \td c_{B,k_2}^\dag \rag_{G_B} c_{A,k_1}^\dag |G_A\rag\lag G_A| c_{A,k_2} \nn\\
&& \phantom{\tr_B ( c_{A,k_1}^\dag c_{B,k_2}^\dag |G\rag \lag G| c_{A,k_2} c_{B,k_1})}
= - \lag c_{B,k_1} c_{B,k_2}^\dag \rag_{G} c_{A,k_1}^\dag |G_A\rag\lag G_A| c_{A,k_2}.
\eea
Note the minus signs due to the anti-commutation relations.
Then one can calculate $\tr_A\r_{A,k_1k_2}^n$ with $n=2,3,4,5,6,7$.
They are the same as those in the fermionic chain presented in the subsection~\ref{singleintervalk1k2}.

\subsubsection{Triple-particle state $|k_1k_2k_3\rag$}

The analytical calculations of the R\'enyi entropy in the triple-particle state $|k_1k_2k_3\rag$ with general $k_1,k_2,k_3$ are similar to the above.
The results are the same as those in the fermionic chain in subsection~\ref{singleintervalk1k2k3}.
We will not show the details here.

\subsection{Double interval}

The double-interval R\'enyi entropies in the XY chain were calculated numerically in \cite{Furukawa:2008uk,Facchi:2008Entanglement,Alba:2009ek,Igloi:2009On,Fagotti:2010yr}. Here we will adopt the efficient method presented in \cite{Fagotti:2010yr} for the numerical calculations.

For the analytical calculations of the double-interval R\'enyi entropy, we define for $i=1,2$
\bea
&& c_{A_i,k} = \f{1}{\sr{L}}\sum_{j\in A_i} \ep^{\ii j \vph_k}a_j, ~~
   c_{A_i,k}^\dag = \f{1}{\sr{L}}\sum_{j\in A_i} \ep^{-\ii j \vph_k}a_j^\dag, \nn\\
&& c_{B_i,k} = \f{1}{\sr{L}}\sum_{j\in B_i} \ep^{\ii j \vph_k}a_j, ~~
   c_{B_i,k}^\dag = \f{1}{\sr{L}}\sum_{j\in B_i} \ep^{-\ii j \vph_k}a_j^\dag.
\eea
We define the strings for $X=A_1,A_2,B_1,B_2$
\be
S_X = \prod_{j \in X} \s_j^z.
\ee
Then we define
\bea
&& \td c_{A_1,k} = c_{A_1,k}, ~~
   \td c_{A_2,k} = S_{B_1} c_{A_2,k}, ~~
   \td c_{B_1,k} = S_{A_1} c_{B_1,k}, ~~
   \td c_{B_2,k} = S_{A_1} S_{A_2} c_{B_2,k}, \nn\\
&& \td c_{A_1,k}^\dag = c_{A_1,k}^\dag, ~~
   \td c_{A_2,k}^\dag = S_{B_1} c_{A_2,k}^\dag, ~~
   \td c_{B_1,k}^\dag = S_{A_1} c_{B_1,k}^\dag, ~~
   \td c_{B_2,k}^\dag = S_{A_1} S_{A_2} c_{B_2,k}^\dag.
\eea
Note that $\td c_{A_i,k}$, $\td c_{A_i,k}^\dag$ are defined locally in $A=A_1\cup A_2$, and $\td c_{B_i,k}$, $\td c_{B_i,k}^\dag$ are defined locally in $B=B_1\cup B_2$. The operators $\td c_{A_i,k}$, $\td c_{A_i,k}^\dag$ commute with the operators $\td c_{B_i,k}$, $\td c_{B_i,k}^\dag$.
For later convenience, we define for $X=A_1,A_2,A,B_1,B_2,B$ the factors
\be
\b_{X,k} = \f{1}{L} \sum_{j\in X} \ep^{\f{2\pi\ii j k}{L}}.
\ee

\subsubsection{Single-particle state $|k\rag$}

The single-particle state can be written as
\be
|k\rag = ( c_{A_1,k}^\dag + c_{A_2,k}^\dag + c_{B_1,k}^\dag + c_{B_2,k}^\dag ) |G\rag
       = ( \td c_{A_1,k}^\dag + \td c_{A_2,k}^\dag + \td c_{B_1,k}^\dag + \td c_{B_2,k}^\dag ) |G\rag,
\ee
and then we get the RDM
\be
\r_{A_1A_2,k} = ( \td c_{A_1,k}^\dag + \td c_{A_2,k}^\dag ) |G_A\rag\lag G_A| ( \td c_{A_1,k} + \td c_{A_2,k} )
               + \lag ( \td c_{B_1,k} + \td c_{B_2,k} ) ( \td c_{B_1,k}^\dag + \td c_{B_2,k}^\dag ) \rag_{G_B} |G_A\rag\lag G_A|,
\ee
and finally we have
\be
\tr_{A_1A_2} \r_{A_1A_2,k}^n = x^n + y^n.
\ee
There are no additional contributions to the universal R\'enyi entropy.

\subsubsection{Double-particle state $|k_1k_2\rag$}

\begin{figure}[p]
  \centering
  \includegraphics[width=0.99\textwidth]{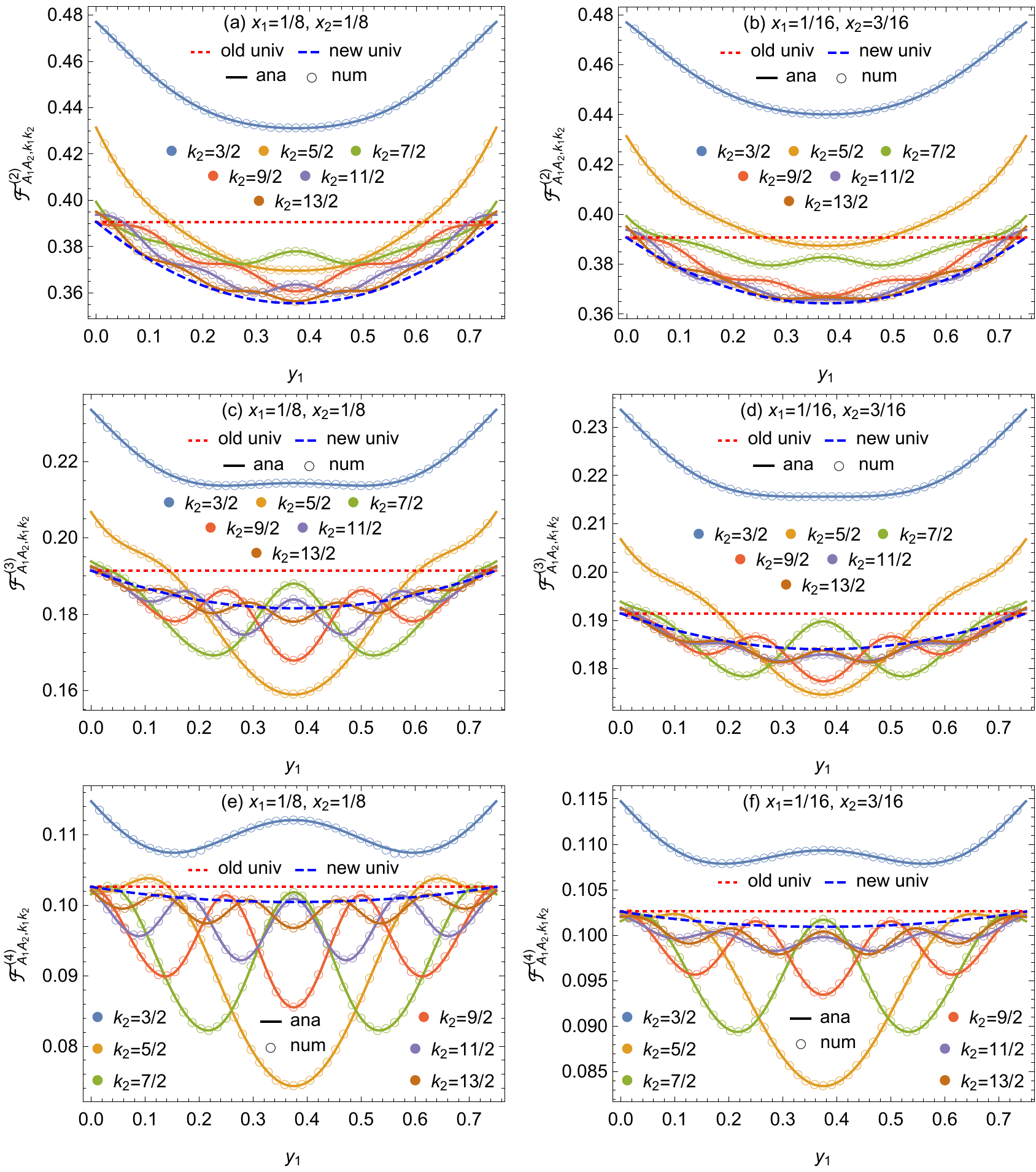}\\
  \caption{The old universal R\'enyi entropy (red dotted lines), the new universal R\'enyi entropy (blue dashed lines), and the analytical(solid lines) and numerical (empty circles) results of the double-interval R\'enyi entropy in the double-particle state $|k_1k_2\rag$ of the extremely gapped XY chain. We have set $\l=+\inf$, $L=64$, $k_1=\f12$.}\label{XYA1A2k1k2}
\end{figure}

The double-particle state can be written as
\bea
&& |k_1k_2\rag = ( c_{A_1,k_1}^\dag + c_{A_2,k_1}^\dag + c_{B_1,k_1}^\dag + c_{B_2,k_1}^\dag )
                 ( c_{A_1,k_2}^\dag + c_{A_2,k_2}^\dag + c_{B_1,k_2}^\dag + c_{B_2,k_2}^\dag ) |G\rag \nn\\
&& \phantom{|k_1k_2\rag}
    = [ ( \td c_{A_1,k_1}^\dag + \td c_{A_2,k_1}^\dag)( \td c_{A_1,k_2}^\dag + \td c_{A_2,k_2}^\dag )
       +( \td c_{A_1,k_1}^\dag - \td c_{A_2,k_1}^\dag) \td c_{B_1,k_2}^\dag \nn\\
&& \phantom{|k_1k_2\rag=}
       +\td c_{B_1,k_1}^\dag( - \td c_{A_1,k_2}^\dag + \td c_{A_2,k_2}^\dag)
       +( \td c_{A_1,k_1}^\dag + \td c_{A_2,k_1}^\dag) \td c_{B_2,k_2}^\dag \nn\\
&& \phantom{|k_1k_2\rag=}
       +\td c_{B_2,k_1}^\dag( - \td c_{A_1,k_2}^\dag - \td c_{A_2,k_2}^\dag)
       +( \td c_{B_1,k_1}^\dag + \td c_{B_2,k_1}^\dag)( \td c_{B_1,k_2}^\dag + \td c_{B_2,k_2}^\dag ) ] |G\rag.
\eea
We get the RDM
\bea
&& \r_{A_1A_2,k_1k_2} = ( \td c_{A_1,k_1}^\dag + \td c_{A_2,k_1}^\dag)( \td c_{A_1,k_2}^\dag + \td c_{A_2,k_2}^\dag ) |G_A\rag\lag G_A|
                        ( \td c_{A_1,k_2} + \td c_{A_2,k_2} )( \td c_{A_1,k_1} + \td c_{A_2,k_1}) \nn\\
&& \phantom{\r_{A_1A_2,k_1k_2} =}
    + \lag \td c_{B_1,k_2} \td c_{B_1,k_2}^\dag \rag_{G_B} ( \td c_{A_1,k_1}^\dag - \td c_{A_2,k_1}^\dag) |G_A\rag\lag G_A| ( \td c_{A_1,k_1} - \td c_{A_2,k_1})\nn\\
&& \phantom{\r_{A_1A_2,k_1k_2} =}
    + \lag \td c_{B_1,k_1} \td c_{B_1,k_1}^\dag \rag_{G_B} ( - \td c_{A_1,k_2}^\dag + \td c_{A_2,k_2}^\dag) |G_A\rag\lag G_A| ( - \td c_{A_1,k_2} + \td c_{A_2,k_2})\nn\\
&& \phantom{\r_{A_1A_2,k_1k_2} =}
    + \lag \td c_{B_1,k_2} \td c_{B_1,k_1}^\dag \rag_{G_B} ( - \td c_{A_1,k_2}^\dag + \td c_{A_2,k_2}^\dag) |G_A\rag\lag G_A| ( \td c_{A_1,k_1} - \td c_{A_2,k_1})\nn\\
&& \phantom{\r_{A_1A_2,k_1k_2} =}
    + \lag \td c_{B_1,k_1} \td c_{B_1,k_2}^\dag \rag_{G_B} ( \td c_{A_1,k_1}^\dag - \td c_{A_2,k_1}^\dag) |G_A\rag\lag G_A| ( - \td c_{A_1,k_2} + \td c_{A_2,k_2})\nn\\
&& \phantom{\r_{A_1A_2,k_1k_2} =}
    + \lag \td c_{B_2,k_2} \td c_{B_2,k_2}^\dag \rag_{G_B} ( \td c_{A_1,k_1}^\dag + \td c_{A_2,k_1}^\dag) |G_A\rag\lag G_A| ( \td c_{A_1,k_1} + \td c_{A_2,k_1})\nn\\
&& \phantom{\r_{A_1A_2,k_1k_2} =}
    + \lag \td c_{B_2,k_1} \td c_{B_2,k_1}^\dag \rag_{G_B} ( - \td c_{A_1,k_2}^\dag - \td c_{A_2,k_2}^\dag) |G_A\rag\lag G_A| ( - \td c_{A_1,k_2} - \td c_{A_2,k_2})\nn\\
&& \phantom{\r_{A_1A_2,k_1k_2} =}
    + \lag \td c_{B_2,k_2} \td c_{B_2,k_1}^\dag \rag_{G_B} ( - \td c_{A_1,k_2}^\dag - \td c_{A_2,k_2}^\dag) |G_A\rag\lag G_A| ( \td c_{A_1,k_1} + \td c_{A_2,k_1})\nn\\
&& \phantom{\r_{A_1A_2,k_1k_2} =}
    + \lag \td c_{B_2,k_1} \td c_{B_2,k_2}^\dag \rag_{G_B} ( \td c_{A_1,k_1}^\dag + \td c_{A_2,k_1}^\dag) |G_A\rag\lag G_A| ( - \td c_{A_1,k_2} - \td c_{A_2,k_2})
     \\
&& \phantom{\r_{A_1A_2,k_1k_2} =}
    + \lag ( \td c_{B_1,k_2} + \td c_{B_2,k_2} )( \td c_{B_1,k_1} + \td c_{B_2,k_1})( \td c_{B_1,k_1}^\dag + \td c_{B_2,k_1}^\dag)( \td c_{B_1,k_2}^\dag + \td c_{B_2,k_2}^\dag )
           \rag_{G_B} |G_A\rag\lag G_A|. \nn
\eea
In the limit of large momentum difference $|k_1-k_2|=+\inf$, we obtain the new universal R\'enyi entropies
\bea
&& \cF_{A_1A_2,p_1p_2}^{(2),\univ} = (x^2 + y^2)^2 - 16 x_1 x_2 y_1 y_2 , \nn\\
&& \cF_{A_1A_2,p_1p_2}^{(3),\univ} = (x^3 + y^3)^2 - 24 x_1 x_2 y_1 y_2 x y , \nn\\
&& \cF_{A_1A_2,p_1p_2}^{(4),\univ} = (x^4 + y^4)^2 -32 x_1 x_2 y_1 y_2 x^2 y^2 +64 x_1^2 x_2^2 y_1^2 y_2^2 , \nn\\
&& \cF_{A_1A_2,p_1p_2}^{(5),\univ} = (x^5 + y^5)^2 -40 x_1 x_2  y_1 y_2  x^3 y^3 +
 160 x_1^2 x_2^2 y_1^2 y_2^2 x y , \nn\\
&& \cF_{A_1A_2,p_1p_2}^{(6),\univ} = (x^6 + y^6)^2 -48 x_1 x_2 y_1 y_2 x^4 y^4 +
 288 x_1^2 x_2^2 y_1^2 y_2^2 x^2 y^2 -
 256 x_1^3 x_2^3 y_1^3 y_2^3 , \nn\\
&& \cF_{A_1A_2,p_1p_2}^{(7),\univ} = (x^7 + y^7)^2 -56 x_1 x_2  y_1 y_2 x y (x^2 y^2 -
   4 x_1 x_2 y_1 y_2)^2 , \nn\\
&& \cF_{A_1A_2,p_1p_2}^{(8),\univ} = (x^8 + y^8)^2 - 64 x_1 x_2 y_1 y_2 x^6 y^6 +
 640 x_1^2 x_2^2 y_1^2 y_2^2 x^4 y^4 \nn\\
&& \phantom{\cF_{A_1A_2,p_1p_2}^{(8),\univ} =} -
 2048 x_1^3 x_2^3 y_1^3 y_2^3 x^2 y^2 +
 1024 x_1^4 x_2^4 y_1^4 y_2^4 , \nn\\
&& \cF_{A_1A_2,p_1p_2}^{(9),\univ} = (x^9 + y^9)^2 -72 x_1 x_2 y_1 y_2 x^7 y^7 +
 864 x_1^2 x_2^2 y_1^2 y_2^2 x^5 y^5 \nn\\
&& \phantom{\cF_{A_1A_2,p_1p_2}^{(9),\univ} =} -
 3840 x_1^3 x_2^3 y_1^3 y_2^3 x^3 y^3 +
 4608 x_1^4 x_2^4  y_1^4 y_2^4 x y.
\eea
Although quite remarkable, this is not surprising, and, in fact as stated in \cite{Castro-Alvaredo:2018dja,Castro-Alvaredo:2018bij}, the validity of the universal R\'enyi entropy therein requires that the quasiparticles are localized quantum excitations, while in the XY chain the spinless fermions are not localized excitations.
We also get $\cF_{A_1A_2,p_1p_2}^{(n),\univ}$ for larger $n$, but we will not show the results in this paper.
Though the results are only for double interval in the XY chain, we still call them universal in the sense that they are not dependent on the explicit values of the momenta.
Moreover, we get the additional contributions to the new universal R\'enyi entropy
\bea
&& \d \cF^{(2)}_{A_1A_2,k_1k_2} =  - 2(x-y) ( |\b_{A_1,k_1-k_2}|^2 + |\b_{A_2,k_1-k_2}|^2 - |\b_{B_1,k_1-k_2}|^2 - |\b_{B_2,k_1-k_2}|^2 ) \nn\\
&& \phantom{\d \cF^{(2)}_{A_1A_2,k_1k_2} =}
      - 2 [ x^2-(y_1-y_2)^2 ] ( \b_{A_1,k_1-k_2} \b_{A_2,k_2-k_1} + cc ) \nn\\
&& \phantom{\d \cF^{(2)}_{A_1A_2,k_1k_2} =}
      - 2 [ y^2-(x_1-x_2)^2 ] ( \b_{B_1,k_1-k_2} \b_{B_2,k_2-k_1} + cc ) \nn\\
&& \phantom{\d \cF^{(2)}_{A_1A_2,k_1k_2} =}
      - 4 ( x y - 2 x_1 y_1 ) ( \b_{A_2,k_1-k_2} \b_{B_2,k_2-k_1} + cc )
      - 4 ( x y - 2 x_2 y_2 ) ( \b_{A_1,k_1-k_2} \b_{B_1,k_2-k_1} + cc ) \nn\\
&& \phantom{\d \cF^{(2)}_{A_1A_2,k_1k_2} =}
      - 4 ( x y - 2 x_1 y_2 ) ( \b_{A_2,k_1-k_2} \b_{B_1,k_2-k_1} + cc )
      - 4 ( x y - 2 x_2 y_1 ) ( \b_{A_1,k_1-k_2} \b_{B_2,k_2-k_1} + cc ) \nn\\
&& \phantom{\d \cF^{(2)}_{A_1A_2,k_1k_2} =}
      + ( \b_{A,k_1-k_2}^2 + \b_{B,k_1-k_2}^2 )( \b_{A,k_2-k_1}^2 + \b_{B,k_2-k_1}^2 )  \nn\\
&& \phantom{\d \cF^{(2)}_{A_1A_2,k_1k_2} =}
      - 8 ( \b_{A_1,k_1-k_2} \b_{A_2,k_1-k_2} \b_{B_1,k_2-k_1} \b_{B_2,k_2-k_1} + cc ),
\eea
where we use ``$cc$'' to denote the complex conjugate terms.
We also get $\d \cF^{(n)}_{A_1A_2,k_1k_2}$ with $n=3,4$, whose forms are too complicated and we will not present them in this paper.
We compare the analytical and numerical results in the figure~\ref{XYA1A2k1k2}.
We see that the results with additional correction terms match perfectly with the numerical results, and in the limit of the large momentum difference the numerical results approach the new universal R\'enyi entropies instead of the old ones.

\subsubsection{Slightly gapped and critical XY chains}

\begin{figure}[p]
  \centering
  \includegraphics[width=0.99\textwidth]{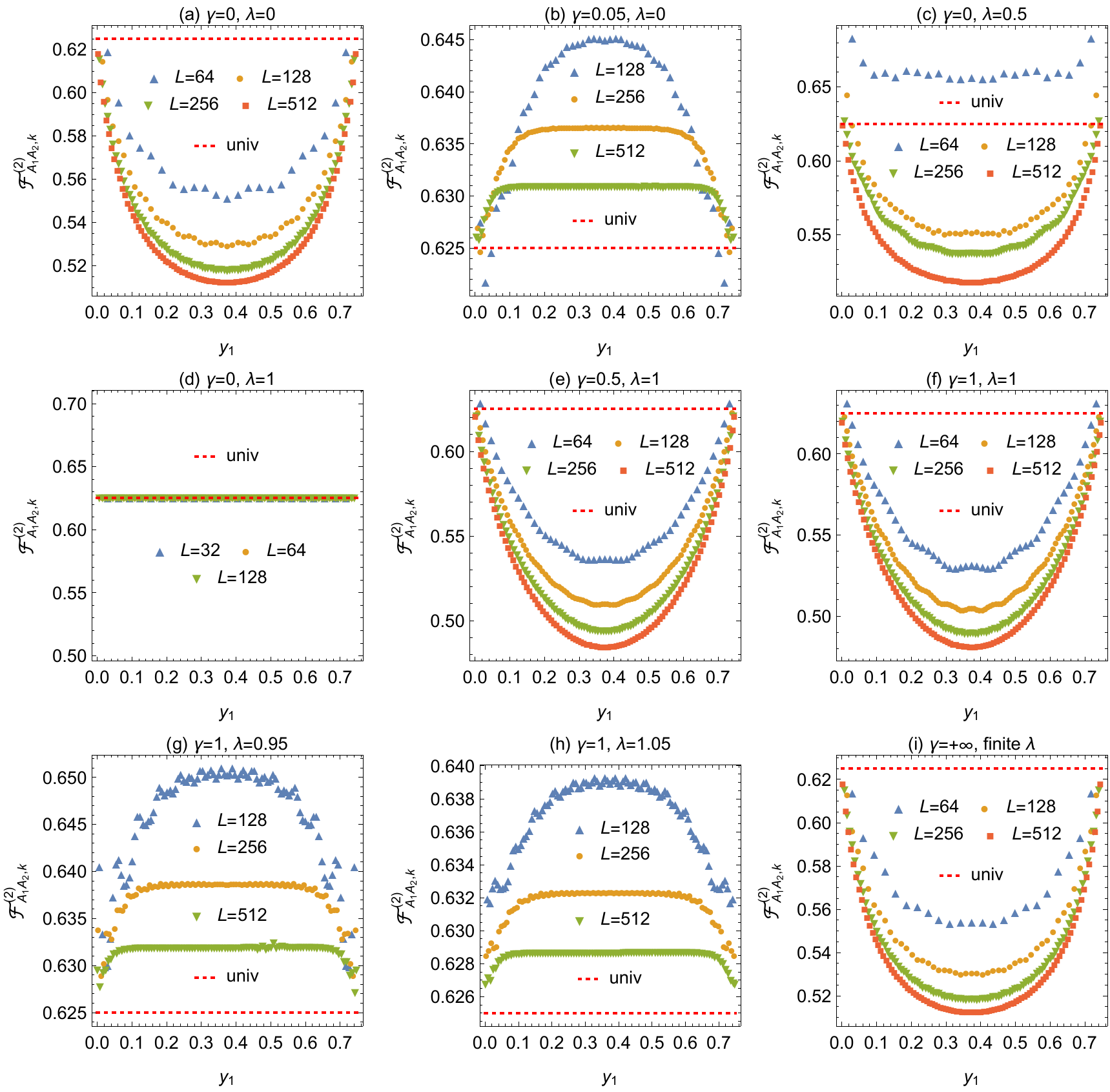}\\
  \caption{The universal R\'enyi entropy (dotted lines) and the numerical double-interval R\'enyi entropy in the slightly gapped and critical XY chains (symbols) in the single-particle state $|k\rag$.
  We have set the momenta $k=\f12+\f{L}{8}$, $x_1=x_2=\f18$.
  For the analytical results with additional terms we have set $L=+\inf$.}\label{XYA1A2kslightlygapped}
\end{figure}

\begin{figure}[p]
  \centering
  \includegraphics[width=0.99\textwidth]{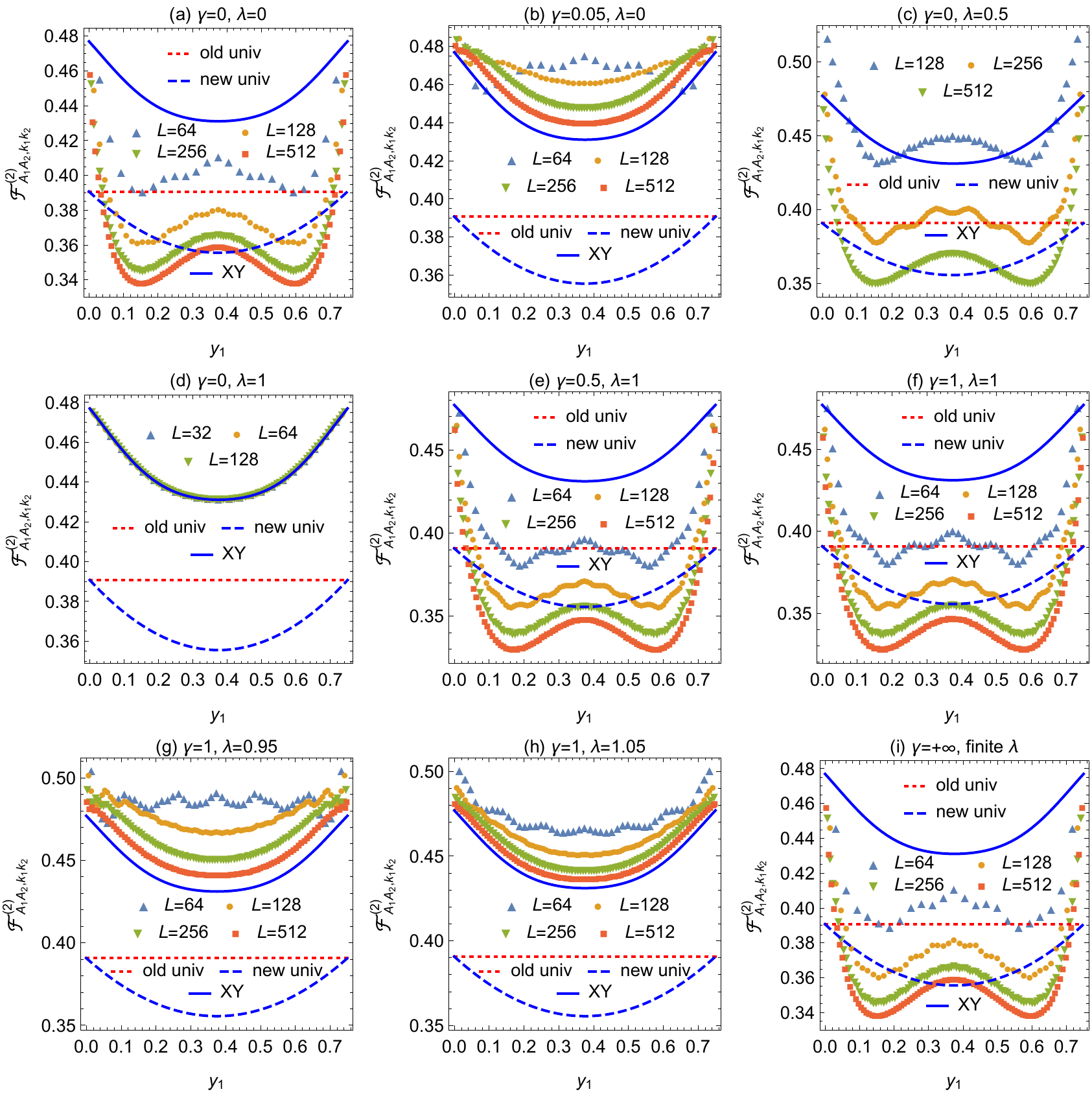}\\
  \caption{The old universal R\'enyi entropy (red dotted lines), the new universal R\'enyi entropy (blue dashed lines), the analytical R\'enyi entropy in the extremely gapped XY chain (solid lines), and the numerical R\'enyi entropy in the slightly gapped and critical XY chains(symbols) in the double-particle state $|k_1k_2\rag$.
  We have set the momenta $(k_1,k_2)=(\f12,\f32)+\f{L}{8}$, $x_1=x_2=\f18$.
  For the analytical results with additional terms we have set $L=+\inf$.}\label{XYA1A2k1k2slightlygapped}
\end{figure}

We calculate the double-interval R\'enyi entropy in the single-particle state $|k\rag$ and the double-particle state $|k_1k_2\rag$ in the slightly gapped and critical XY chains, as shown respectively in the figures~\ref{XYA1A2kslightlygapped} and \ref{XYA1A2k1k2slightlygapped}.
In the slightly gapped XY chains, the results with additional correction terms are still valid in the limit of large momenta.
As in the fermionic chain, the R\'enyi entropy is exact in the XY chain with $(\g,\l)=(0,1)$, and this is because the Bogoliubov angle $\th_k$ defined in (\ref{sjpmvphkthk}) is vanishing.
However, in the critical XY chains with $(\g,\l)=(0,0)$, $(\g,\l)=(0,0.5)$, $(\g,\l)=(0.5,1)$, $(\g,\l)=(1,1)$ and $(\g,\l)=(+\inf,\rm{finite})$ there are significant mismatches, for which we do not have a good understanding.

\subsection{Multiple intervals}

Unlike those in the extremely gapped fermionic and bosonic chains, the generalization to multiple intervals in the extremely gapped XY chain is more complicated.
The analytical and numerical calculations of the multi-interval R\'enyi entropy in the extremely gapped XY chain are straightforward but tedious.
We will not consider them in this paper.

\section{Discussions} \label{secDis}

The main results that we have obtained in this paper are summarized in the section~\ref{secSum}.
We have calculated the R\'enyi entropies in single-particle, double-particle and triple-particle excited states of fermionic, bosonic, and XY models that depend on the model, momenta of the excited quasiparticles, and the connectedness of the subsystem.
Although they are derived in the extremely gapped limit they are still valid in the slightly gapped and critical models as long as all the momenta of the excited quasiparticles are large.
The R\'enyi entropy approaches the universal R\'enyi entropy in the limit that all the momentum differences among the excited quasiparticles are large.

\begin{figure}[th]
  \centering
  \includegraphics[height=0.6\textwidth]{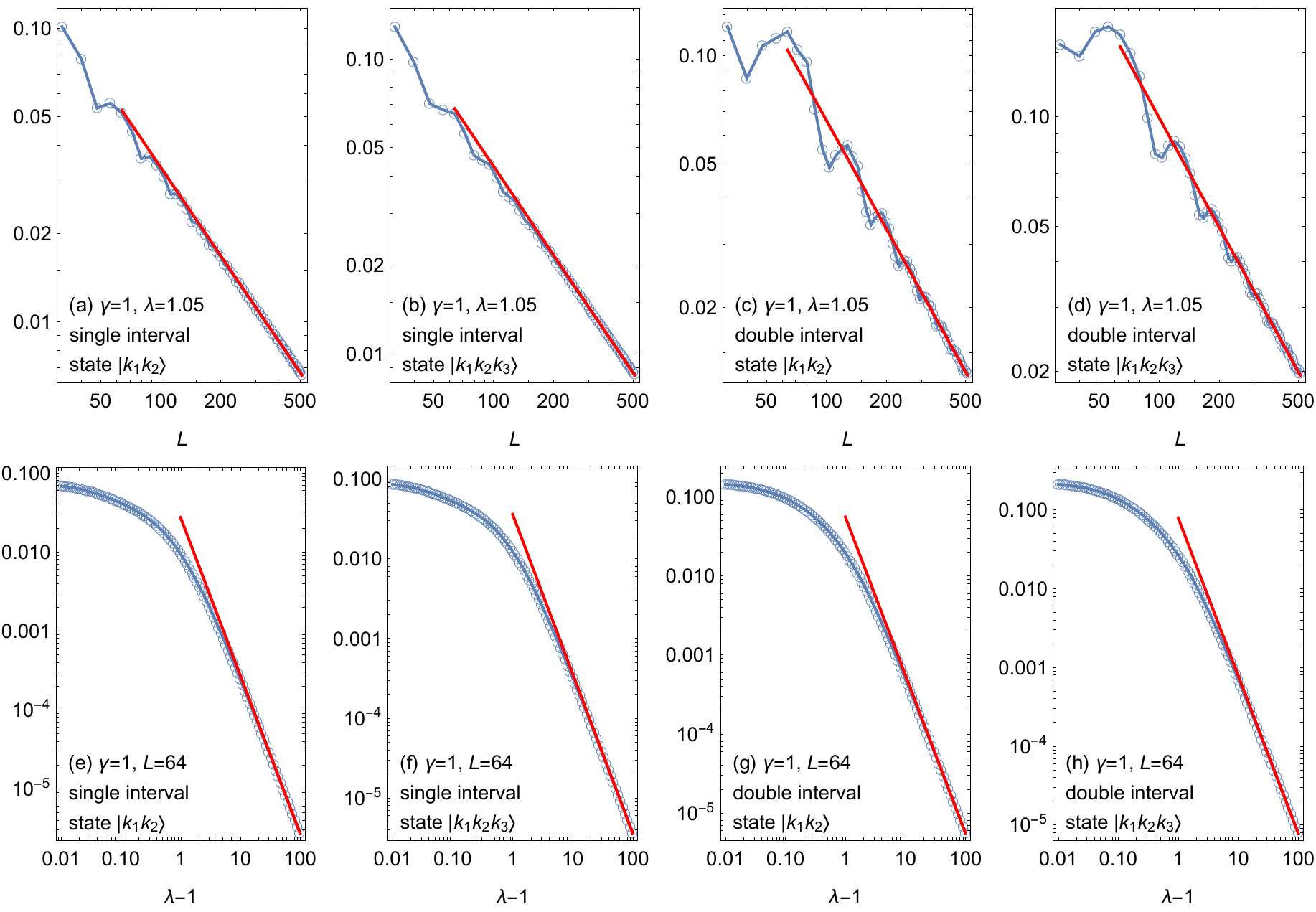}\\
  \caption{The ratio $\big|1-\f{\cF_{A,K}^{(2)}({\rm finite}~\l)}{\cF_{A,K}^{(2)}(\l=+\inf)}\big|$ in the fermionic chain (joined empty circles). For single interval we set $x=\f14$ and for double interval we set $x_1=x_2=y_1=\f18$. We have set the momenta $(k_1,k_2,k_3)=(\f12,\f32,\f52)+\f{L}{8}$. We add red straight lines in each figure serving as guidance to the eye to demonstrate the asymptotic behaviors of the R\'enyi entropy. In the first row the red lines are proportional to $\f1L$, and in the second row the red lines are proportional to $\f{1}{(\l-1)^2}$.}
  \label{Fermionslightlygapped}
\end{figure}

\begin{figure}[tph]
  \centering
  \includegraphics[height=0.6\textwidth]{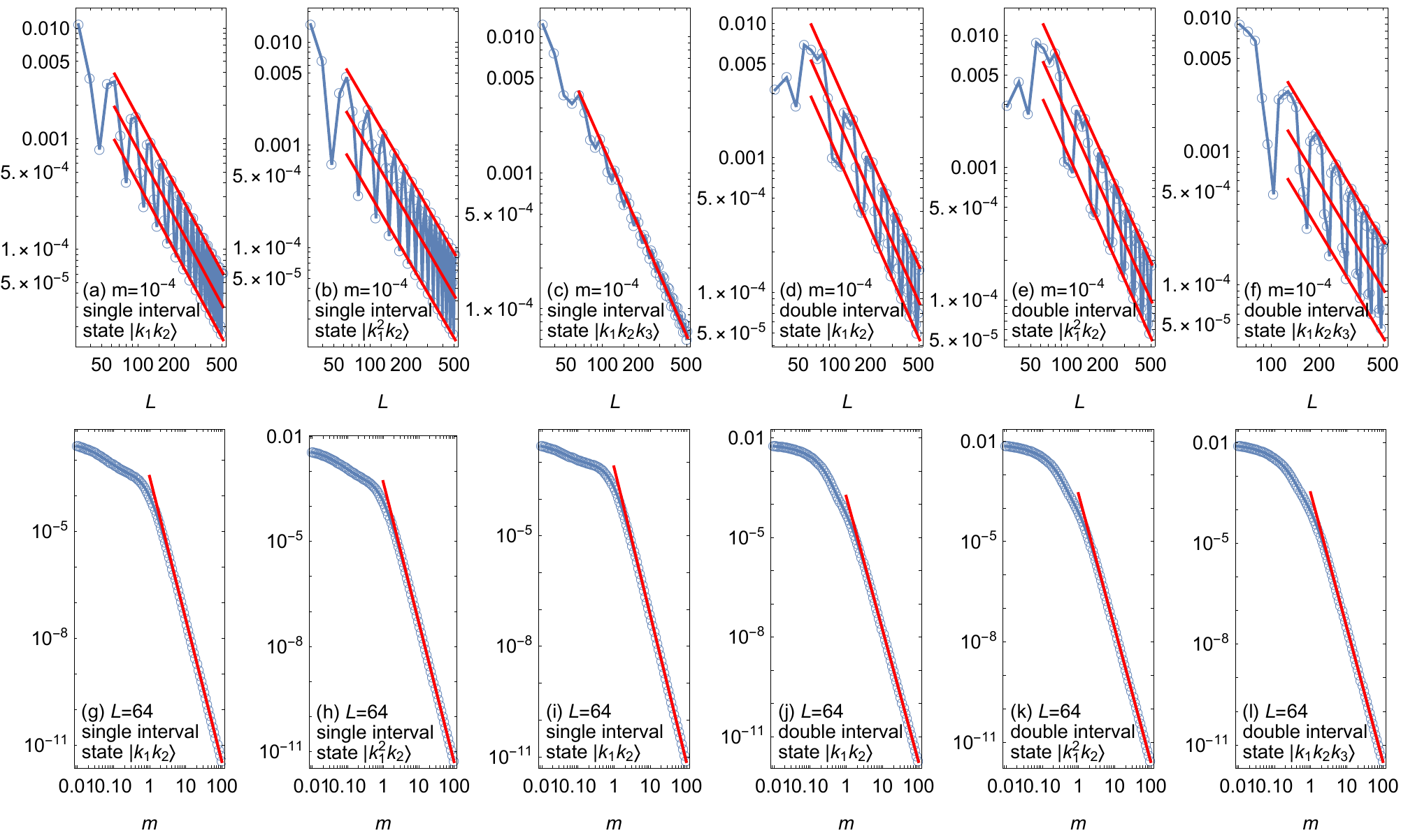}\\
  \caption{The ratio $\big|1-\f{\cF_{A,K}^{(2)}({\rm finite}~m)}{\cF_{A,K}^{(2)}(m=+\inf)}\big|$ in the bosonic chain (joined empty circles). For single interval we set $x=\f14$ and for double interval we set $x_1=x_2=y_1=\f18$. We have set the momenta $(k_1,k_2,k_3)=(1,2,3)+\f{L}{8}$. We add the red straight lines as guidance to the eye. In the panels (a), (b), (d), (e) and (f) there are oscillations, and we add multiple red straight lines. In the first row the red lines are proportional to $\f{1}{L^2}$, and in the second row the red lines are proportional to $\f{1}{m^4}$.}
  \label{Bosonslightlygapped}
\end{figure}

\begin{figure}[tph]
  \centering
  \includegraphics[height=0.3\textwidth]{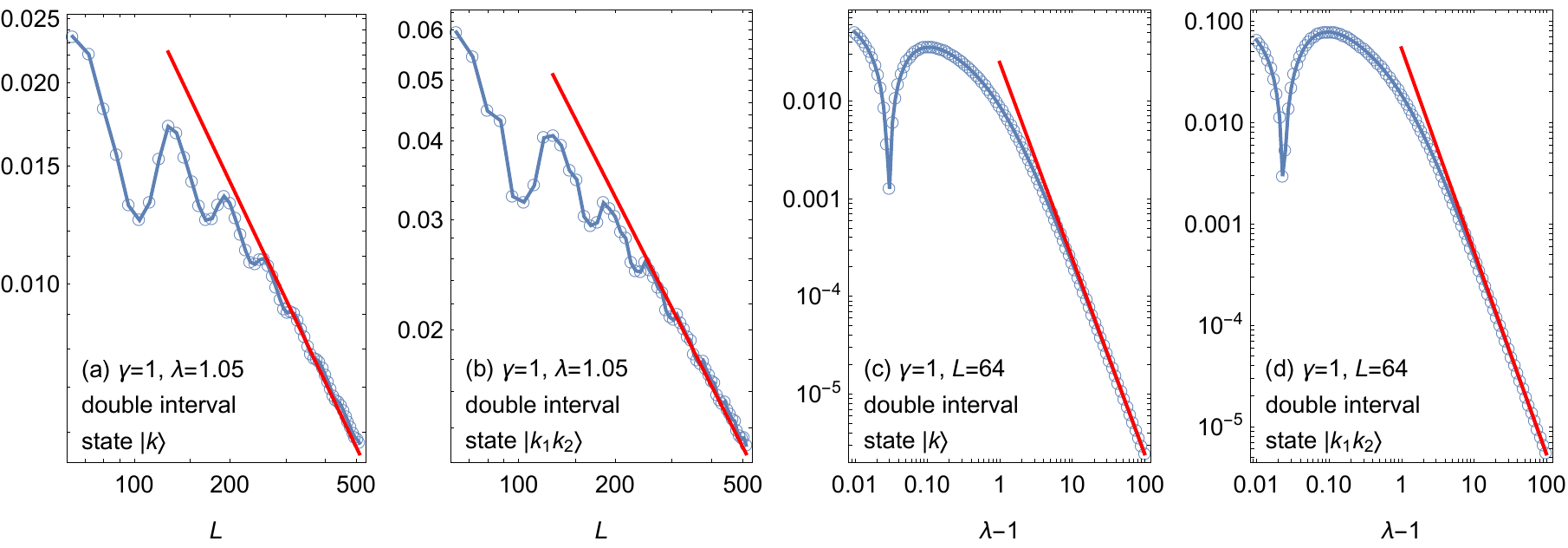}\\
  \caption{The ratio $\big|1-\f{\cF_{A,K}^{(2)}({\rm finite}~\l)}{\cF_{A,K}^{(2)}(\l=+\inf)}\big|$ in the XY chain (joined empty circles). For single interval we set $x=\f14$ and for double interval we set $x_1=x_2=y_1=\f18$. We have set the momenta $k=\f12+\f{L}{8}$ and $(k_1,k_2)=(\f12,\f32)+\f{L}{8}$. The red straight lines are guidance to the eye. In the left two panels the red lines are proportional to $\f1L$, and in the right two panels the red lines are proportional to $\f{1}{(\l-1)^2}$.}
  \label{XYslightlygapped}
\end{figure}

In figures~\ref{FermionAk1k2slightlygapped}, \ref{FermionAk1k2k3slightlygapped}, \ref{FermionA1A2k1k2slightlygapped}, \ref{FermionA1A2k1k2k3slightlygapped}, \ref{BosonAslightlygapped}, \ref{BosonA1A2slightlygapped}, \ref{XYA1A2kslightlygapped}, \ref{XYA1A2k1k2slightlygapped} we showed that the R\'enyi entropy in the slightly gapped fermionic, bosonic and XY chains approach to the results in the extremely gapped chain in the large system size and the large momentum limit, i. e. $L\to+\inf$ and $k_i\to+\inf$. It is interesting to quantify the convergence using numerical calculations.%
\footnote{We thank the anonymous referee for stimulating the discussions in this paragraph.}
We show the result of the fermionic chain in the first row of figure~\ref{Fermionslightlygapped} and find
\be
\Big| 1 - \f{\cF_{A,K}^{(2)}({\rm finite}~\l)}{\cF_{A,K}^{(2)}(\l=+\inf)} \Big| \sim \f{1}{L}.
\ee
We show the result of the bosonic chain in the first row of figure~\ref{Bosonslightlygapped} and find
\be
\Big| 1 - \f{\cF_{A,K}^{(2)}({\rm finite}~m)}{\cF_{A,K}^{(2)}(m=+\inf)} \Big| \sim \f{1}{L^2}.
\ee
We show the result of the XY chain in the left two panels of figure~\ref{XYslightlygapped} and find
\be
\Big| 1 - \f{\cF_{A,K}^{(2)}({\rm finite}~\l)}{\cF_{A,K}^{(2)}(\l=+\inf)} \Big| \sim \f{1}{L}.
\ee
It is also interesting to see how the R\'enyi entropy approaches to the expected analytical result with the increase of the gap $\D\to+\inf$.
In the fermionic chain (\ref{fermionicchain}) with fixed $\g=1$, the gap is $\D=\l-1$ and in the bosonic chain (\ref{bosonicchain}) the gap is $\D=m$.
We show the results of the fermionic chain in the second row of figure~\ref{Fermionslightlygapped} and find
\be
\Big| 1 - \f{\cF_{A,K}^{(2)}({\rm finite}~\l)}{\cF_{A,K}^{(2)}(\l=+\inf)} \Big| \sim \f{1}{(\l-1)^2}.
\ee
We show the results of the bosonic chain in the second row of figure~\ref{Bosonslightlygapped} and find
\be
\Big| 1 - \f{\cF_{A,K}^{(2)}({\rm finite}~m)}{\cF_{A,K}^{(2)}(m=+\inf)} \Big| \sim \f{1}{m^4}.
\ee
We show the results of the XY chain in the right two panels of figure~\ref{XYslightlygapped} and find
\be
\Big| 1 - \f{\cF_{A,K}^{(2)}({\rm finite}~\l)}{\cF_{A,K}^{(2)}(\l=+\inf)} \Big| \sim \f{1}{(\l-1)^2}.
\ee
It is interesting to note the different scaling behaviors in different models. We get the same results for other values of the R\'enyi index $n$.

It is remarkable, though not surprising, that exact calculations reveals a new universal double-interval double-particle state R\'enyi entropy in the XY chain, which does not depend on the values of the momenta. In the limit of large gap and large momentum difference, the double-interval double-particle state R\'enyi entropies approach to the new universal R\'enyi entropies, instead of the old ones.
The analytical double-interval double-particle state R\'enyi entropy we have obtained in XY chain is exact in the extremely gapped limit, i.e.\ that $\g$ is finite and $\l\to+\inf$, as well as at the critical and non-relativistic point $\g=0,\l=1$.
We checked that it is still valid in the nearly gapped XY chains as long as all the momenta of the quasiparticles are large.
However, it is not valid in the critical chains even when all the momenta of the quasiparticles are large (except at the special non-relativistic point $\g=0,\l=1$), for which we do not have a good explanation. We hope to come back to the problem in the future.

We calculated analytically the R\'enyi entropy with a relatively small R\'enyi index $n$ in an excited state with a rather small number of quasiparticles in the extremely gapped fermionic, bosonic, and XY chains by writing the excited states in terms of local excitations.
In the extremely gapped fermionic and bosonic chains, we established respectively the formulas (\ref{TheFormulaF}) and (\ref{TheFormulaB}), which are much more efficient in the analytical calculations.
These formulas are also very efficient in numerical calculations.
We anticipate that they would be useful to calculate the R\'enyi entropy for generic $n$, make the analytical continuation $n\to1$, and get the entanglement entropy.
In subsection (\ref{FermionEE}) we have made some preliminary investigations and obtained the entanglement entropy in the double-particle and triple-particle states in the extremely gapped fermionic chain.
This is especially intriguing for the excited state R\'enyi entropy in the bosonic chain, for which as far as we know there is no even a numerical method to calculate the excited state entanglement entropy.

In this paper we have only considered the models where there are only one kind of quasiparticles, and the ``interactions'' between quasiparticles of different momenta lead to the additional contributions to the universal R\'enyi entropy.
It would be interesting to consider the excited state R\'enyi entropy in the models with more than one kind of quasiparticles, and the interactions between different kinds of quasiparticles can in principle lead to the additional terms.

It would be interesting to compare quantitatively the difference of a subsystem density matrices of two different quasiparticle excited states. The Schatten and trace distances between the two RDMs can do this job.
Then one expects the universal subsystem Schatten and trace distances and their corrections, similar to the universal R\'enyi and entanglement entropies whose validity requires both large momenta and large momentum difference and their corrections that are valid in the limit of large momenta.
We have reported some preliminary results in \cite{Zhang:2020ouz,Zhang:2020txb}, and we will report more details in \cite{ZRDistance}.

\section*{Acknowledgements}

We thank Pasquale Calabrese for reading of a previous version of the draft, encouragement, and valuable discussions, comments and suggestions.
We also thank D\'avid Horv\'ath, Gianluca Lagnese and Sara Murciano for helpful discussions.
We are indebted to Olalla Castro-Alvaredo, Cecilia De Fazio, Benjamin Doyon and Istv\'an Sz\'ecs\'enyi for important comments and clarifications.
MAR thanks CNPq and FAPERJ (grant number 210.354/2018) for partial support.
JZ acknowledges support from ERC under Consolidator grant number 771536 (NEMO).

\appendix

\section{Wave function method in bosonic chain} \label{appMWF}

We review the method of wave function to calculate the excited state R\'enyi entropy in the bosonic chain \cite{Castro-Alvaredo:2018dja,Castro-Alvaredo:2018bij}, and especially we also show how it can be used to calculate analytically the R\'enyi entropy in the extremely massive limit.
We only give the formulas we used to calculate numerically and analytically the R\'enyi entropy. One can see more details in \cite{Castro-Alvaredo:2018bij}, as well as in \cite{Zhang:2020txb}.
We follow the convention in \cite{Zhang:2020txb}.

We focus on the case of one single interval $A=[1,\ell]$ and its complement $B=[\ell+1,L]$.
It is similar for the double-interval and multi-interval cases, one just needs to relabel the sites.
We will not show the details here.

\subsection{General gapped bosonic chain}

The ground state wave function is related to the $L\times L$ matrix
\be \label{Wj1j2fj}
W_{j_1j_2}=w_{j_1-j_2}, ~~ w_j = \f1L \sum_{k=1-\f{L}2}^{\f{L}2} \ve_k \ep^{\f{2\pi\ii j k}{L}},
\ee
where $\ve_k$ is the single-particle energy (\ref{vek}).
One can  write the above matrix as a $2\times2$ block matrix as follows:
\be
W = \lt( \ba{cc} \cA & \cB \\ \cC & \cD \ea \rt),
\ee
with the matrices $\cA$, $\cB$, $\cC$, $\cD$ of respectively $\ell\times\ell$, $\ell\times(L-\ell)$, $(L-\ell)\times\ell$, $(L-\ell)\times(L-\ell)$ entries.
By replica trick, one can construct the $nL\times nL$ matrix $\cM$ with $2n\times2n$ blocks
\be \label{cM}
\cM = \lt( \ba{cc|cc|cc|cc}
2\cA & \cB  &        & & & & & \cB \\
\cC  & 2\cD & \cC    & & & & & \\ \hline
     & \cB  & 2\cA   & \cB    & & & & \\
     &      & \cC    & 2\cD   & \cC    & & & \\ \hline
     &      &        & \cB    & \ddots & \ddots &      & \\
     &      &        &        & \ddots & \ddots & \cC  & \\ \hline
     &      &        &        &        & \cB    & 2\cA & \cB \\
\cC  &      &        &        &        &        & \cC  & 2\cD
\ea \rt).
\ee

We consider a general excited state $|K\rag$, and its wave function is related to a function $f_K(\{u_k\})$ that depends on
\be
u_k = \sr{\f{2\ve_k}{L}} \sum_{j=1}^L \ep^{-\f{2\pi\ii j k}{L}} q_j.
\ee
The wave functions $f_K(\{u_{a,k}\})$ and $f_K(\{v_{a,k}\})$ of the replicated ket state $|K\rag$ and bra state $\lag K|$ depend respectively on
\bea \label{uakvak}
&& u_{a,k} = \sr{\f{2\ve_k}{L}} \sum_{j=1}^L \ep^{-\f{2\pi\ii j k}{L}} q_{a,j}, \nn\\
&& v_{a,k} = \sr{\f{2\ve_k}{L}} \Big( \sum_{j=1}^\ell \ep^{\f{2\pi\ii j k}{L}} q_{a+1,j}
                                    + \sum_{j=\ell+1}^L \ep^{\f{2\pi\ii j k}{L}} q_{a,j} \Big),
\eea
with $q_{a,j}$ being the replicated coordinates and $a=1,2,\cdots,n$ being the replica indices. It is understood that $q_{n+1,j}=q_{1,j}$.

The R\'enyi entropy in the state $|K\rag$ can be written as the expectation value
\be \label{cFAKn}
\cF_{A,K}^{(n)} = \Big\lag\!\!\Big\lag \prod_{a=1}^n [ f_K(\{u_{a,k}\}) f_K(\{v_{a,k}\}) ] \Big\rag\!\!\Big\rag.
\ee
The definition of $\lag\!\lag \cdots \rag\!\rag$ could be found in Eq.\ (3.30) of \cite{Zhang:2020txb} and it could be evaluated by the bosonic Wick contractions
\bea
&& \wick{\c u_{a_1,k_1} \c u_{a_2,k_2}} = U_{a_1,k_1}^T \cM^{-1} U_{a_2,k_2}, \nn\\
&& \wick{\c v_{a_1,k_1} \c u_{a_2,k_2}} = V_{a_1,k_1}^T \cM^{-1} U_{a_2,k_2}, \nn\\
&& \wick{\c v_{a_1,k_1} \c v_{a_2,k_2}} = V_{a_1,k_1}^T \cM^{-1} V_{a_2,k_2}.
\eea
The vectors $U_{a,k}$, $V_{a,k}$ with $nL$ components have the nonvanishing entries
\bea \label{UakVakVakVak}
&& [ U_{a,k} ]_{(a-1)L+j} = \sr{\f{2\ve_k}{L}} \ep^{-\f{2\pi\ii j k}{L}}, ~ a=1,\cdots,n, ~ j=1,\cdots,L, \nn\\
&& [ V_{a,k} ]_{(a-1)L+j} = \sr{\f{2\ve_k}{L}} \ep^{\f{2\pi\ii j k}{L}}, ~ a=1,\cdots,n, ~ j=\ell+1,\cdots,L, \nn\\
&& [ V_{a,k} ]_{a L+j} = \sr{\f{2\ve_k}{L}} \ep^{\f{2\pi\ii j k}{L}}, ~ a=1,\cdots,n-1, ~ j=1,\cdots,\ell, \nn\\
&& [ V_{n,k} ]_{j} = \sr{\f{2\ve_k}{L}} \ep^{\f{2\pi\ii j k}{L}}, ~ j=1,\cdots,\ell.
\eea

\subsection{Extremely gapped bosonic chain}

In the extremely gapped limit $m\to+\inf$, it is convenient to rescale the matrix (\ref{cM}), the coordinates (\ref{uakvak}), and the vectors (\ref{UakVakVakVak}) as
\be
\td \cM = \f{\cM}{2m}, ~~
\td u_{a,k} = \f{u_{a,k}}{\sr{2m}}, ~~
\td v_{a,k} = \f{v_{a,k}}{\sr{2m}}, ~~
\td U_{a,k} = \f{U_{a,k}}{\sr{2m}}, ~~
\td V_{a,k} = \f{V_{a,k}}{\sr{2m}}.
\ee
The function $w_j$ defined in (\ref{Wj1j2fj}) becomes trivially
\be
w_j = m \d_j,
\ee
and then the matrix $\td \cM=I_{nL}$ is just the $nL\times nL$ identity matrix.
We obtain the vectors
\bea
&& [\td U_{a,k} ]_{(a-1)L+j} = \f1{\sr{L}} \ep^{-\f{2\pi\ii j k}{L}}, ~ a=1,\cdots,n, ~ j=1,\cdots,L, \nn\\
&& [\td V_{a,k} ]_{(a-1)L+j} = \f1{\sr{L}} \ep^{\f{2\pi\ii j k}{L}}, ~ a=1,\cdots,n, ~ j=\ell+1,\cdots,L, \nn\\
&& [\td V_{a,k} ]_{aL+j} = \f1{\sr{L}} \ep^{\f{2\pi\ii j k}{L}}, ~ a=1,\cdots,n-1, ~ j=1,\cdots,\ell, \nn\\
&& [\td V_{n,k} ]_{j} = \f1{\sr{L}} \ep^{\f{2\pi\ii j k}{L}}, ~ j=1,\cdots,\ell.
\eea
Then we get the R\'enyi entropy in the extremely gapped bosonic chain
\be
\cF_{A,K}^{(n)} = \Big\lag\!\!\Big\lag \prod_{a=1}^n [ f_K(\{\td u_{a,k}\}) f_K(\{\td v_{a,k}\}) ] \Big\rag\!\!\Big\rag,
\ee
which is evaluated by the bosonic Wick contractions
\bea \label{wicktdutdutdvtdutdvtdv}
&& \wick{\c {\td u}_{a_1,k_1} \c {\td u}_{a_2,k_2}} = \td U_{a_1,k_1}^T \td U_{a_2,k_2}, \nn\\
&& \wick{\c {\td v}_{a_1,k_1} \c {\td u}_{a_2,k_2}} = \td V_{a_1,k_1}^T \td U_{a_2,k_2}, \nn\\
&& \wick{\c {\td v}_{a_1,k_1} \c {\td v}_{a_2,k_2}} = \td V_{a_1,k_1}^T \td V_{a_2,k_2}.
\eea
Note that
\be
\td U_{a_1,k_1}^T \td U_{a_2,k_2} = \td V_{a_1,k_1}^T \td V_{a_2,k_2} = \d_{a_1a_2} \d_{k_1+k_2},
\ee
with $\d_{k_1+k_2}=1$ for $k_1+k_2=0$ or $k_1=k_2=\f{L}{2}$ and $\d_{k_1+k_2}=0$ otherwise.
We get the product
\be \label{tdVa1k1TtdUa2k2}
\td V_{a_1,k_1}^T \td U_{a_2,k_2} =
\lt\{
\ba{cl}
\lt\{ \ba{cl} 1-x & k_1=k_2 \\ -\a_{k_1-k_2} & k_1 \neq k_2 \ea \rt. & a_1 = a_2 \\
\lt\{ \ba{cl} x   & k_1=k_2 \\ \a_{k_1-k_2}  & k_1 \neq k_2 \ea \rt. & a_1 = a_2-1\!\!\!\!\mod n \\
0                                                                    & \rm{otherwise}
\ea
\rt.\!\!\!,
\ee
with the definition of the function $\a_k$ given in (\ref{alphakdefinition}).

We consider a special state $|k_1^{r_1}k_2^{r_2}\cdots k_s^{r_s}\rag$ that only the modes with momenta $k_1,k_2,\cdots,k_s\in\{1,2,\cdots,\f{L}{2}-1\}$ are excited.
We have
\be
f_K(\{\td u_{a,k}\}) = \prod_{i=1}^s \f{\td u_{a,k}^{r_i}}{\sr{r_i!}}, ~~
f_K(\{\td v_{a,k}\}) = \prod_{i=1}^s \f{\td v_{a,k}^{r_i}}{\sr{r_i!}}.
\ee
Then we get the R\'enyi entropy
\be \label{cFAKnextremelygapped}
\cF_{A,K}^{(n)} = \f{1}{\prod_{i=1}^s ( r_i! )^n} \Big\lag\!\!\Big\lag \prod_{a=1}^n \prod_{i=1}^s [ \td u_{a,k_i}^{r_i} \td v_{a,k_i}^{r_i} ] \Big\rag\!\!\Big\rag,
\ee
with the only non-vanishing contractions
\be
\wick{\c {\td v}_{a_1,k_1} \c {\td u}_{a_2,k_2}} = \td V_{a_1,k_1}^T \td U_{a_2,k_2},
\ee
which is just (\ref{tdVa1k1TtdUa2k2}).
The total number of excited modes is
\be \label{Rdefinition}
R = \sum_{i=1}^s r_i,
\ee
and in replica trick it becomes $n R$.
On the RHS of (\ref{cFAKnextremelygapped}) there are $n R$ number of $\td u_{a,k}$ and $nR$ number of $\td v_{a,k}$, and we organize them as $\td u_I$, $\td v_I$ with $I=1,2,\cdots,n R$.
Note that each of $\td u_{a,k_i}$ and $\td v_{a,k_i}$ has a repetition of $r_i$ times.
For $\td u_I$, $\td v_I$, there are corresponding vectors $\td U_I$, $\td V_I$.
We define the $n R\times n R$ matrix $\O_{A,K}^{(n)}$ with entries
\be \label{OAKnIJdefinition}
[\O_{A,K}^{(n)}]_{IJ} = \td V_I^T \td U_J.
\ee
The excited state R\'enyi entropy in the extremely gapped bosonic chain (\ref{cFAKnextremelygapped}) is just the permanent%
\footnote{Naively, we would expect that the formula
\be \label{cFAknnaive}
\cF_{A,K}^{(n)} = \det \O_{A,K}^{(n)},
\ee
applies to a general excited state $|K\rag=|k_1k_2\cdots k_s\rag$ in the extremely gapped fermionic chain.
We compared the analytical results coming from (\ref{cFAknnaive}) with the results we get by writing the excited states in terms of local excitations in the section~\ref{secF}, and found that (\ref{cFAknnaive}) leads to a correct result for an odd integer $n=3,5,\cdots$ but leads to a wrong result when $n$ is an even integer.
In fact for an even integer $n$ the formula (\ref{cFAknnaive}) cannot even give the correct universal part of the R\'enyi entropy let alone the additional part.
It would be interesting to have a better understanding of the formula (\ref{cFAknnaive}) in the extremely gapped fermionic chain and the formula (\ref{TheFormulaBapp}) in the extremely gapped bosonic chain.}
\be \label{TheFormulaBapp}
\cF_{A,K}^{(n)} = \f{\per\O_{A,K}^{(n)}}{\prod_{i=1}^s ( r_i! )^n}.
\ee

Although we have derived the final formula (\ref{TheFormulaBapp}) for the special state $|K\rag=|k_1^{r_1}k_2^{r_2}\cdots k_s^{r_s}\rag$ that only the modes with momenta $k_1,k_2,\cdots,k_s\in\{1,2,\cdots,\f{L}{2}-1\}$ are excited, it can be shown that the formula applies to a general excited state $|k_1^{r_1}k_2^{r_2}\cdots k_s^{r_s}\rag$ with momenta $k_1,k_2,\cdots,k_s\in\{-\f{L}{2}+1,\cdots,-1,0,1,\cdots,\f{L}{2}\}$ as long as a property of the Hermite polynomials and the complex Hermite polynomials, which we elaborate below, is correct.

As it is already shown in \cite{Zhang:2020txb}, in the wave function with the modes $k=0,\f{L}{2}$,  appear the Hermite polynomials
\be
H_r(x) = \ep^{\f{x^2}{2}} ( x - \p_x )^r \ep^{-\f{x^2}{2}}=(-)^r\ep^{x^2}\p_x^r \ep^{-x^2},
\ee
and for other modes there appear the complex Hermite polynomials
\be
H_{r,s}(z,\bar z) = \ep^{z\bar z} ( z - \p_{\bar z} )^r ( \bar z - \p_{z} )^s \ep^{-z\bar z}
                  = (-)^{r+s} \ep^{2z\bar z} \p_{\bar z}^r\p_z^s \ep^{-2z\bar z}.
\ee
The Hermite polynomials are orthogonal
\be
\lag H_r(x) H_s(x) \rag = \d_{rs}2^{r} r!,
\ee
with the expectation value evaluated by the bosonic Wick contraction
\be
\wick{\c x \c x} = \f12.
\ee
Note that $H_0(x)=1$ and $H_1(x)=2x$.
We claim that
\be \label{claimlagHrxHsxrag}
\lag H_r(x) H_s(x) \rag = \lag \overbrace{(2x)^r} H_s(x) \rag,
\ee
where we impose the rule that the Wick contractions between the variables under the same curly bracket are not allowed.
Using the recursive formula for $s\geq1$
\be
H_s(x) = 2x H_{s-1}(x) - H_{s-1}'(x),
\ee
we get for $r\geq1$
\be
\lag \overbrace{(2x)^r} H_s(x) \rag = 2 r \lag \overbrace{(2x)^{r-1}} H_{s-1}(x) \rag,
\ee
which proves the claim (\ref{claimlagHrxHsxrag}).
As the Hermite polynomials are complete, we get
\be
\lag H_r(x) P(x) \rag = \lag \overbrace{(2x)^r} P(x) \rag,
\ee
for an arbitrary polynomial $P(x)$ of $x$.

The complex Hermite polynomials are also complete and orthogonal
\be
\lag H_{r,s}(z,\bar z) H_{p,q}(z,\bar z) \rag = \d_{rq} \d_{sp} 2^{r+s} r!s!,
\ee
with the expectation value evaluated by the bosonic Wick contractions
\be
\wick{\c z \c z} = \wick{\c {\bar z} \c {\bar z}} = 0, ~~
\wick{\c z \c {\bar z}} = \f12.
\ee
Note that $H_{0,0}(z,\bar z)=1$, $H_{1,0}(z,\bar z)=2z$, $H_{0,1}(z,\bar z)=2\bar z$.
Note also $[H_{r,s}(z,\bar z)]^* = H_{s,r}(z,\bar z)$.
From the recursive formulas
\bea
&& H_{r,s}(z,\bar z) = 2z H_{r-1,s}(z,\bar z) - \p_{\bar z} H_{r-1,s}(z,\bar z), ~ r \geq 1, \nn\\
&& H_{r,s}(z,\bar z) = 2\bar z H_{r,s-1}(z,\bar z) - \p_{z} H_{r,s-1}(z,\bar z), ~ s \geq 1,
\eea
we get
\bea
&& \lag \overbrace{(2z)^r(2\bar z)^s} H_{p,q}(z,\bar z) \rag = 2r \lag \overbrace{(2z)^{r-1}(2\bar z)^s} H_{p,q-1}(z,\bar z) \rag, ~ r,q\geq1, \nn\\
&& \lag \overbrace{(2z)^r(2\bar z)^s} H_{p,q}(z,\bar z) \rag = 2s \lag \overbrace{(2z)^{r}(2\bar z)^{s-1}} H_{p-1,q}(z,\bar z) \rag, ~ s,p\geq1.
\eea
This leads to
\be
\lag H_{r,s}(z,\bar z) H_{p,q}(z,\bar z) \rag = \lag \overbrace{(2z)^r(2\bar z)^s} H_{p,q}(z,\bar z) \rag,
\ee
and then for an arbitrary polynomial $P(z,\bar z)$ of $z,\bar z$ we further get
\be
\lag H_{r,s}(z,\bar z) P(z,\bar z) \rag = \lag \overbrace{(2z)^r(2\bar z)^s} P(z,\bar z) \rag.
\ee

This proves the formula (\ref{TheFormulaBapp}) with (\ref{tdVa1k1TtdUa2k2}), which applies to a general excited state $|k_1^{r_1}k_2^{r_2}\cdots k_s^{r_s}\rag$ in the extremely gapped bosonic chain.
It can be used to calculate analytically the R\'enyi entropy in the extremely gapped bosonic chain when the number of excited quasiparticles $R$ is small.
We check that the analytical calculations with the wave function method always lead to the same results as we get by the subsystem mode method in the section~\ref{secB}, but unfortunately, we cannot prove the equivalence of the two methods for general states.
When $R$ is large, although the analytical calculations are cumbersome, the numerical calculations are still very efficient.


\begin{thebibliography}{10}

\bibitem{Amico:2007ag}
L.~Amico, R.~Fazio, A.~Osterloh and V.~Vedral, \textit{{Entanglement in
  many-body systems}}, \href{http://dx.doi.org/10.1103/RevModPhys.80.517}{Rev.
  Mod. Phys. {\bfseries 80}, 517 (2008)},
  [\href{https://arxiv.org/abs/quant-ph/0703044}{{\ttfamily
  arXiv:quant-ph/0703044}}].

\bibitem{Eisert:2008ur}
J.~Eisert, M.~Cramer and M.~B. Plenio, \textit{{Area laws for the entanglement
  entropy - a review}}, \href{http://dx.doi.org/10.1103/RevModPhys.82.277}{Rev.
  Mod. Phys. {\bfseries 82}, 277--306 (2010)},
  [\href{https://arxiv.org/abs/0808.3773}{{\ttfamily arXiv:0808.3773}}].

\bibitem{calabrese2009entanglement}
P.~Calabrese, J.~Cardy and B.~Doyon, \textit{{Entanglement entropy in extended
  quantum systems}}, \href{http://dx.doi.org/10.1088/1751-8121/42/50/500301}{J.
  Phys. A: Math. Gen. {\bfseries 42}, 500301 (2009)}.

\bibitem{Laflorencie:2015eck}
N.~Laflorencie, \textit{{Quantum entanglement in condensed matter systems}},
  \href{http://dx.doi.org/10.1016/j.physrep.2016.06.008}{Phys. Rept. {\bfseries
  646}, 1 (2016)}, [\href{https://arxiv.org/abs/1512.03388}{{\ttfamily
  arXiv:1512.03388}}].

\bibitem{Bombelli:1986rw}
L.~Bombelli, R.~K. Koul, J.~Lee and R.~D. Sorkin, \textit{{A quantum source of
  entropy for black holes}},
  \href{http://dx.doi.org/10.1103/PhysRevD.34.373}{Phys. Rev. D {\bfseries 34},
  373 (1986)}.

\bibitem{Srednicki:1993im}
M.~Srednicki, \textit{{Entropy and area}},
  \href{http://dx.doi.org/10.1103/PhysRevLett.71.666}{Phys. Rev. Lett.
  {\bfseries 71}, 666 (1993)},
  [\href{https://arxiv.org/abs/hep-th/9303048}{{\ttfamily
  arXiv:hep-th/9303048}}].

\bibitem{Callan:1994py}
C.~G. Callan~Jr. and F.~Wilczek, \textit{{On geometric entropy}},
  \href{http://dx.doi.org/10.1016/0370-2693(94)91007-3}{Phys. Lett. B
  {\bfseries 333}, 55 (1994)},
  [\href{https://arxiv.org/abs/hep-th/9401072}{{\ttfamily
  arXiv:hep-th/9401072}}].

\bibitem{Holzhey:1994we}
C.~Holzhey, F.~Larsen and F.~Wilczek, \textit{{Geometric and renormalized
  entropy in conformal field theory}},
  \href{http://dx.doi.org/10.1016/0550-3213(94)90402-2}{Nucl. Phys. B
  {\bfseries 424}, 443 (1994)},
  [\href{https://arxiv.org/abs/hep-th/9403108}{{\ttfamily
  arXiv:hep-th/9403108}}].

\bibitem{peschel1999density}
I.~Peschel, M.~Kaulke and {\"O}.~Legeza, \textit{Density-matrix spectra for
  integrable models},
  \href{http://dx.doi.org/10.1002/(SICI)1521-3889(199902)8:2<153::AID-ANDP153>3.0.CO;2-N}{Ann.
  der Phys. {\bfseries 8}, 153--164 (1999)},
  [\href{https://arxiv.org/abs/cond-mat/9810174}{{\ttfamily
  arXiv:cond-mat/9810174}}].

\bibitem{Peschel:1999DensityMatrices}
I.~{Peschel} and M.-C. {Chung}, \textit{{Density matrices for a chain of
  oscillators}}, \href{http://dx.doi.org/10.1088/0305-4470/32/48/305}{J. Phys.
  A: Math. Gen. {\bfseries 32}, 8419--8428 (1999)},
  [\href{https://arxiv.org/abs/cond-mat/9906224}{{\ttfamily
  arXiv:cond-mat/9906224}}].

\bibitem{Chung2000Densitymatrix}
M.-C. {Chung} and I.~{Peschel}, \textit{{Density-matrix spectra for
  two-dimensional quantum systems}},
  \href{http://dx.doi.org/10.1103/PhysRevB.62.4191}{Phys. Rev. B {\bfseries
  62}, 4191--4193 (2000)},
  [\href{https://arxiv.org/abs/cond-mat/0004222}{{\ttfamily
  arXiv:cond-mat/0004222}}].

\bibitem{chung2001density}
M.-C. Chung and I.~Peschel, \textit{Density-matrix spectra of solvable
  fermionic systems}, \href{http://dx.doi.org/10.1103/PhysRevB.64.064412}{Phys.
  Rev. B {\bfseries 64}, 064412 (2001)},
  [\href{https://arxiv.org/abs/cond-mat/0103301}{{\ttfamily
  arXiv:cond-mat/0103301}}].

\bibitem{Vidal:2002rm}
G.~Vidal, J.~I. Latorre, E.~Rico and A.~Kitaev, \textit{{Entanglement in
  Quantum Critical Phenomena}},
  \href{http://dx.doi.org/10.1103/PhysRevLett.90.227902}{Phys. Rev. Lett.
  {\bfseries 90}, 227902 (2003)},
  [\href{https://arxiv.org/abs/quant-ph/0211074}{{\ttfamily
  arXiv:quant-ph/0211074}}].

\bibitem{peschel2003calculation}
I.~Peschel, \textit{{Calculation of reduced density matrices from correlation
  functions}}, \href{http://dx.doi.org/10.1088/0305-4470/36/14/101}{J. Phys. A:
  Math. Gen. {\bfseries 36}, L205 (2003)},
  [\href{https://arxiv.org/abs/cond-mat/0212631}{{\ttfamily
  arXiv:cond-mat/0212631}}].

\bibitem{Latorre:2003kg}
J.~I. Latorre, E.~Rico and G.~Vidal, \textit{{Ground state entanglement in
  quantum spin chains}}, \href{http://dx.doi.org/10.26421/QIC4.1}{Quant. Inf.
  Comput. {\bfseries 4}, 48 (2004)},
  [\href{https://arxiv.org/abs/quant-ph/0304098}{{\ttfamily
  arXiv:quant-ph/0304098}}].

\bibitem{jin2004quantum}
B.-Q. Jin and V.~E. Korepin, \textit{{Quantum spin chain, Toeplitz determinants
  and the Fisher-Hartwig conjecture}},
  \href{http://dx.doi.org/10.1023/B:JOSS.0000037230.37166.42}{J. Stat. Phys.
  {\bfseries 116}, 79--95 (2004)},
  [\href{https://arxiv.org/abs/quant-ph/0304108}{{\ttfamily
  arXiv:quant-ph/0304108}}].

\bibitem{Korepin:2004zz}
V.~Korepin, \textit{{Universality of Entropy Scaling in One Dimensional Gapless
  Models}}, \href{http://dx.doi.org/10.1103/PhysRevLett.92.096402}{Phys. Rev.
  Lett. {\bfseries 92}, 096402 (2004)},
  [\href{https://arxiv.org/abs/cond-mat/0311056}{{\ttfamily
  arXiv:cond-mat/0311056}}].

\bibitem{Plenio:2004he}
M.~Plenio, J.~Eisert, J.~Dreissig and M.~Cramer, \textit{{Entropy,
  entanglement, and area: analytical results for harmonic lattice systems}},
  \href{http://dx.doi.org/10.1103/PhysRevLett.94.060503}{Phys. Rev. Lett.
  {\bfseries 94}, 060503 (2005)},
  [\href{https://arxiv.org/abs/quant-ph/0405142}{{\ttfamily
  arXiv:quant-ph/0405142}}].

\bibitem{Calabrese:2004eu}
P.~Calabrese and J.~L. Cardy, \textit{{Entanglement entropy and quantum field
  theory}}, \href{http://dx.doi.org/10.1088/1742-5468/2004/06/P06002}{J. Stat.
  Mech. (2004) P06002}, [\href{https://arxiv.org/abs/hep-th/0405152}{{\ttfamily
  arXiv:hep-th/0405152}}].

\bibitem{Cramer:2005mx}
M.~Cramer, J.~Eisert, M.~Plenio and J.~Dreissig, \textit{{An Entanglement-area
  law for general bosonic harmonic lattice systems}},
  \href{http://dx.doi.org/10.1103/PhysRevA.73.012309}{Phys. Rev. A {\bfseries
  73}, 012309 (2006)},
  [\href{https://arxiv.org/abs/quant-ph/0505092}{{\ttfamily
  arXiv:quant-ph/0505092}}].

\bibitem{Casini:2005rm}
H.~Casini, C.~Fosco and M.~Huerta, \textit{{Entanglement and alpha entropies
  for a massive Dirac field in two dimensions}},
  \href{http://dx.doi.org/10.1088/1742-5468/2005/07/P07007}{J. Stat. Mech.
  (2005) P07007}, [\href{https://arxiv.org/abs/cond-mat/0505563}{{\ttfamily
  arXiv:cond-mat/0505563}}].

\bibitem{Casini:2005zv}
H.~Casini and M.~Huerta, \textit{{Entanglement and alpha entropies for a
  massive scalar field in two dimensions}},
  \href{http://dx.doi.org/10.1088/1742-5468/2005/12/P12012}{J. Stat. Mech.
  (2005) P12012}, [\href{https://arxiv.org/abs/cond-mat/0511014}{{\ttfamily
  arXiv:cond-mat/0511014}}].

\bibitem{Casini:2009sr}
H.~Casini and M.~Huerta, \textit{{Entanglement entropy in free quantum field
  theory}}, \href{http://dx.doi.org/10.1088/1751-8113/42/50/504007}{J. Phys. A:
  Math. Gen. {\bfseries 42}, 504007 (2009)},
  [\href{https://arxiv.org/abs/0905.2562}{{\ttfamily arXiv:0905.2562}}].

\bibitem{Calabrese:2009qy}
P.~Calabrese and J.~Cardy, \textit{{Entanglement entropy and conformal field
  theory}}, \href{http://dx.doi.org/10.1088/1751-8113/42/50/504005}{J. Phys. A:
  Math. Gen. {\bfseries 42}, 504005 (2009)},
  [\href{https://arxiv.org/abs/0905.4013}{{\ttfamily arXiv:0905.4013}}].

\bibitem{peschel2009reduced}
I.~Peschel and V.~Eisler, \textit{Reduced density matrices and entanglement
  entropy in free lattice models},
  \href{http://dx.doi.org/10.1088/1751-8113/42/50/504003}{J. Phys. A: Math.
  Gen. {\bfseries 42}, 504003 (2009)},
  [\href{https://arxiv.org/abs/0906.1663}{{\ttfamily arXiv:0906.1663}}].

\bibitem{peschel2012special}
I.~Peschel, \textit{{Special review: Entanglement in solvable many-particle
  models}}, \href{http://dx.doi.org/10.1007/s13538-012-0074-1}{Braz. J. Phys.
  {\bfseries 42}, 267--291 (2012)},
  [\href{https://arxiv.org/abs/1109.0159}{{\ttfamily arXiv:1109.0159}}].

\bibitem{Casini:2004bw}
H.~Casini and M.~Huerta, \textit{{A Finite entanglement entropy and the
  c-theorem}}, \href{http://dx.doi.org/10.1016/j.physletb.2004.08.072}{Phys.
  Lett. B {\bfseries 600}, 142--150 (2004)},
  [\href{https://arxiv.org/abs/hep-th/0405111}{{\ttfamily
  arXiv:hep-th/0405111}}].

\bibitem{Furukawa:2008uk}
S.~Furukawa, V.~Pasquier and J.~Shiraishi, \textit{{Mutual Information and
  Compactification Radius in a c=1 Critical Phase in One Dimension}},
  \href{http://dx.doi.org/10.1103/PhysRevLett.102.170602}{Phys. Rev. Lett.
  {\bfseries 102}, 170602 (2009)},
  [\href{https://arxiv.org/abs/0809.5113}{{\ttfamily arXiv:0809.5113}}].

\bibitem{Casini:2008wt}
H.~Casini and M.~Huerta, \textit{{Remarks on the entanglement entropy for
  disconnected regions}},
  \href{http://dx.doi.org/10.1088/1126-6708/2009/03/048}{JHEP {\bfseries 03}
  (2009) 048}, [\href{https://arxiv.org/abs/0812.1773}{{\ttfamily
  arXiv:0812.1773}}].

\bibitem{Facchi:2008Entanglement}
P.~{Facchi}, G.~{Florio}, C.~{Invernizzi} and S.~{Pascazio},
  \textit{{Entanglement of two blocks of spins in the critical Ising model}},
  \href{http://dx.doi.org/10.1103/PhysRevA.78.052302}{Phys. Rev. A {\bfseries
  78}, 052302 (2008)}, [\href{https://arxiv.org/abs/0808.0600}{{\ttfamily
  arXiv:0808.0600}}].

\bibitem{Caraglio:2008pk}
M.~Caraglio and F.~Gliozzi, \textit{{Entanglement Entropy and Twist Fields}},
  \href{http://dx.doi.org/10.1088/1126-6708/2008/11/076}{JHEP {\bfseries 11}
  (2008) 076}, [\href{https://arxiv.org/abs/0808.4094}{{\ttfamily
  arXiv:0808.4094}}].

\bibitem{Casini:2009vk}
H.~Casini and M.~Huerta, \textit{{Reduced density matrix and internal dynamics
  for multicomponent regions}},
  \href{http://dx.doi.org/10.1088/0264-9381/26/18/185005}{Class. Quant. Grav.
  {\bfseries 26}, 185005 (2009)},
  [\href{https://arxiv.org/abs/0903.5284}{{\ttfamily arXiv:0903.5284}}].

\bibitem{Calabrese:2009ez}
P.~Calabrese, J.~Cardy and E.~Tonni, \textit{{Entanglement entropy of two
  disjoint intervals in conformal field theory}},
  \href{http://dx.doi.org/10.1088/1742-5468/2009/11/P11001}{J. Stat. Mech.
  (2009) P11001}, [\href{https://arxiv.org/abs/0905.2069}{{\ttfamily
  arXiv:0905.2069}}].

\bibitem{Alba:2009ek}
V.~Alba, L.~Tagliacozzo and P.~Calabrese, \textit{{Entanglement entropy of two
  disjoint blocks in critical Ising models}},
  \href{http://dx.doi.org/10.1103/PhysRevB.81.060411}{Phys. Rev. B {\bfseries
  81}, 060411 (2010)}, [\href{https://arxiv.org/abs/0910.0706}{{\ttfamily
  arXiv:0910.0706}}].

\bibitem{Igloi:2009On}
F.~{Igl{\'o}i} and I.~{Peschel}, \textit{{On reduced density matrices for
  disjoint subsystems}},
  \href{http://dx.doi.org/10.1209/0295-5075/89/40001}{EPL {\bfseries 89}, 40001
  (2010)}, [\href{https://arxiv.org/abs/0910.5671}{{\ttfamily
  arXiv:0910.5671}}].

\bibitem{Fagotti:2010yr}
M.~Fagotti and P.~Calabrese, \textit{{Entanglement entropy of two disjoint
  blocks in XY chains}},
  \href{http://dx.doi.org/10.1088/1742-5468/2010/04/P04016}{J. Stat. Mech.
  (2010) P04016}, [\href{https://arxiv.org/abs/1003.1110}{{\ttfamily
  arXiv:1003.1110}}].

\bibitem{Headrick:2010zt}
M.~Headrick, \textit{{Entanglement R\'enyi entropies in holographic theories}},
  \href{http://dx.doi.org/10.1103/PhysRevD.82.126010}{Phys. Rev. D {\bfseries
  82}, 126010 (2010)}, [\href{https://arxiv.org/abs/1006.0047}{{\ttfamily
  arXiv:1006.0047}}].

\bibitem{Calabrese:2010he}
P.~Calabrese, J.~Cardy and E.~Tonni, \textit{{Entanglement entropy of two
  disjoint intervals in conformal field theory II}},
  \href{http://dx.doi.org/10.1088/1742-5468/2011/01/P01021}{J. Stat. Mech.
  (2011) P01021}, [\href{https://arxiv.org/abs/1011.5482}{{\ttfamily
  arXiv:1011.5482}}].

\bibitem{Alba:2011fu}
V.~Alba, L.~Tagliacozzo and P.~Calabrese, \textit{{Entanglement entropy of two
  disjoint intervals in c=1 theories}},
  \href{http://dx.doi.org/10.1088/1742-5468/2011/06/P06012}{J. Stat. Mech.
  (2011) P06012}, [\href{https://arxiv.org/abs/1103.3166}{{\ttfamily
  arXiv:1103.3166}}].

\bibitem{Rajabpour:2011pt}
M.~A. Rajabpour and F.~Gliozzi, \textit{{Entanglement Entropy of Two Disjoint
  Intervals from Fusion Algebra of Twist Fields}},
  \href{http://dx.doi.org/10.1088/1742-5468/2012/02/P02016}{J. Stat. Mech.
  (2012) P02016}, [\href{https://arxiv.org/abs/1112.1225}{{\ttfamily
  arXiv:1112.1225}}].

\bibitem{Coser:2013qda}
A.~Coser, L.~Tagliacozzo and E.~Tonni, \textit{{On R\'enyi entropies of
  disjoint intervals in conformal field theory}},
  \href{http://dx.doi.org/10.1088/1742-5468/2014/01/P01008}{J. Stat. Mech.
  (2014) P01008}, [\href{https://arxiv.org/abs/1309.2189}{{\ttfamily
  arXiv:1309.2189}}].

\bibitem{DeNobili:2015dla}
C.~De~Nobili, A.~Coser and E.~Tonni, \textit{{Entanglement entropy and
  negativity of disjoint intervals in CFT: Some numerical extrapolations}},
  \href{http://dx.doi.org/10.1088/1742-5468/2015/06/P06021}{J. Stat. Mech.
  (2015) P06021}, [\href{https://arxiv.org/abs/1501.04311}{{\ttfamily
  arXiv:1501.04311}}].

\bibitem{Coser:2015dvp}
A.~Coser, E.~Tonni and P.~Calabrese, \textit{{Spin structures and entanglement
  of two disjoint intervals in conformal field theories}},
  \href{http://dx.doi.org/10.1088/1742-5468/2016/05/053109}{J. Stat. Mech.
  (2016) 053109}, [\href{https://arxiv.org/abs/1511.08328}{{\ttfamily
  arXiv:1511.08328}}].

\bibitem{Dupic:2017hpb}
T.~Dupic, B.~Estienne and Y.~Ikhlef, \textit{{Entanglement entropies of minimal
  models from null-vectors}},
  \href{http://dx.doi.org/10.21468/SciPostPhys.4.6.031}{SciPost Phys.
  {\bfseries 4}, 031 (2018)},
  [\href{https://arxiv.org/abs/1709.09270}{{\ttfamily arXiv:1709.09270}}].

\bibitem{Ruggiero:2018hyl}
P.~Ruggiero, E.~Tonni and P.~Calabrese, \textit{{Entanglement entropy of two
  disjoint intervals and the recursion formula for conformal blocks}},
  \href{http://dx.doi.org/10.1088/1742-5468/aae5a8}{J. Stat. Mech. (2018)
  113101}, [\href{https://arxiv.org/abs/1805.05975}{{\ttfamily
  arXiv:1805.05975}}].

\bibitem{Arias:2018tmw}
R.~E. Arias, H.~Casini, M.~Huerta and D.~Pontello, \textit{{Entropy and modular
  Hamiltonian for a free chiral scalar in two intervals}},
  \href{http://dx.doi.org/10.1103/PhysRevD.98.125008}{Phys. Rev. D {\bfseries
  98}, 125008 (2018)}, [\href{https://arxiv.org/abs/1809.00026}{{\ttfamily
  arXiv:1809.00026}}].

\bibitem{Brightmore:2019Entanglement}
L.~{Brightmore}, G.~P. {Geh{\'e}r}, A.~R. {Its}, V.~E. {Korepin},
  F.~{Mezzadri}, M.~Y. {Mo} and J.~A. {Virtanen}, \textit{{Entanglement entropy
  of two disjoint intervals separated by one spin in a chain of free fermion}},
  \href{http://dx.doi.org/10.1088/1751-8121/ab9cf2}{J. Phys. A {\bfseries 53},
  345303 (2020)}, [\href{https://arxiv.org/abs/1912.08658}{{\ttfamily
  arXiv:1912.08658}}].

\bibitem{Alba:2009th}
V.~Alba, M.~Fagotti and P.~Calabrese, \textit{{Entanglement entropy of excited
  states}}, \href{http://dx.doi.org/10.1088/1742-5468/2009/10/P10020}{J. Stat.
  Mech. (2009) P10020}, [\href{https://arxiv.org/abs/0909.1999}{{\ttfamily
  arXiv:0909.1999}}].

\bibitem{Alcaraz:2011tn}
F.~C. Alcaraz, M.~I. Berganza and G.~Sierra, \textit{{Entanglement of
  low-energy excitations in Conformal Field Theory}},
  \href{http://dx.doi.org/10.1103/PhysRevLett.106.201601}{Phys. Rev. Lett.
  {\bfseries 106}, 201601 (2011)},
  [\href{https://arxiv.org/abs/1101.2881}{{\ttfamily arXiv:1101.2881}}].

\bibitem{Berganza:2011mh}
M.~I. Berganza, F.~C. Alcaraz and G.~Sierra, \textit{{Entanglement of excited
  states in critical spin chians}},
  \href{http://dx.doi.org/10.1088/1742-5468/2012/01/P01016}{J. Stat. Mech.
  (2012) P01016}, [\href{https://arxiv.org/abs/1109.5673}{{\ttfamily
  arXiv:1109.5673}}].

\bibitem{Pizorn2012Universality}
I.~{Pizorn}, \textit{{Universality in entanglement of quasiparticle
  excitations}},  \href{https://arxiv.org/abs/1202.3336}{{\ttfamily
  arXiv:1202.3336}}.

\bibitem{Essler2013ShellFilling}
F.~H.~L. {Essler}, A.~M. {L{\"a}uchli} and P.~{Calabrese},
  \textit{{Shell-Filling Effect in the Entanglement Entropies of Spinful
  Fermions}}, \href{http://dx.doi.org/10.1103/PhysRevLett.110.115701}{Phys.
  Rev. Lett. {\bfseries 110}, 115701 (2013)},
  [\href{https://arxiv.org/abs/1211.2474}{{\ttfamily arXiv:1211.2474}}].

\bibitem{Berkovits2013Twoparticle}
R.~{Berkovits}, \textit{{Two-particle excited states entanglement entropy in a
  one-dimensional ring}},
  \href{http://dx.doi.org/10.1103/PhysRevB.87.075141}{Phys. Rev. B {\bfseries
  87}, 075141 (2013)}, [\href{https://arxiv.org/abs/1302.4031}{{\ttfamily
  arXiv:1302.4031}}].

\bibitem{Taddia:2013Entanglement}
L.~{Taddia}, J.~C. {Xavier}, F.~C. {Alcaraz} and G.~{Sierra},
  \textit{{Entanglement entropies in conformal systems with boundaries}},
  \href{http://dx.doi.org/10.1103/PhysRevB.88.075112}{{Phys. Rev. B} {\bfseries
  88}, 075112 (2013)}, [\href{https://arxiv.org/abs/1302.6222}{{\ttfamily
  arXiv:1302.6222}}].

\bibitem{Storms2014Entanglement}
M.~{Storms} and R.~R.~P. {Singh}, \textit{{Entanglement in ground and excited
  states of gapped free-fermion systems and their relationship with Fermi
  surface and thermodynamic equilibrium properties}},
  \href{http://dx.doi.org/10.1103/PhysRevE.89.012125}{Phys. Rev. E {\bfseries
  89}, 012125 (2014)}, [\href{https://arxiv.org/abs/1308.6257}{{\ttfamily
  arXiv:1308.6257}}].

\bibitem{Palmai:2014jqa}
T.~P\'almai, \textit{{Excited state entanglement in one dimensional quantum
  critical systems: Extensivity and the role of microscopic details}},
  \href{http://dx.doi.org/10.1103/PhysRevB.90.161404}{Phys. Rev. B {\bfseries
  90}, 161404 (2014)}, [\href{https://arxiv.org/abs/1406.3182}{{\ttfamily
  arXiv:1406.3182}}].

\bibitem{Calabrese:2014Entanglement}
P.~Calabrese, F.~H.~L. Essler and A.~M. Lauchli, \textit{{Entanglement
  Entropies of the quarter filled Hubbard model}},
  \href{http://dx.doi.org/10.1088/1742-5468/2014/09/P09025}{J. Stat. Mech.
  (2014) P09025}, [\href{https://arxiv.org/abs/1406.7477}{{\ttfamily
  arXiv:1406.7477}}].

\bibitem{Molter2014Bound}
J.~{M{\"o}lter}, T.~{Barthel}, U.~{Schollw{\"o}ck} and V.~{Alba},
  \textit{{Bound states and entanglement in the excited states of quantum spin
  chains}}, \href{http://dx.doi.org/10.1088/1742-5468/2014/10/P10029}{J. Stat.
  Mech. (2014) 10029}, [\href{https://arxiv.org/abs/1407.0066}{{\ttfamily
  arXiv:1407.0066}}].

\bibitem{Taddia:2016dbm}
L.~Taddia, F.~Ortolani and T.~P\'almai, \textit{{R\'enyi entanglement entropies
  of descendant states in critical systems with boundaries: conformal field
  theory and spin chains}},
  \href{http://dx.doi.org/10.1088/1742-5468/2016/09/093104}{J. Stat. Mech.
  (2016) 093104}, [\href{https://arxiv.org/abs/1606.02667}{{\ttfamily
  arXiv:1606.02667}}].

\bibitem{Castro-Alvaredo:2018dja}
O.~A. Castro-Alvaredo, C.~De~Fazio, B.~Doyon and I.~M. Sz\'ecs\'enyi,
  \textit{{Entanglement Content of Quasiparticle Excitations}},
  \href{http://dx.doi.org/10.1103/PhysRevLett.121.170602}{Phys. Rev. Lett.
  {\bfseries 121}, 170602 (2018)},
  [\href{https://arxiv.org/abs/1805.04948}{{\ttfamily arXiv:1805.04948}}].

\bibitem{Castro-Alvaredo:2018bij}
O.~A. Castro-Alvaredo, C.~De~Fazio, B.~Doyon and I.~M. Sz\'ecs\'enyi,
  \textit{{Entanglement content of quantum particle excitations. Part I. Free
  field theory}}, \href{http://dx.doi.org/10.1007/JHEP10(2018)039}{JHEP
  {\bfseries 10} (2018) 039},
  [\href{https://arxiv.org/abs/1806.03247}{{\ttfamily arXiv:1806.03247}}].

\bibitem{Murciano:2018cfp}
S.~Murciano, P.~Ruggiero and P.~Calabrese, \textit{{Entanglement and relative
  entropies for low-lying excited states in inhomogeneous one-dimensional
  quantum systems}}, \href{http://dx.doi.org/10.1088/1742-5468/ab00ec}{J. Stat.
  Mech. (2019) 034001}, [\href{https://arxiv.org/abs/1810.02287}{{\ttfamily
  arXiv:1810.02287}}].

\bibitem{Castro-Alvaredo:2019irt}
O.~A. Castro-Alvaredo, C.~De~Fazio, B.~Doyon and I.~M. Sz\'ecs\'enyi,
  \textit{{Entanglement content of quantum particle excitations. Part II.
  Disconnected regions and logarithmic negativity}},
  \href{http://dx.doi.org/10.1007/JHEP11(2019)058}{JHEP {\bfseries 11} (2019)
  058}, [\href{https://arxiv.org/abs/1904.01035}{{\ttfamily
  arXiv:1904.01035}}].

\bibitem{Castro-Alvaredo:2019lmj}
O.~A. Castro-Alvaredo, C.~De~Fazio, B.~Doyon and I.~M. Sz\'ecs\'enyi,
  \textit{{Entanglement Content of Quantum Particle Excitations III. Graph
  Partition Functions}}, \href{http://dx.doi.org/10.1063/1.5098892}{J. Math.
  Phys. {\bfseries 60}, 082301 (2019)},
  [\href{https://arxiv.org/abs/1904.02615}{{\ttfamily arXiv:1904.02615}}].

\bibitem{Jafarizadeh:2019xxc}
A.~Jafarizadeh and M.~Rajabpour, \textit{{Bipartite entanglement entropy of the
  excited states of free fermions and harmonic oscillators}},
  \href{http://dx.doi.org/10.1103/PhysRevB.100.165135}{Phys. Rev. B {\bfseries
  100}, 165135 (2019)}, [\href{https://arxiv.org/abs/1907.09806}{{\ttfamily
  arXiv:1907.09806}}].

\bibitem{Capizzi:2020jed}
L.~Capizzi, P.~Ruggiero and P.~Calabrese, \textit{{Symmetry resolved
  entanglement entropy of excited states in a CFT}},
  \href{http://dx.doi.org/10.1088/1742-5468/ab96b6}{J. Stat. Mech. (2020)
  073101}, [\href{https://arxiv.org/abs/2003.04670}{{\ttfamily
  arXiv:2003.04670}}].

\bibitem{You:2020osa}
Y.~You, E.~Wybo, F.~Pollmann and S.~Sondhi, \textit{{Observing Quasiparticles
  through the Entanglement Lens}},
  \href{https://arxiv.org/abs/2007.04318}{{\ttfamily arXiv:2007.04318}}.

\bibitem{Haque:2020Entanglement}
M.~{Haque}, P.~A. {McClarty} and I.~M. {Khaymovich}, \textit{{Entanglement of
  mid-spectrum eigenstates of chaotic many-body systems -- deviation from
  random ensembles}},  \href{https://arxiv.org/abs/2008.12782}{{\ttfamily
  arXiv:2008.12782}}.

\bibitem{Wybo:2020fiz}
E.~Wybo, F.~Pollmann, S.~L. Sondhi and Y.~You, \textit{{Visualizing
  quasiparticles from quantum entanglement for general one-dimensional
  phases}}, \href{http://dx.doi.org/10.1103/PhysRevB.103.115120}{Phys. Rev. B
  {\bfseries 103}, 115120 (2021)},
  [\href{https://arxiv.org/abs/2010.15137}{{\ttfamily arXiv:2010.15137}}].
  
\bibitem{Jurcevic:2014qfa}
P.~{Jurcevic}, B.~P. {Lanyon}, P.~{Hauke}, C.~{Hempel}, P.~{Zoller}, R.~{Blatt}
  and C.~F. {Roos}, \textit{{Quasiparticle engineering and entanglement
  propagation in a quantum many-body system}},
  \href{http://dx.doi.org/10.1038/nature13461}{Nature {\bfseries 511}, 202--205
  (2014)}, [\href{https://arxiv.org/abs/1401.5387}{{\ttfamily
  arXiv:1401.5387}}].

\bibitem{Jurcevic:2015inc}
P.~{Jurcevic}, P.~{Hauke}, C.~{Maier}, C.~{Hempel}, B.~P. {Lanyon}, R.~{Blatt}
  and C.~F. {Roos}, \textit{{Spectroscopy of Interacting Quasiparticles in
  Trapped Ions}}, \href{http://dx.doi.org/10.1103/PhysRevLett.115.100501}{Phys.
  Rev. Lett. {\bfseries 115}, 100501 (2015)},
  [\href{https://arxiv.org/abs/1505.02066}{{\ttfamily arXiv:1505.02066}}].

\bibitem{Deutsch:2013Microscopic}
J.~Deutsch, H.~Li and A.~Sharma, \textit{Microscopic origin of thermodynamic
  entropy in isolated systems},
  \href{http://dx.doi.org/10.1103/PhysRevE.87.042135}{Phys. Rev. E {\bfseries
  87}, 042135 (2013)}, [\href{https://arxiv.org/abs/1202.2403}{{\ttfamily
  arXiv:1202.2403}}].

\bibitem{Santos:2012Weak}
L.~F. {Santos}, A.~{Polkovnikov} and M.~{Rigol}, \textit{{Weak and strong
  typicality in quantum systems}},
  \href{http://dx.doi.org/10.1103/PhysRevE.86.010102}{Phys. Rev. E {\bfseries
  86}, 010102 (2012)}, [\href{https://arxiv.org/abs/1202.4764}{{\ttfamily
  arXiv:1202.4764}}].

\bibitem{Beugeling:2015Global}
W.~{Beugeling}, A.~{Andreanov} and M.~{Haque}, \textit{{Global characteristics
  of all eigenstates of local many-body Hamiltonians: participation ratio and
  entanglement entropy}},
  \href{http://dx.doi.org/10.1088/1742-5468/2015/02/P02002}{J. Stat. Mech.
  02002}, [\href{https://arxiv.org/abs/1410.7702}{{\ttfamily
  arXiv:1410.7702}}].

\bibitem{Garrison:2015lva}
J.~R. Garrison and T.~Grover, \textit{{Does a single eigenstate encode the full
  Hamiltonian?}}, \href{http://dx.doi.org/10.1103/PhysRevX.8.021026}{Phys. Rev.
  X {\bfseries 8}, 021026 (2018)},
  [\href{https://arxiv.org/abs/1503.00729}{{\ttfamily arXiv:1503.00729}}].

\bibitem{Page:1993df}
D.~N. Page, \textit{{Average entropy of a subsystem}},
  \href{http://dx.doi.org/10.1103/PhysRevLett.71.1291}{Phys. Rev. Lett.
  {\bfseries 71}, 1291--1294 (1993)},
  [\href{https://arxiv.org/abs/gr-qc/9305007}{{\ttfamily
  arXiv:gr-qc/9305007}}].

\bibitem{Vidmar:2017uux}
L.~Vidmar, L.~Hackl, E.~Bianchi and M.~Rigol, \textit{{Entanglement Entropy of
  Eigenstates of Quadratic Fermionic Hamiltonians}},
  \href{http://dx.doi.org/10.1103/PhysRevLett.119.020601}{Phys. Rev. Lett.
  {\bfseries 119}, 020601 (2017)},
  [\href{https://arxiv.org/abs/1703.02979}{{\ttfamily arXiv:1703.02979}}].

\bibitem{Vidmar:2017pak}
L.~Vidmar and M.~Rigol, \textit{{Entanglement Entropy of Eigenstates of Quantum
  Chaotic Hamiltonians}},
  \href{http://dx.doi.org/10.1103/PhysRevLett.119.220603}{Phys. Rev. Lett.
  {\bfseries 119}, 220603 (2017)},
  [\href{https://arxiv.org/abs/1708.08453}{{\ttfamily arXiv:1708.08453}}].

\bibitem{Huang:2019dxk}
Y.~Huang, \textit{{Universal eigenstate entanglement of chaotic local
  Hamiltonians}},
  \href{http://dx.doi.org/10.1016/j.nuclphysb.2018.09.013}{Nucl. Phys. B
  {\bfseries 938}, 594--604 (2019)},
  [\href{https://arxiv.org/abs/1708.08607}{{\ttfamily arXiv:1708.08607}}].

\bibitem{Vidmar:2018rqk}
L.~Vidmar, L.~Hackl, E.~Bianchi and M.~Rigol, \textit{{Volume Law and Quantum
  Criticality in the Entanglement Entropy of Excited Eigenstates of the Quantum
  Ising Model}}, \href{http://dx.doi.org/10.1103/PhysRevLett.121.220602}{Phys.
  Rev. Lett. {\bfseries 121}, 220602 (2018)},
  [\href{https://arxiv.org/abs/1808.08963}{{\ttfamily arXiv:1808.08963}}].

\bibitem{Hackl:2018tyl}
L.~Hackl, L.~Vidmar, M.~Rigol and E.~Bianchi, \textit{{Average eigenstate
  entanglement entropy of the XY chain in a transverse field and its
  universality for translationally invariant quadratic fermionic models}},
  \href{http://dx.doi.org/10.1103/PhysRevB.99.075123}{Phys. Rev. B {\bfseries
  99}, 075123 (2019)}, [\href{https://arxiv.org/abs/1812.08757}{{\ttfamily
  arXiv:1812.08757}}].

\bibitem{Zhang:2020vtc}
J.~Zhang and M.~A. Rajabpour, \textit{{Universal R\'enyi entanglement entropy
  of quasiparticle excitations}},
  \href{https://doi.org/10.1209/0295-5075/ac130e}{EPL in press},
  [\href{https://arxiv.org/abs/2010.13973}{{\ttfamily arXiv:2010.13973}}].

\bibitem{Lieb:1961fr}
E.~H. Lieb, T.~Schultz and D.~Mattis, \textit{{Two soluble models of an
  antiferromagnetic chain}},
  \href{http://dx.doi.org/10.1016/0003-4916(61)90115-4}{Annals Phys. {\bfseries
  16}, 407 (1961)}.

\bibitem{katsura1962statistical}
S.~Katsura, \textit{{Statistical mechanics of the anisotropic linear Heisenberg
  model}}, \href{http://dx.doi.org/10.1103/PhysRev.127.1508}{{Phys. Rev.}
  {\bfseries 127}, 1508 (1962)}.

\bibitem{pfeuty1970one}
P.~Pfeuty, \textit{{The one-dimensional Ising model with a transverse field}},
  \href{http://dx.doi.org/10.1016/0003-4916(70)90270-8}{Annals Phys. {\bfseries
  57}, 79 (1970)}.

\bibitem{Lee:2014nra}
C.~H. Lee, P.~Ye and X.-L. Qi, \textit{{Position-momentum duality in the
  entanglement spectrum of free fermions}},
  \href{http://dx.doi.org/10.1088/1742-5468/2014/10/P10023}{J. Stat. Mech.
  (2014) P10023}, [\href{https://arxiv.org/abs/1403.1039}{{\ttfamily
  arXiv:1403.1039}}].

\bibitem{Carrasco:2017eul}
J.~A. Carrasco, F.~Finkel, A.~Gonzalez-Lopez and P.~Tempesta, \textit{{A
  duality principle for the multi-block entanglement entropy of free fermion
  systems}}, \href{http://dx.doi.org/10.1038/s41598-017-09550-1}{Sci. Rep.
  {\bfseries 7}, 11206 (2017)},
  [\href{https://arxiv.org/abs/1701.05355}{{\ttfamily arXiv:1701.05355}}].

\bibitem{Zhang:2020ouz}
J.~Zhang and M.~A. Rajabpour, \textit{{Excited state R\'enyi entropy and
  subsystem distance in two-dimensional non-compact bosonic theory. Part I.
  Single-particle states}},
  \href{http://dx.doi.org/10.1007/JHEP12(2020)160}{JHEP {\bfseries 12} (2020)
  160}, [\href{https://arxiv.org/abs/2009.00719}{{\ttfamily
  arXiv:2009.00719}}].

\bibitem{Zhang:2020txb}
J.~Zhang and M.~A. Rajabpour, \textit{{Excited state R\'enyi entropy and
  subsystem distance in two-dimensional non-compact bosonic theory. Part II.
  Multi-particle states}},
  \href{http://dx.doi.org/10.1007/JHEP08(2021)106}{JHEP {\bfseries 08} (2021)
  106}, [\href{https://arxiv.org/abs/2011.11006}{{\ttfamily
  arXiv:2011.11006}}].
  
\bibitem{ZRDistance}
J.~Zhang and M.~A. Rajabpour, \textit{{Universal and nonuniversal subsystem
  distances in quasiparticle excited states}}, work in progress.

\end{thebibliography}

\providecommand{\href}[2]{#2}\begingroup\raggedright\endgroup

\end{document}